\documentclass[twocolumn,epjc3]{svjour3}          

\RequirePackage[T1]{fontenc}

\smartqed  

\RequirePackage{graphicx}
\RequirePackage{mathptmx}      
\RequirePackage{flushend}
\RequirePackage[numbers,sort&compress]{natbib}
\RequirePackage[colorlinks,citecolor=blue,urlcolor=blue,linkcolor=blue]{hyperref}

\journalname{Eur. Phys. J. A}
\usepackage{amsmath,enumitem,graphicx,epsfig,epstopdf}
\usepackage{gensymb}
\newcommand{\beqy}{\begin{eqnarray}}
\newcommand{\eeqy}{\end{eqnarray}}
\newcommand{\bmlet}{\begin{subequations}}
\newcommand{\emlet}{\end{subequations}}

\begin{document}

\def\gsimeq{\,\,\raise0.14em\hbox{$>$}\kern-0.76em\lower0.28em\hbox
{$\sim$}\,\,}
\def\lsimeq{\,\,\raise0.14em\hbox{$<$}\kern-0.76em\lower0.28em\hbox
{$\sim$}\,\,}
\def\ga{\,\,\raise0.14em\hbox{$>$}\kern-0.76em\lower0.28em\hbox
{$\sim$}\,\,}
\def\la{\,\,\raise0.14em\hbox{$<$}\kern-0.76em\lower0.28em\hbox
{$\sim$}\,\,}
\def\mass#1{${\mathrm{#1\, M}_\odot}$}
\def\Msun{$M_{\odot}$}
\def\be{\begin{equation}} 
\def\ee{\end{equation}}
\def\beqy{\begin{eqnarray}}
\def\eeqy{\end{eqnarray}}
\def\bmlet{\begin{mathletters}}
\def\emlet{\end{mathletters}}

\title{Reference Database for Photon Strength Functions}

\author{
S. Goriely \thanksref{iaa} \and P. Dimitriou \thanksref{iaea}  \and M. Wiedeking \thanksref{ithemba}
\and T. Belgya  \thanksref{has}
\and R. Firestone \thanksref{berk}
\and J. Kopecky \thanksref{juko}
\and M. Krti\v{c}ka \thanksref{charles}
\and V. Plujko \thanksref{kiev}
\and R. Schwengner  \thanksref{hzdr}
\and S. Siem \thanksref{oslo}
\and H. Utsunomiya  \thanksref{konan}
\and S. Hilaire \thanksref{bruyeres}
\and S. P\'eru \thanksref{bruyeres}
%
\and Y. S. Cho \thanksref{korea}
\and D. M. Filipescu \thanksref{elinp}
\and N. Iwamoto \thanksref{jaea}
\and T. Kawano \thanksref{lanl}
\and V. Varlamov \thanksref{lom}
\and R. Xu \thanksref{china}
}
\institute{
Institut d'Astronomie et d'Astrophysique, Universit\'e Libre de Bruxelles, Campus de la Plaine, CP 226, 1050 Brussels, Belgium\label{iaa}
\and
International Atomic Energy Agency, Wagramerstrasse 5, A-1400 Vienna, Austria\label{iaea}
\and
iThemba LABS, P.O. Box 722, Somerset West, 7129, South Africa\label{ithemba}
\and
Centre for Energy Research, Hungarian Academy of Sciences, Konkoly Thege Miklos 29-33, 1525 Budapest, Hungary \label{has}
\and
University of California, Berkeley CA 94720, USA \label{berk}
\and
JUKO Research, Kalmanstraat 4, Alkmaar 1817, The Netherlands \label{juko}
\and
Charles University, V Hole\v{s}ovi\v{c}k\'{a}ch 2, 18000 Prague, Czech Republic \label{charles}
\and
Taras Shevchenko National University, Faculty of Physics, Volodymyrska str. 60, Kyiv 01601, Ukraine \label{kiev}
\and
Helmholtz Zentrum Dresden-Rossendorf, Bautzner Landstrasse 400, 01328 Dresden, Germany \label{hzdr}
\and
University of Oslo, Sem Saelands vel 24, P.O. Box 1048, Oslo 0316, Norway \label{oslo}
\and
Konan University, Department of Physics, 8-9-1 Okamoto, Higashinada,  Kobe 658-8501, Japan \label{konan}
\and
CEA, DAM, DIF, F-91297 Arpajon, France \label{bruyeres}
\and
Nuclear Data Center, Korea Atomic Energy Research Institute, Daedeok-Daero 989-111, Yuseong-gu, Daejeon, Korea \label{korea}
\and
Horia Hulubei National Institute, 30 Reactorului St., P.O. Box MG-6 , 76900 Bucuresti-Magurele, Romania \label{elinp}
\and
Japan Atomic Energy Agency, 2-4 Shirakata, Tokai-mura, Naka-gun, Ibaraki 319-1195, Japan \label{jaea}
\and
Los Alamos National Laboratory, P.O. Box 1663 , Los Alamos NM 87545, USA \label{lanl}
\and
Lomonosov Moscow State University, Skobeltsyn Institute of Nuclear Physics, Leninskie Gory 1/2 , 119991 Moscow, Russian Federation \label{lom}
\and
China Institute of Atomic Energy, China Nuclear Data Center, P.O. Box 275-41, 102413 Beijing, China \label{china}
}

\date{\today}

\maketitle

\begin{abstract}
Photon strength functions describing the average response of the nucleus to an electromagnetic probe are key input information in the theoretical modelling of nuclear reactions. Consequently they are important for a wide range of fields such as nuclear structure, nuclear astrophysics, medical isotope production, fission and fusion reactor technologies. They are also sources of information for widely used reaction libraries such as the IAEA Reference Input Parameter Library and evaluated data files such as EGAF. 

In the past two decades, the amount of reaction gamma-ray data measured to determine photon strength functions has grown rapidly. Different experimental techniques have led to discrepant results and users are faced with the dilemma which (if any) of the divergent data to adopt.

We report on a coordinated effort to compile and assess the existing experimental data on photon strength functions from the giant dipole resonance region to energies below the neutron separation energy. The assessment of the discrepant data at energies around or below the neutron separation energy has been possible only in a few cases where adequate information on the model-dependent analysis and estimation of uncertainties was available. In the giant dipole resonance region, we adopt the recommendations of the new IAEA photonuclear data library.
We also present global empirical and semi-microscopic models that describe the photon strength functions in the entire energy region and reproduce reasonably well most of the experimental data. 

The compiled experimental photon strengths and recommended model calculations are available from the PSF database hosted at the IAEA (URL:www-nds.iaea.org/PSFdatabase).

\end{abstract}

%

\section{Introduction}
\label{sec_intro}

Photon strength functions (PSFs) describe the average response of the nucleus to an electromagnetic probe. They are  important quantities for the 
theoretical modelling of nuclear reactions. The PSF describing both the photoexcitation and deexcitation of the atomic nucleus by $\gamma$-ray 
absorption or emission plays a key role in all kinds of nuclear reactions where the electromagnetic interaction may compete with the strong or weak 
interactions. As a consequence, PSFs are also relevant sources of input information for other databases such as the photonuclear data library \cite{Photo00}, the Reference 
Input Parameter Library (RIPL) \cite{Capote09},  evaluated data files such as Evaluated Gamma Activation File (EGAF) \cite{EGAF}, Evaluated Nuclear 
Structure and Decay File (ENSDF) \cite{ENSDF}, and transport files in ENDF-6 format \cite{Trkov11}, which are supported by the International Atomic Energy Agency (IAEA). 
The concept of PSF stems from statistical physics and is based on the assumption that at high excitation energies the number of excited states, hence the nuclear 
level densities (NLDs), is high enough, so that  the nuclear decay properties can be treated statistically. In this respect, the PSF describes the 
average probability to absorb or emit a $\gamma$-ray of a given energy $E_\gamma$. Reaction theory relates the PSF to the photoabsorption cross 
section that is known to be dominated by the electric dipole ($E1$) radiation, at least in the high $\gamma$-ray  energy region of 10--20~MeV 
characterizing the well-known Giant Dipole Resonance (GDR) \cite{Photo00,Harakeh01}. Outside this energy region, especially below the particle 
separation energies, the magnetic dipole ($M1$) contribution may become significant.

Most of the PSF studies, be it experimental or theoretical, make the assumption that the average electromagnetic decay process ({\it i.e.} the photo-deexcitation) can be directly related to the inverse photoexcitation and essentially depends only on the energy of the emitted $\gamma$-ray, and not on the absolute excitation 
energy of the initial or final states, or the specific nuclear properties (such as the spin and parity) of the nuclear states involved. This 
assumption is known as the Brink-Axel hypothesis \cite{Brink55,Axel62} that has played a key role in the description of the photo-deexcitation
process, especially in reaction theory. 
While the Brink-Axel hypothesis is well established in the GDR energy region, at low energies, in particular 
below the neutron threshold, its validity is still open to debate and is under both theoretical as well as experimental investigation. 
For example, theoretically studies within the Fermi liquid theory \cite{Kadmenskii83} have found that photo-deexcitation PSF, traditionally denoted as $\overleftarrow{f}$, is a function of the excitation energy of the final state, which in turn depends on the excitation energy of the initial state and the $\gamma$-ray energy $E_\gamma$. In contrast, the photo-excitation process, with the PSF denoted as $\overrightarrow{f}$, only depends on the  $\gamma$-ray energy. 
At low excitation energies, such a temperature effect in the photo-deexcitation PSF was shown to be rather small \cite{Plujko99}, so that, in this energy regime, $\overleftarrow{f}(E_\gamma) \simeq \overrightarrow{f}(E_\gamma)$.
Experimentally, the Brink-Axel hypothesis was investigated and shown to be valid to a good approximation, for $\gamma$-ray transitions between states in the 
quasicontinuum region below the particle separation energy, from a variety of experiments, including those measuring average intensities of primary transitions from 
(n,$\gamma$) \cite{Bollinger67,Bollinger70}, (p,$\gamma$)  \cite{Szeflinska79,Szeflinski83,Erlandsson79}, and ($\gamma$,$\gamma_0$)  reactions \cite{Alarcon87}, or using data from two-step $\gamma$ cascades \cite{Krticka04} or charged-particle-induced reactions \cite{Guttormsen11,Guttormsen16,Martin17}. 
However, different experimental studies exploiting the photon scattering ($\gamma$,$\gamma^\prime$) technique have found indications that the Brink-Axel hypothesis is at least partially violated below the neutron separation energy (see {\it e.g.} \cite{Angell12,Isaak13}), 
including novel methods, using a combination of quasi-monochromatic photon beams and a $\gamma-\gamma$ coincidence setup which 
allows for the simultaneous determination of the photo-absorption and photo-deexcitation PSFs \cite{Isaak19}.

As already mentioned above, a large number of experiments have been devoted over the past decades to unraveling the electromagnetic response of the atomic nucleus providing a wealth of information on the total PSFs  and the relative contributions of the various components of given multipolarities ($L$) and types (electric or magnetic, 
 $X=E$ or $M$).
 Starting from the early 60s, significant effort was made to study PSFs using particle reactions such as (p,$\gamma$), (d,p) and other charge-exchange reactions. 
 A comprehensive review of all this work was published by Bartholomew {\it et al.}~\cite{Bartholomew73} which sets the principles of the 
 method for extracting PSFs from charged-particle-induced reactions.  
Ref.~\cite{Bartholomew73} also describes additional techniques yielding information on PSFs including ($\gamma$,$\gamma^\prime$) and (n,$\gamma$) measurements. Significant contribution from the (n,$\gamma$) reaction came from analysis of primary transitions following the decay of neutron resonances. Information on PSFs from these experiments were later detailed in Ref. \cite{Kopecky90}.
 Subsequently, a series of  coordinated international efforts to develop reliable PSFs for reaction modelling on the basis of all the available experimental information at the time, were  conducted at the IAEA.  The recommended PSFs (generically denoted as $f_{XL}(E_\gamma)$) for both the photoexcitation and photo-deexcitation processes were included in RIPL~\cite{Capote09} which has been widely used by the scientific community. 
In the past two decades, there has been considerable growth in the amount of reaction data measured to determine integrated PSFs using photon, neutron and 
charged-particle beams, with each method probing different or overlapping energy ranges and revealing interesting phenomena such as pygmy resonance strength, $M1$ scissors mode, 
and low-energy strength enhancement, often referred to as ``upbend''. Quite often the different experimental 
techniques used to extract PSFs lead to discrepant results and users are faced with the dilemma of trying to decide which (if any) amongst the divergent data they should adopt. It is therefore important that all these experimental data are compiled and assessed by experts who would then recommend the most reliable data for 
use in the various applications. 

To address these growing needs in PSFs, the IAEA held a consultants' meeting where experts reviewed the experimental methods and currently available PSF data \cite{Dimitriou13}, and recommended a coordinated effort to compile, assess and make recommendations to the user community. As a result, the IAEA organised a Coordinated Research Project (CRP) on ``Generating a Reference Database for Photon Strength 
Functions'' (2013-2019) \cite{Dimitriou15, Dimitriou18}. The objective of the project was to create a dedicated database for PSFs at relatively low excitation energies (typically below 30~MeV) which would include all available experimental data, a critical analysis of the discrepant data and recommendations to user community supported by 
global theoretical calculations. Three meetings were held during the CRP to monitor progress and revise the assignments in order to achieve the final objective \cite{Goriely16a,Goriely18c,Wiedeking19}.
The scope of the CRP included the following activities:
\begin{itemize}
\item	measurements,
\item	compilation of existing data,
\item	assessment and recommendation of data,
\item	global theoretical calculations,
\item comparison of models with the bulk of data,
\item	dissemination through an online data library.
\end{itemize}

In the present review paper we report on the work that was performed and the results that were obtained for all the above items. Specifically, 
Sec.~\ref{sec_exp} includes a description of the various experimental methods sensitive to PSFs and used to extract or test PSFs. 
Sec.~\ref{sec_ass} shows how, for each of the experimental methods, the PSFs have been assessed and extracted to be included in the final PSF library. Statistical and systematic uncertainties are also discussed in Sec.~\ref{sec_ass} in view of existing discrepancies between different experimental techniques. 
Sec.~\ref{sec_th} provides theoretical recommendations for a detailed and large-scale description of the dipole $E1$ and $M1$ strength functions. Two different theoretical approaches are considered, namely the phenomenological Lorentzian-type model and the more fundamental semi-microscopic quasi-particle random phase approximation (QRPA). 
A detailed comparison between experiments and theory is presented in Sec.~\ref{sec_comp}. Such a 
comparison allows us to test systematically the recommended models and validate their predictive power in the various energy regions of interest in applications, namely in the zero-energy limit, below the particle separation energy and in the GDR regime.
 Finally, in Sec.~\ref{sec_data}, the IAEA PSF database (URL: www-nds.iaea.org/PSFdatabase), including both experimental data (Sec.~\ref{sec_database_exp}) and theoretical predictions 
 (Sec.~\ref{sec_database_th}), is described and final recommendations are given. Both recommended PSF models are also compared for exotic neutron-deficient and neutron-rich nuclei in Sec.~\ref{sec_sys}. Conclusions are drawn in Sec.~\ref{sec_conc}.

\section{Experimental methods}
 \label{sec_exp}
Many experimental techniques have been used to obtain information on PSFs which are included in the database described in Secs.~\ref{sec_ass} and \ref{sec_data}. 
This section gives a short description of those  experimental methods as well as other 
techniques that were used to verify the PSF models (see Secs.~\ref{sec_comp}).

\subsection{Nuclear Resonance Fluorescence}
\label{sec_exp_nrf}

Photon scattering from nuclei, also called nuclear resonance fluorescence
(NRF), is a suitable tool to study dipole PSFs below the neutron-separation
energy. Nuclear states are excited from the ground state via absorption of
dipole ($L=1$) and, to a lesser extent, quadrupole ($L=2$) photons. NRF
experiments aim at the determination of the photoabsorption cross section
$\sigma_\gamma$ on an absolute scale. The PSF $f_{XL}$ is connected with $\sigma_\gamma$  via the relation 
\begin{equation}
\overrightarrow{f_{XL}} (E_\gamma) = \frac{\sigma_\gamma(J_x^\pi)}
                                       {g_J (\pi \hbar c)^2 E_\gamma^{2L-1}}, 
\label{eq_nrf1}
\end{equation}
with $g_J = (2J_x+1)/(2J_0+1)$, where $J_0$ and $J_x$ are the spins of the
ground and the excited states, respectively, and $\sigma_\gamma(J_x^\pi)$ corresponding to $\sigma_\gamma$ for states with specific spin and parity $J_x^\pi$. In photon scattering, the
energy-integrated scattering cross section $I_s = \int \sigma_{\gamma\gamma}~dE$
of an excited state at an energy $E_x$ can directly be deduced from the
intensity of the respective transition to the ground state: 
\begin{equation}
\label{eq:sigs}
I_s(E_x) = I_\gamma(E_\gamma,\theta) / \left[W(E_\gamma,\theta) \Phi_\gamma(E_x)  
N_{\rm at}\right] .
\end{equation}
Here, $I_\gamma(E_\gamma,\theta)$ denotes the intensity of a considered
ground-state transition at $E_\gamma$, observed at an angle $\theta$ relative
to the beam direction. $W(E_\gamma,\theta)$ describes the angular distribution
of this transition. The quantity $N_{\rm at}$ is the areal density of the atoms
in the target and $\Phi_\gamma(E_x)$ stands for the photon flux through the
target area at the energy of the considered level.

Spins of excited states can be deduced by comparing ratios of $\gamma$-ray
intensities, measured at two angles, with theoretical predictions. The optimum
combination comprises angles of 90$^\circ$ and 127$^\circ$ because the
respective ratios for the spin sequences $0-1-0$ and $0-2-0$ in even-even
nuclei differ most at these angles. The parities of excited states can be
derived from the polarizations of the ground-state transitions from experiments
using polarized $\gamma$-beams or Compton polarimeters.

The integrated scattering cross section is related to the partial radiative
width of the ground-state transition $\Gamma_0$ according to
\begin{equation}
\label{eq:gam}
I_s = \left(\frac{\pi \hbar c}{E_x}\right)^2 g_J
            \frac{\Gamma_0^2}{\Gamma_\gamma},
\end{equation}
where $\Gamma_\gamma$ is the total radiative width of the excited level. The
partial radiative width $\Gamma_0$ is proportional to the reduced transition
strength $B(XL)$ of a ground-state transition. These reduced
transition strengths can be deduced from spectra including well-isolated
transitions from low-lying states and have been the basis for the study of 
phenomena appearing up to excitation energies of about 3 MeV, such as couplings
of quadrupole and octupole states \cite{Kneissl06} and the scissors
mode \cite{Heyde10}.

In experiments with high-energy $\gamma$ beams, the determination of the
absorption cross section $\sigma_\gamma$ and consequently the PSF is 
complicated by the following problems. First, a high-lying excited state can
deexcite to low-lying excited states (inelastic scattering) in addition to the
direct deexcitation to the ground state (elastic scattering). In the case of
inelastic scattering, inelastic and subsequent cascade transitions appear in
the measured spectrum in addition to ground-state transitions. To deduce the
absorption cross section from the elastic scattering cross section, which is
proportional to the measured intensity, one needs to know the branching ratio
$b_0$ of the ground-state transition:
$\sigma_{\gamma} =  \sigma_{\gamma\gamma}/b_0$. 
This branching ratio is also needed, if one is interested in the partial
radiative width of a ground-state transition $\Gamma_0$ to deduce $E1$ or $M1$
transition strengths.
The branching ratio appears as the quantity $b_0 = \Gamma_0 / \Gamma_\gamma$ in
Eq.~\ref{eq:gam}. 

In experiments using quasi-monoenergetic photons, which have mainly 
been performed at the High-Intensity $\gamma$-ray Source (HI$\gamma$S) 
\cite{Weller09} of the Triangle Universities Nuclear Laboratory (TUNL) in
Durham, the branching ratios $b_0$ may be estimated from the intensities of the
ground-state transitions in the excited energy window and the intensities of
the transitions depopulating the lowest-lying states ($2^+$ states in 
even-even nuclei), which collect the intensities of most of the inelastic
transitions \cite{Tonchev10,Massarczyk14}. Alternatively, $\gamma-\gamma$
coincidence measurements performed at HI$\gamma$S have been used to determine
relative branching ratios $\Gamma_i/\Gamma_0$, where $i$ denotes an inelastic
transition (see, for example Refs.~\cite{Loeher16,Wilhelmy18}). 

In experiments using broad-band bremsstrahlung up to high energy, such as the
ones performed at the $\gamma$ELBE facility \cite{Schwengner05} of the 
Helmholtz-Zentrum Dresden-Rossendorf (HZDR), a great number of levels is excited
in a wide energy range. The inelastic transitions from high-lying levels can
feed a considered level well below the end-point energy of the bremsstrahlung,
which is a further complication. In the case of such a feeding, the measured
intensity of the ground-state transition is greater than the one resulting from
a direct excitation only. As a consequence, the integrated scattering cross
section $I_{s+f}$ deduced from this intensity contains a portion $I_f$
originating from feeding in addition to the true integrated scattering cross
section $I_s$, (Eq.~\ref{eq:gam}). The problem of feeding can partly be solved
by measuring at several bremsstrahlung end-point energies and considering
transitions close to the respective end-point energies only. 
An alternative is the estimate of intensities of inelastic transitions using
codes for the simulation of statistical $\gamma$ cascades, namely MCGCS
\cite{Rusev08} and $\gamma$DEX \cite{Massarczyk12}, which are analogous to the DICEBOX code
\cite{Becvar98, Krticka19a} used mainly for neutron-capture reactions.

In experiments that populate states at high excitation energy and therefore
high level density, a number of weak transitions may not be resolved, but they 
rather form a quasicontinuum. To take into account the full intensity of all
transitions in the determination of the cross sections, various attempts have
been made to estimate the intensity in the quasicontinuum, which has to be
separated from the intensity appearing from atomic scattering processes in the
target. This ``atomic background'' can for example be simulated using codes in
GEANT4~\cite{Agostinelli03} and subtracted from the experimental spectrum.
The remaining nuclear spectrum including resolved peaks and quasicontinuum is
then used for further analysis. Such analyses are described for example in
Refs.~\cite{Rusev08,Massarczyk12} for experiments with bremsstrahlung at the 
$\gamma$ELBE facility and in Refs.~\cite{Tonchev10,Massarczyk14} for experiments
with quasi-monoenergetic beams at the HI$\gamma$S facility.

\subsection{The Oslo method}
\label{sec_exp_oslo}

The Oslo method is a technique which allows for the simultaneous extraction of the NLD and PSF from particle-$\gamma$ coincidence data and is described in detail in Ref.~\cite{Schiller00}. This method probes the PSF below the neutron separation energy.

Until recently, all experiments have been performed at the Oslo Cyclotron Laboratory (OCL) using proton, deuteron, $^3$He or alpha beams on isotopically-enriched targets and 
the experimental setup at OCL. 
{\bf
There, the energy of the outgoing charged particles are measured with the Silicon Ring particle-detector array \cite{Guttormsen11}  which can be placed in forward or backward angles with respect to the beam direction. The emitted $\gamma$-rays are measured with the CACTUS array \cite{Guttormsen96}, consisting of 28 collimated 5" by 5" NaI(Tl) detectors and more recently with the upgraded OSCAR array consisting of 30 large-volume (89 mm x 203 mm) LaBr$_3$:Ce detectors.

The particle-$\gamma$ coincidence data are sorted into a matrix of initial excitation energy $E_i$ versus $\gamma$-ray energy $E_\gamma$. 
}
For each excitation energy bin the $\gamma$-ray spectra are unfolded \cite{Guttormsen96} using the response functions of the detectors. From these unfolded $\gamma$-ray spectra, the distribution of primary $\gamma$-rays was obtained for each excitation energy bin by means of an iterative subtraction technique, known as the first-generation method \cite{Guttormsen87}. Here, the main assumptions are that the Brink hypothesis \cite{Brink55} is valid and that the $\gamma$-decay routes from a given excitation energy are independent on whether it was populated directly in the reaction, or through $\gamma$-ray decay from above-lying states (see Ref.~\cite{Larsen11} for a discussion on the uncertainties for the subtraction technique). 

\begin{figure}
\begin{center}
\includegraphics[scale=0.5]{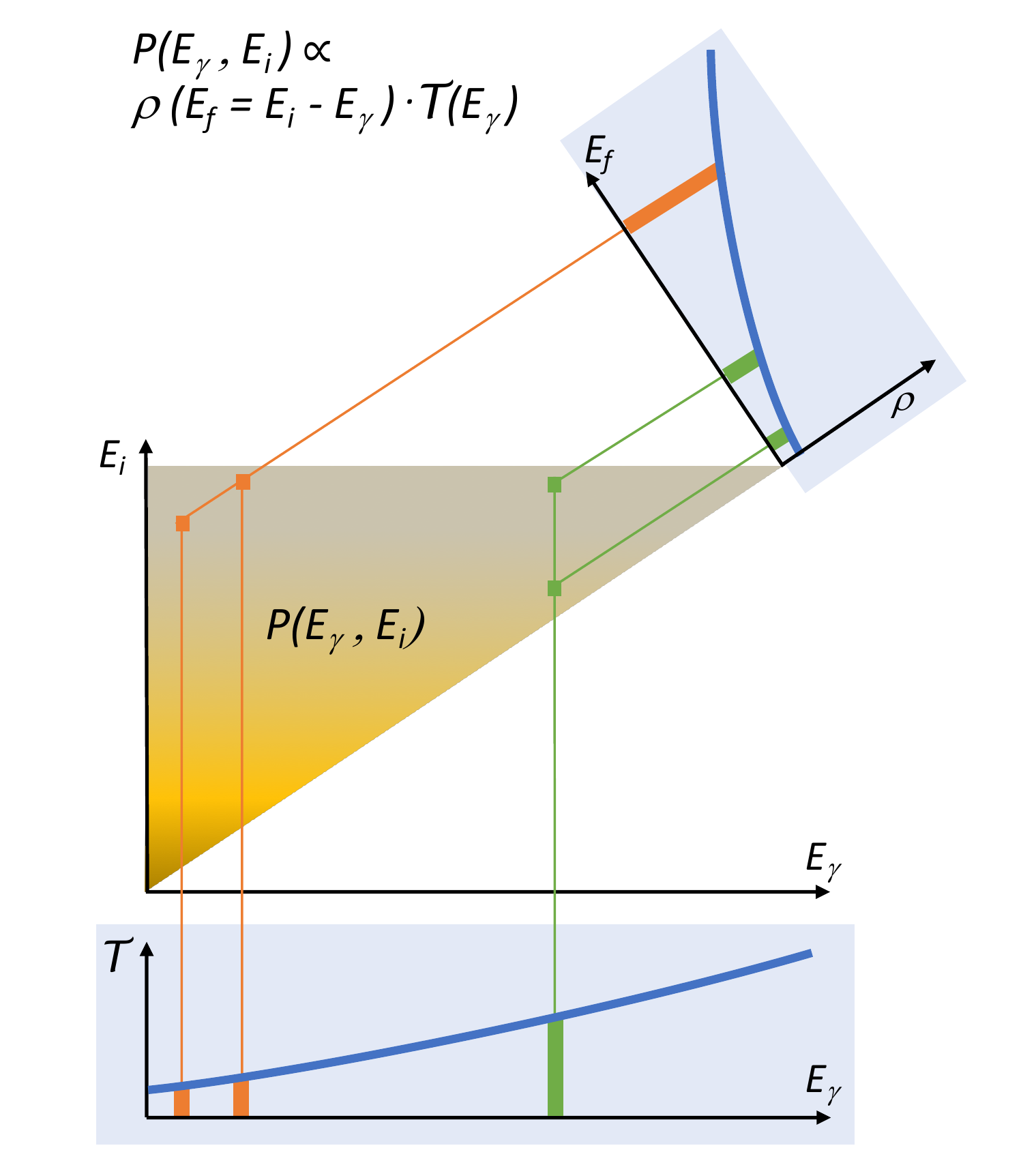}  
\caption{(Color online) Schematic representation on how  NLDs and PSFs are extracted from the primary $\gamma$-ray spectrum. The first-generation $\gamma$-ray distribution (yellow triangle) is given by the product of the level density $\rho(E_i-E_\gamma)$ and the $\gamma$-ray transmission coefficient $T_\gamma(E_\gamma)$. All values of the elements of the $\rho(E_i-E_\gamma)$ and  $T_\gamma(E_\gamma)$ vectors are allowed to vary in order to give the best fit to the $P(E_\gamma,E_i)$  landscape.}
\label{fig_exp_oslo}
\end{center}
\end{figure}

The NLD $\rho(E_f)$ at the excitation energy $E_f=E_i-E_\gamma$ and the total $\gamma$-ray transmission coefficient, $T_\gamma(E_\gamma) = T_{M1}+T_{E1}$ (assuming dipole transitions dominate), are related to the primary $\gamma$-ray spectrum by
\begin{equation}
P(E_\gamma,E_i) \propto \rho(E_i - E_\gamma) T_\gamma(E_\gamma),
\label{eq:PrimaryMat}
\end{equation}
as also illustrated in Fig.~\ref{fig_exp_oslo}, where the energy distribution $P(E_\gamma,E_i)$ of the first-generation $\gamma$-transitions is shown. Since no $\gamma$-rays are emitted with energy larger than the initial excitation energy, the matrix appears as a triangle. The $P(E_\gamma,E_i)$ landscape is assumed to be described by the product of the level density and the $\gamma$ transmission coefficient. Thus, one value for the level density (orange) is based on the values of the transmission coefficient in a certain $\gamma$-ray energy range. Analogous, one value of the transmission coefficient (green) is determined by the NLD in a corresponding excitation-energy range.
$T_\gamma(E_\gamma)$ and $\rho(E_f)$ are extracted with a $\chi^2$ fit \cite{Schiller00} yielding the unique solution of the functional shape of $\rho(E_f)$ and $T_\gamma(E_\gamma)$. Furthermore, a normalization to known experimental level data is performed to establish the correct slope and absolute values of the NLD and total dipole PSF.

The extraction is limited to the $E_\gamma$ and the excitation energy region of the primary $\gamma$-ray matrix where the decay is assumed to be statistical. 
The NLD is normalized by comparing with the known discrete levels at low excitation energy, and then by extrapolation using a 
constant temperature (CT) form \cite{Gilbert65} from the highest excitation energy deduced from the Oslo method to the neutron separation energy $S_n$. 
The NLD at $S_n$ is determined from the average neutron resonance spacing $D_0$ and the spin cutoff parameter $\sigma$ in a process detailed in Ref. \cite{Schiller00}. 
The transmission coefficients are normalized to the average total radiative width $\langle \Gamma_\gamma \rangle$ of neutron resonances, as described in 
Ref.~\cite{Voinov01}, and converted to the total dipole PSF $f_1$, which includes  both the $E1$ and $M1$ contributions, by
\begin{equation}
\overleftarrow{f_1}(E_\gamma) = T_\gamma(E_\gamma)/(2\pi E_\gamma^3).
\label{eq_dipole_psf}
\end{equation}
$f_1$ in Eq.~\ref{eq_dipole_psf}, is a special case of the more general $f_{XL}$, since in this method $L=1$ dominance is assumed and $X$ cannot be distinguished, therefore what is measured is the sum $f_1=f_{E1}+f_{M1}$. In cases where $\langle \Gamma_\gamma \rangle$ and/or $D_0$ are not available, systematics from a suitable mass region or neighboring nuclei is often used, see for example Refs. \cite{Brits19,Kheswa17,Larsen16}.

Recently, the Oslo method has been further developed to allow for the study of the NLDs and PSFs in more neutron-rich nuclei, either 
via the analysis of experimental data following beta-decay, the so-called beta-Oslo method~\cite{Spyrou14,Spyrou17,Larsen18} or using inverse kinematic 
reactions ~\cite{Ingeberg16}. It is important to note that both of these newly developed methods measure particle-$\gamma$ coincidences and use these coincidences to obtain excitation energy $E_i$ versus $\gamma$-ray energy matrices to which the Oslo method is applied. 

\subsection{Neutron resonance capture data}
\label{sec_exp_arc}

During the 1960's-1990's the resonance behaviour of neutron interaction with matter was studied in many laboratories worldwide using both, the white neutron spectra and time-of-flight (TOF) techniques, allowing to measure properties of individual neutron resonances. A significant fraction 
of these studies measured $\gamma$-ray spectra that were primarily used as a spectroscopy tool for determining properties of neutron resonances as well as levels at low-excitation energy of the residual nuclides. 
However, in some cases, the $\gamma$-decay properties of different radiation types $XL$ were exploited for obtaining information on $f_{XL}$ in the $E_\gamma$ range between $S_n$ and $S_n - 2$ or 3 MeV.  

The compound nucleus mechanism for neutron capture is a dominant process up to several MeV of incident neutron energy. Therefore, the statistical model is generally used to describe $\gamma$-ray decay at these energies. An exception to this can occur in thermal or resonance capture in certain mass regions, where non-statistical processes may become important.

The derived $ f_{XL}(E_\gamma)$ data are based on the experimental determination of the partial radiative width $\Gamma_{\gamma f}$ from measured primary $\gamma$-ray intensities. Two types of experiments are usually considered, {\it i)} the capture on isolated resonances using TOF spectrometry, known as the Discrete Resonance Capture (DRC), and {\it ii)} the average resonance capture (ARC) with filtered neutron beams. Three filter materials, $^{10}$B, $^{45}$Sc or $^{56}$Fe, have been used for ARC experiments. The beams are produced by transmission through filter materials, which yield neutron beams with bell-shaped energy distributions at mean neutron energies of about 150~eV, 2~keV and 24~keV, respectively. The boron-filtered beam primarily removes the thermal component, while Sc and Fe yield quasi-monoener\-getic beams of a 1--2~keV width as a result of the presence of a maxima in the transmission of neutrons through these elements/isotopes. Such facilities were built in four laboratories in the US, namely Argonne National Laboratory (ANL), the National Bureau of Standards, the Idaho Nuclear Engineering Laboratory INEL and Brook\-haven National Laboratory (BNL)  between 1970 and 1980. Three other laboratories, IAEP/PPEI Obninsk (Russia),  Kiev (Ukraine) and KFK Karlsruhe (Germany), have also published ARC data. 
The majority of all adopted data originates from BNL due to its high neutron fluence and efficient processing tools.


Common to all these experiments is the necessity to average over Porter-Thomas fluctuations \cite{Porter56} which are expected to govern the distribution of 
partial radiative widths. In the DRC experiments the differential data are numerically averaged over measured isolated resonances 
to decrease the influence of these Porter-Thomas fluctuations, while in the ARC experiments the averaging is inherent in the experiment since what is 
measured is the capture on neutron resonances present in the filtered beam neutron window. The DRC are given 
in the absolute PSF scale and can be used for the absolute normalization of the ARC data (which are in all cases given only in relative units).
 
The individual strength corresponding to primary transitions from resonances with a given parity to individual final levels with the same parity connected via transitions of $XL$ type, $f_{XL,f}$, was determined for a number of different energies as
\begin{equation}
 f_{XL,f}(E_\gamma) = \frac{ \langle \Gamma_{\gamma f} / E_\gamma^{(2L+1)} \rangle}{D},
\label{eq_t_to_f}
\end{equation}
where $\Gamma_{\gamma f}$ is a partial radiation width of a transition with $E_\gamma$ corresponding to the energy difference between the 
initial state and a final level $f$. The symbol $\langle \,\rangle$ stands for unweighted averaging over included resonances and 
$D$ is the $s$- or $p$-wave resonance spacing for resonances with given spin and parity. The spread of individual $f_{XL,f}$ values is assumed to be primarily but not complete suppression of the Porter-Thomas fluctuations, or other effects such as $p$-wave contributions, and is taken care of in the data processing \cite{Kopecky16,Kopecky17,Kopecky17b,Kopecky18}.

\begin{figure}
\includegraphics[scale=0.3]{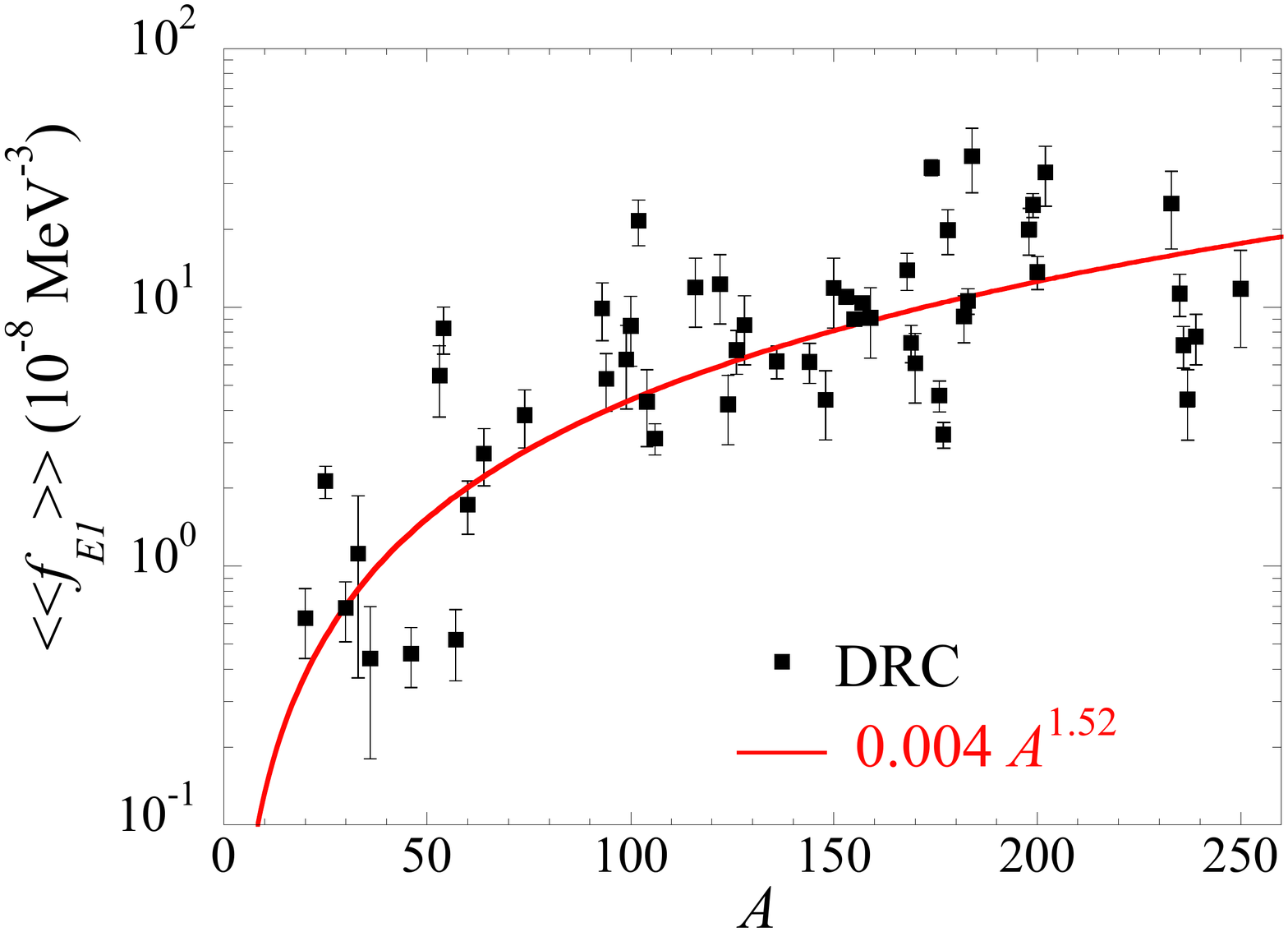}  
\caption{\label{fig_DRCsys}(Color online) Comparison between the DRC quasi-monoenergetic doubly average strength functions $\langle \langle f_{E1} \rangle \rangle$ at $\langle E_\gamma \rangle= 6.5 \pm 0.5$~MeV (squares) with the least-square analysis systematic (Eq.~\ref{eq_drc1}) as a function of the atomic mass $A$ (for details see Ref.~\cite{Kopecky18}). The quoted errors are the statistical error increased by $\Gamma_\gamma$ and $D$ uncertainty estimates of 10\%.}
\end{figure}

In order to increase the statistical accuracy of DRC data, the averaged quasi-monoenergetic strength function was introduced, involving 
an additional averaging step, and was implemented in all the previous compilations \cite{McCullagh79,McCullagh81,Kopecky94}. The average is 
applied over a selected number of primary transitions in the narrow energy region, neglecting the additional energy dependence above the 
phase factor of partial widths. For an energy range of about 1 MeV, this is an acceptable assumption. The average strength function can therefore be expressed as
\begin{equation}
\langle f_{XL}(E_{\gamma)} \rangle=\frac{\langle \langle \Gamma_{\gamma f}/ E_{\gamma }^{(2L+1)} \rangle \rangle}{D}, 
\label{eq_darc}
\end{equation}
where $\langle \langle \Gamma_{\gamma f}/ E_{\gamma}^{(2L+1)} \rangle \rangle$  is an unweighted mean over the used primary transitions and 
included resonances. Eq.~\ref{eq_darc} is valid for both DRC and ARC data, though the averaging over resonances is implicit in the
experimental process in the ARC case but needs to be performed explicitly in the DRC case.
 These estimates of $\langle f(E_\gamma) \rangle$ obtained from DRC are then used in the absolute normalization of the ARC data. In order to have such information 
 also for nuclides without available DRC measurements, mass-dependent systematics (as power function of $A$) have been derived for both $E1$ and $M1$ 
 transitions. The case of the $E1$ radiation (used for normalization) is shown in Fig.~\ref{fig_DRCsys}. The absolute majority of data lies between $6 -7$~MeV, except for the low-mass and actinide nuclides. The least-square-fit based on a power dependence on the mass $A$ leads to the following systematics for the $E1$ and $M1$ PSFs  
\begin{eqnarray}
& & \langle \langle f_{E1} \rangle \rangle = 0.004~A^{1.52\pm0.21} \label{eq_drc1} \\
& & \langle \langle f_{M1} \rangle \rangle = 0.012~A^{0.49\pm 0.21}, \label{eq_drc2} 
\end{eqnarray}
where the measure of the goodness of the fit can be estimated as $\chi^2=0.6$ for the $E1$ and 0.13 for the $M1$. More details can be found in Refs.~\cite{Kopecky17,Kopecky17b}.

\subsection{Primaries from thermal neutron capture}
\label{sec_exp_egaf}

Thermal/cold neutron beams are produced by nickel lined guides that transport the neutrons to low-background counting stations far from the neutron source.  If the guides are curved, no fast neutrons and $\gamma$-rays coming from the source reach the target area. At the Budapest reactor facility, where many experiments have been performed, cold and thermal neutron beams are transported to the Prompt Gamma-ray Activation Analysis (PGAA) target station approximately 35~m from the reactor wall~\cite{Rosta02}. 
A similar experimental configuration has been constructed at the Garching FRM-2 reactor \cite{Revay15}.

Gamma rays following decay of the thermal neutron capture are measured with the help of semiconductor detectors and relative $\gamma$-ray intensities per neutron capture can be often deduced from these measurements. At the Budapest reactor, the relative intensities are obtained from a comparison to known absolute cross sections of individual transitions -- that are obtained using stoichiometric compounds or mixtures containing $\gamma$-ray cross section standards such as H, N, Cl, S, Na, Ti, or Au \cite{Revay03} -- and from the thermal cross sections for (n,$\gamma$) reactions. 
The partial $\gamma$-ray cross section values have been compiled in the EGAF library \cite{EGAF,Firestone14} for all elements with $Z = 1-83$, 90, 92 except for He and Pm. 

In nuclei with sufficiently high level density, the total radiative width of individual neutron resonances show very small fluctuations. Relative $\gamma$-ray intensities per neutron capture can thus be converted to partial radiative widths to individual final levels $\Gamma_{\gamma f}$ via $\Gamma_{\gamma f} = P_\gamma  \langle \Gamma_{\gamma} \rangle$. Partial radiative widths of primary transitions can be used to obtain information on PSFs similarly as in the DRC approach. The difference between the DRC data and thermal neutron capture is that in the latter the averaging can be done only for different final levels $f$ in a selected range $\Delta E_\gamma$, but not over different initial resonances. It means that the individual $f_{XL,f}$ values are obtained only from a single value of $\Gamma_{\gamma f}$ and not from averaging indicated by $\langle \rangle$ in Eq.\ref{eq_darc}. 

Absence of averaging over different resonances leads to smaller suppression of the Porter-Thomas fluctuations than in the case of DRC/ARC data. In addition, these fluctuations prevent observation of primary transitions to all levels in the range $\Delta E_\gamma$ and a correction for unobserved transitions has to applied. This correction can be done under the assumption that the observed transitions are the strongest ones in the $\Delta E_\gamma$ energy range and that the number of final levels $f$ accessible via transitions of $XL$ type in this range is known. However, as thermal/cold neutron capture proceeds purely via $s$-wave neutrons, the capturing state has a unique parity and the $XL$ types of primary $\gamma$-rays populating final levels of known spin and parity can be directly inferred.

\subsection{Average resonance proton capture}
\label{sec_exp_pg}

Measurements from (p,$\gamma$) reactions to deduce the PSFs are similar to the ARC method introduced in 
Sec.~\ref{sec_exp_arc} where many different resonances are populated in the capture reaction. Using high-resolution $\gamma$-ray detectors 
to detect $\gamma$-rays deexciting these resonances allows one to identify the primary transitions connecting to 
low-lying final levels. There are also similarities between the extraction of data from (p,$\gamma$) measurements and from those in the Ratio method shown in Fig.\ref{fig_exp_RM}.  However, for (p,$\gamma$) reactions the excitation-energy is determined by the 
proton beam energy.

The measurements were typically performed for proton energies $E_p$ ranging between 1 and 4 MeV. The number of resonances in each measurement 
is usually determined by the thickness of the target. Typically, targets that cause $10-50$~keV energy loss of the incoming 
proton energies were used. The proton energy loss in the target also determines the width of the $\gamma$ line observed in the detector. The excitation energy 
resolution limits the applicability of the method to nuclei for which the spacing of individual final low-lying levels is at least several tens 
of keV. Another factor that needs to be considered is that with an increasing Coulomb barrier, (p,$\gamma$) cross sections decrease. Hence, the method is suitable for nuclei 
with $A \lsimeq 90$ for which (p,$\gamma)$ cross sections can be measured with good statistics. On the other hand, the need for sufficiently high NLD in the resonance region to suppress expected 
Porter-Thomas fluctuations of individual transition intensities requires that nuclei with masses $A \gsimeq 50$ are used. 
Another factor to consider is that the method works when the neutron separation energy in the product nucleus is much higher than the proton separation energy, {\it i.e.} for nuclei such that $Q_{p,n}<-E_p$, where $E_p$ is the maximum proton energy.

To suppress the influence of the Porter-Thomas fluctuations, the average intensities of primary $\gamma$ transitions to a specified final low-lying level 
were extracted for proton energies within (typically) $\Delta E_p = 0.5-1.2$ MeV wide interval. 
Intensities of transitions to the same final level were summed together for all proton energies in the $\Delta E_p$ range and this sum of intensities was attributed to the $\gamma$ energy at the middle of the $\Delta E_p$ range. Using transitions to levels with the same spin and parity, the relative $E_\gamma$ dependence of the PSFs was thus obtained; for details see Refs. \cite{Szeflinska79,Erlandsson79}. 

Levels of different spins have often been considered. However, to get the same normalization of data sets for different spins,  a correction 
is needed for feeding from resonances with different spins. These different contributions were usually calculated within the Hauser-Feshbach 
formalism. Absolute normalization of measured intensities to the PSF is determined from a comparison of measured cross sections for the direct population of selected low-lying states 
(one or a few) using the Hauser-Feshbach calculations; the $\gamma$-ray transmission coefficient $T_\gamma(E_\gamma)$ (for a single $\gamma$-ray energy) 
was the only quantity in the simulation of the cross section which was assumed to be unknown and its value needed for reproducing the cross 
section yielded the absolute PSF \cite{Szeflinska79,Erlandsson79}. Obtained PSF values correspond to the total dipole PSF $f_1$.

\subsection{The Ratio method}
\label{sec_exp_ratio}

The Ratio method \cite{Wiedeking12} is a model-independent approach to obtain the energy dependence of the PSF. The method relies on the detection and 
extraction of correlated particle-$\gamma$-$\gamma$ events from reactions for which the excitation energy of the residual nucleus can be experimentally determined. 
Charged particles and their energies are detected in particle detectors ({\it e.g.} silicon particle telescopes \cite{Wiedeking12}, phoswhich detectors \cite{Jones18} or similar) and their kinematics and knowledge of energy losses allows for the determination of the excitation energy of the residual nucleus which is produced in the reaction.

\begin{figure}
\begin{center}
\includegraphics[scale=0.33]{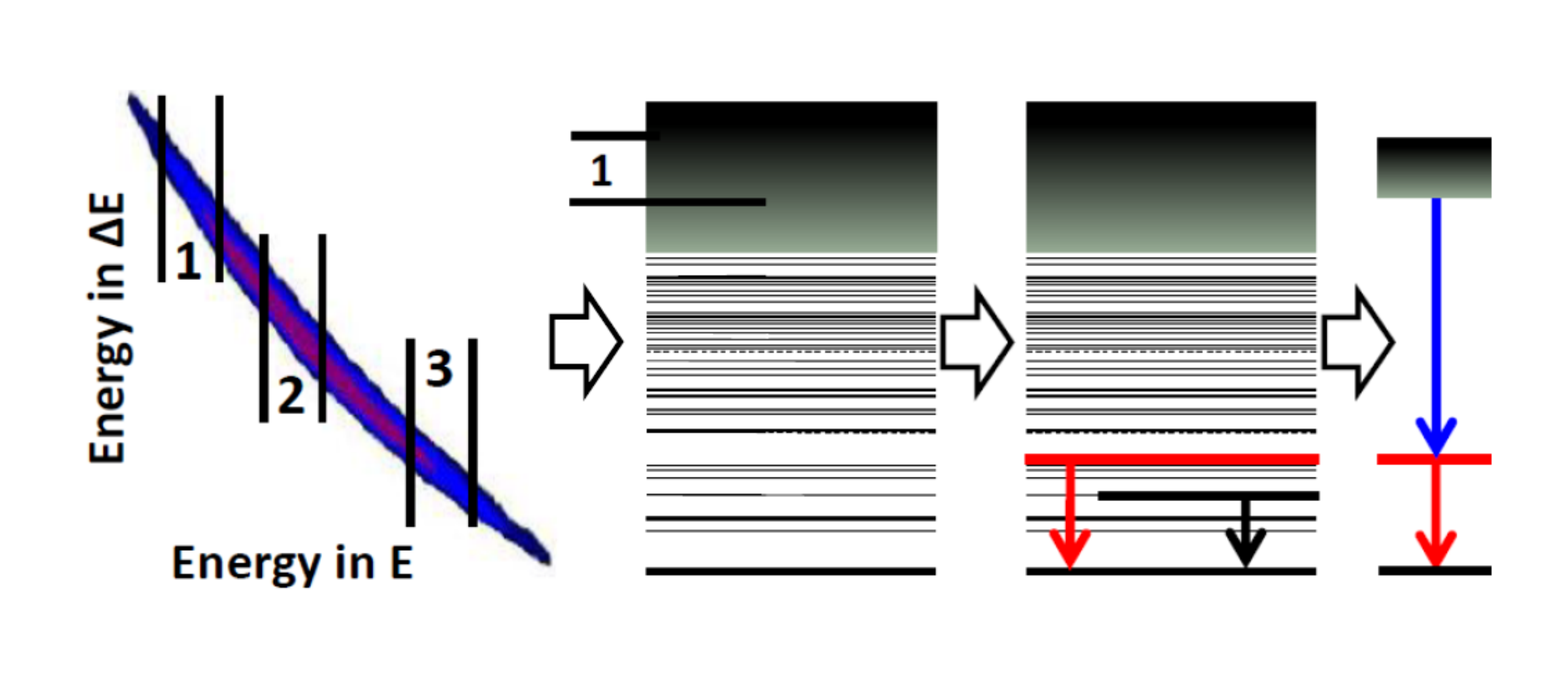}  
\caption{(Color online) Schematic representation of the procedure to identify the primary transitions in the Ratio method. The particle energies from the $\Delta E - E$ particle detectors (left) determine the excitation energy of the residual nucleus (second panel from left). Well-known transitions deexciting low-lying levels (third panel from left) are used for the identification of these levels. Only coincidence events where the energy of the second detected transition fits the difference between the region "1" and the energy of a selected low-lying level (right) are considered in the analysis.}
\label{fig_exp_RM}
\end{center}
\end{figure}

The $\gamma$-rays are detected with high-resolution, high-purity Germanium detectors, possibly in combination with high-efficiency detectors, such as LaBr$_3$:Ce. Only events for which the energy sum of the two detected $\gamma$-rays, one of them being the primary transition feeding a well resolved low-lying level from $E_i$ and the second transition originating from a known decay of a low-lying level, equals to the initial energy $E_i$ within the energy resolutions of the detectors are considered in the subsequent analysis, see schematic sketch in Fig.\ref{fig_exp_RM}. Any particle-$\gamma$-$\gamma$ event satisfying these conditions provides an unambiguous determination of the origin and destination of the observed primary transition, as shown sche\-matically in Fig.\ref{fig_exp_RM}. The data is extracted on an event-by-event basis and each $\gamma$-ray transition is corrected for their efficiency as well as for the branching ratio in the case of transitions from discrete states. The Ratio method can be applied as long as the primary $\gamma$-ray transitions feed discrete states of the same spin and parity and is independent of model input and eliminates systematic uncertainties. 

The energy dependence of the PSF is obtained from the ratio $R$ of intensities $I(E_\gamma)$ for two different primary $\gamma$-ray energies from the same initial excitation energy $E_i$ to discrete low-lying levels of same spin and parity at energies $E_{f_1}$ and $E_{f_2}$ as
\begin{equation}
R = \frac{f_1(E_i - E_{f_1})}  {f_1(E_i - E_{f_2})} = \frac{I(E_i - E_{f_1}) (E_i - E_{f_1})^3}{I(E_i - E_{f_2}) (E_i - E_{f_2})^3}.
\label{eq:ratio_method}
\end{equation}

Data on primary $\gamma$-ray intensities of transitions from an excitation energy bin to different discrete levels can be used to obtain the $E_\gamma$ dependence in a broad energy range by a $\chi^2$ minimization procedure \cite{Wiedeking12, Krticka16, Jones18}. Data corresponding to different spins and parities of final low-lying levels can be normalized in the same way.
The absolute value of the PSF must be determined independently and an attempt to normalize relative PSF values to that from the ($\gamma$,n) reaction was made in Ref. \cite{Krticka16}.

\subsection{Inelastic proton scattering}
\label{sec_exp_pp}

Inelastic proton scattering experiments, {\it i.e.} (p,p$^\prime$) reactions, with polarized proton beams at energies of 295 MeV have been recently performed at the Research Center for Nuclear Physics (RCNP) at Osaka University (Japan). The energy distribution of scattered protons is measured with the high-resolution GRAND RAI\-DEN magnetic spectrometer at various forward laboratory angles, typically between $0\degree$ and $10\degree$. The measured spectra provide information on the electromagnetic excitation probability of a nucleus from the ground state to excitation energies in the range of approximately $5-20$~MeV.   

The contribution of $E1$ and $M1$ transitions to this process can be separated by two independent methods, using {\it i)} polarization transfer observables that can be determined from the measurement of the polarization of scattered neutrons using a carbon polarimeter \cite{Tamii09}, and {\it ii)} multipole decomposition analysis that exploits the angular distribution of scattered protons \cite{Tamii11,Poltoratska12}. Both methods give consistent results \cite{Martin17}.
When separating $E1$ and $M1$ transitions using the polarization method, it is assumed that the spin-flip transitions originate from $M1$ transitions for $E_\gamma\approx 5-15$~MeV.

The $E1$ PSF $\overleftarrow{f_{E1}}$ is obtained from the measured cross section under the assumption that it comes solely from the Coulomb excitation process via a virtual photon \cite{Bertulani88}. The $M1$ cross sections are converted to reduced transition strengths and corresponding $M1$ photoabsorption cross sections with the approach described in detail in Refs. \cite{Birkhan16,Mathy17}.

It should be mentioned that only the spin part of the $M1$ transition operator is expected to play a role in the small-angle proton scattering. Strictly speaking, only part of the $M1$ PSF is determined. As it is expected that the orbital part of the $M1$ operator does not significantly contribute to transitions with $E_\gamma\approx 5-15$ MeV, the $M1$ PSF determined in this $E_\gamma$ range should be a very good approximation of the actual $M1$ PSF.  
Similar experiments aiming at extracting the spin-flip part of the $M1$ transitions have been performed previously \cite{Frekers90} but the PSFs were not determined.

\subsection{Photonuclear data}
\label{sec_exp_photo}

The dipole PSFs were calculated on the basis of all the experimental data on photoreaction cross sections compiled in the EXFOR database \cite{EXFOR}.  
The photoneutron cross sections have been measured as a function of the photon energy by means of monochromatic beams produced predominantly by 
annihilation-in-flight of positrons ({\it e.g.} measurements at the Lawrence Livermore National Laboratory, USA, and the CEA-Saclay, France) 
as well as using Brems\-strahlung beams ({\it e.g.} the experiments at the Max-Planck-Institute for Chemistry (Germany), Melbourne University (Australia), Moscow State University (Russia)). 
For partial photoneutron reactions, ($\gamma$ ,n), ($\gamma$ ,2n), \dots cross sections were determined through direct neutron detection and counting of residual $\gamma $-ray activity. 
Additionally, various methods were used to obtain cross sections with protons in the outgoing reaction channels which are needed for the determination of the total photoabsorption cross section \cite{Shoda62,Ishkhanov70b,Varlamov03}. Photoneutron cross sections have also been measured at GDR peak energies and 
below in experiments based on laser-induced Compton backscattered $\gamma$-rays ({\it e.g.} at the NewSUBARU facility of Konan University, Japan). Partial and total photoneutron cross sections have been revised using the experi\-mental-theoretical re-eva\-luation method of the partial photoneutron reaction cross sections based on objective physical criteria of the data reliability \cite{Varlamov14}.  It should however be noted that open questions on the 
determination of the effective neutron detection efficiency, may impact the determination of the photoabsorption cross sections, as discussed in Ref.~\cite{Banu19}. Details about the adopted 
photoreaction data, the experimental conditions as well as the recommendation in case of conflicting data (including in particular discrepancies between the Livermore and Saclay data) are given in the CRP review paper on ``Updated IAEA Photonuclear Data Library'' ~\cite{Kawano19}.

\subsection{Additional methods for PSF comparisons}
\label{sec_exp_add}

Here we describe some additional methods that do not allow for the extraction of absolute values or energy dependences of the PSF but are sensitive to the PSF and therefore can provide information on the compatibility or validity of existing PSF models.

\subsubsection{Singles $\gamma$-ray spectra from (n,$\gamma$) reaction}
\label{sec_exp_ncap2}

The validity of various PSF models can be checked using unfolded (or detector response corrected) $\gamma$-ray spectra from (n,$\gamma$) reactions. Predicted spectra can be obtained from any 
code that can generate a $\gamma$-ray spectrum using the statistical model. Transitions between levels below a critical excitation energy can not be treated in the statistical model, therefore, in these cases the relative intensities of these transitions, which are known experimentally, are adopted. Internal conversion coefficients are also considered, which are importantespecially in heavy nuclei.

The methodology of this testing technique has been reported in Ref. \cite{Belgya17} for the $^{113}$Cd(n,$\gamma$)$^{114}$Cd reaction. The data analysis is performed with the statistical model code, Bin Type Simulation (BITS), which uses as input different PSF and NLD models and can only be used with unfolded experimental spectra which are corrected for the detector response function. The method for unfolding the spectra detailed in Ref. \cite{Belgya14} is based on the prescription described in Sec. 3.2 of Ref.~\cite{Bartholomew73}. 

The BITS code solves the sequential integration numerically by setting up 100 keV bins from the critical energy up to the separation energy of the daughter nucleus. The number of levels with different spins and parities in the bins is calculated from the corresponding NLD and their summed populations are calculated from thee feeding from the levels in the bins above.
 The starting level is the capture state with definite spin and parity and is given a population of 1. The program starts 
with this initial condition and distributes the intensity to final levels or bins using the average decay widths $\langle \Gamma_{\gamma,XL}(E-E_\gamma) \rangle  $ to calculate the electromagnetic branching ratios which are corrected for 
internal conversion.  The sum of the branching ratios is normalized to 1. Repeating the process downwards, in decreasing excitation energy, 
the decay-scheme is built up and the decay strengths are stored in a decay-scheme matrix of 6 dimensions which are indexed by the initial and final levels and their spins and 
parities. Below a given critical energy, the experimentally known (discrete) part of the decay scheme of the nucleus is taken into account to describe discrete electromagnetic transitions using 
internal conversion corrections. The $\gamma$-ray spectrum of full energy peaks is collected from the decay-scheme matrix from which single, two-step and higher multiplicity spectra can also be 
collected. It is useful to collect spectra versus multipolarities for electric and magnetic types of transitions to learn about their relative contributions.  In addition, visual comparison of the calculated and experimental spectra are provided with contributions from the calculated $E1$ and $M1$ decays.

\subsubsection{Two-step and Multi-step Cascade Spectra}
\label{sec_exp_msc}

\begin{figure*}[t]
\includegraphics[clip,width=\textwidth]{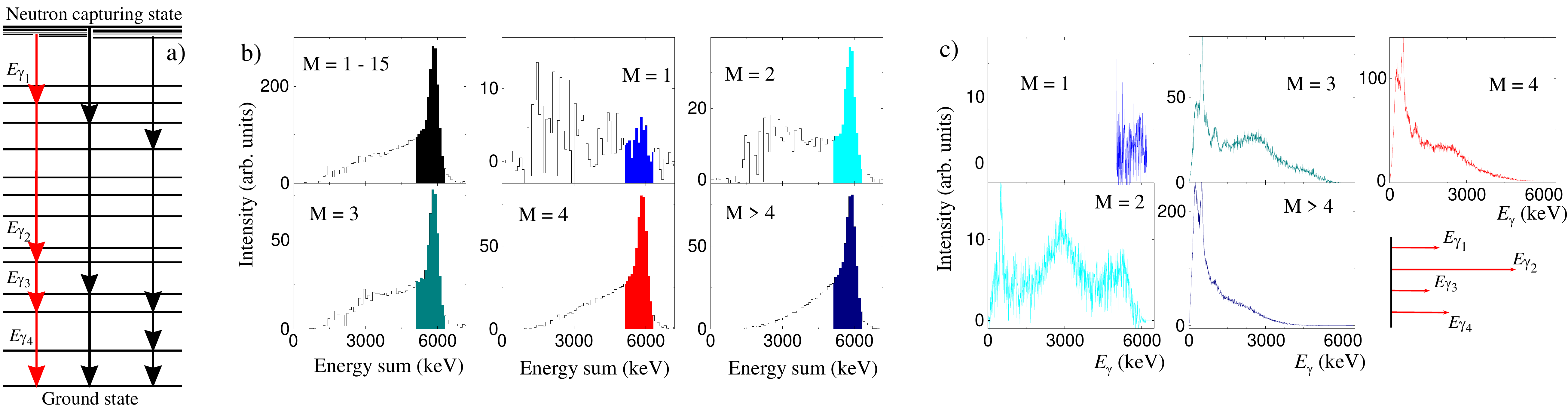}
\caption{(Color online) Illustration of the production of MSC spectra for a nucleus with the neutron separation energy close to 6 MeV. The left part of the figure (a) shows three possible decays of the nucleus. The sum-energy spectra (b) can be obtained for individual multiplicities. Using events in the highlighted areas in the sum-energy spectra, the MSC spectra (c) for these multiplicities can be obtained. The red cascade from a) will contribute to the multiplicity $M = 4$ MSC spectrum at points indicated by red arrows in bottom right part of c).}
\label{fig:scheme_msc}
\end{figure*}

Another method to validate PSFs is via comparison of coincidence $\gamma$-ray spectra with predictions of the statistical model of the nucleus. Two different experimental setups are used 
for these measurements. 

The first setup consists of a pair of high-resolution Ge detectors which allows for measurements of so-called two-step $\gamma$ cascades (TSCs), see {\it e.g.} \cite{Honzatko96,Krticka08,Valenta15}.
These experiments have been mainly performed at Dubna (Russia) \cite{Bogdzel82} and \v{R}e\v{z} near Prague (Czech Republic) \cite{Honzatko96}.

The analysis of experimental data allows to get background-free $\gamma$-ray spectra corresponding to decays that connect the capture state (very often just above $S_n$) with 
preselected, well-separated low-lying levels of the nucleus via two $\gamma$-rays. The spectra can be obtained by applying a cut on the detected energy sum deposited in 
the two Ge detectors. 

Due to the high resolution power of Ge detectors, narrow peaks corresponding to TSCs to the preselected levels depositing the total energy of the cascade, are observed in the sum-energy 
spectra. Only events contributing to these peaks are then analyzed. Spectra of deposited $\gamma$-ray energies from one or both detectors can then be constructed. The analysis method 
\cite{Honzatko96} excludes detected TSCs populating other levels than those of interest and efficiently rules out the accidental coincidence and Compton-related background. A contribution of a 
TSC via an intermediate level in a TSC spectrum is almost exclusively given by a pair of narrow, symmetrically located lines.
Typically, spectra for several pre-selected final levels are available which allows analysis of not only ``true'' two-step  cascades but also of more-step $\gamma$ cascades. The influence of the detection system on spectra is relatively simple and can be applied if efficiencies of the two detectors are taken into account \cite{Krticka08}.   

The second experimental setup exploits a highly-segmen\-ted array of lower resolution scintillation detectors that allows for measurements of cascades for different detected multiplicities $M$. Different measured spectra can be used for comparison with statistical model predictions. They include the sum-energy spectra, multiplicity-distribution (MD) spectra and spectra of individual deposited $\gamma$ energies for individual $M$, often called multi-step $\gamma$ cascade (MSC) spectra. The MSC spectra, constructed only from those $\gamma$ decays that deposit the energy sum corresponding to the $Q$-value of the reaction in the detection system, allow to get more information on PSFs than spectra constructed from all detected events. In addition, a cut on $Q$-value usually allows also for very efficient subtraction of the background \cite{Sheets09,Valenta17}. The proximity of individual detectors requires simulation of the response of the detection setup. This response is usually applied to simulated cascades. These spectra have been so far obtained mainly from measurements at the DANCE detector at Los Alamos \cite{Heil01,Reifarth04}. The DANCE detector is a highly-segmented high-efficiency array consisting of 160 BaF$_2$ crystals that cover a solid angle of approximately $3.5\pi$. Fig. \ref{fig:scheme_msc} illustrates the way how MSC spectra are produced.

With the exception of the TSC spectra in Ref. \cite{Voinov10} that used the (p,$\gamma$) reaction, all other available data come from the capture of slow-energy neutrons. Specifically, TSC spectra, that suffer from the low efficiency of Ge detectors are measured using thermal neutrons while MSC spectra typically use spectra from isolated neutron resonances. The advantage of neutron capture is that the spin and parity of the capturing state are known, which is important as spectra from resonances with different $J^\pi$ can significantly differ \cite{Sheets09,Walker15}. 

The simulations of $\gamma$ cascades were usually performed with the Monte Carlo code DICEBOX \cite{Becvar98}, which allows for the consideration of Porter-Thomas fluctuations of individual transitions. 


\subsubsection{Average radiative widths}
\label{sec_exp_gamgam}

The total average radiative width $\langle \Gamma_{\gamma} \rangle$ is also a quantity containing global information on the PSFs. Theoretically, $\langle \Gamma_{\gamma} \rangle$ represents a folding of the deexcitation PSFs (or equivalently transmission coefficients $T_{XL}(E_\gamma)$) and NLD (see {\it e.g.} \cite{Capote09}), {\it i.e.} 
\begin{equation}
\langle\Gamma_{\gamma} \rangle=\frac{D}{2\pi}\sum_{X,L,J,\pi} \int_0^{S_n+E_n}T_{XL}(E_\gamma) \rho(S_n+E_n-E_{\gamma},J,\pi)dE_\gamma ~,
\label{eq_gamgam}
\end{equation}
where  the summation runs over all spins $J$, parities $\pi$ and transition types $XL$, $E_n$ is the neutron incident energy and  $\rho(E,J,\pi)$ the energy-, spin- and parity-dependent NLD.
The $\gamma$-ray transmission coefficient, $T_{XL}(E_\gamma)$  is related to the PSF $f_{XL}(E_\gamma)$ as
\begin{equation}
T_{XL}(E_\gamma)=2\pi E_\gamma^{2L+1} f_{XL}(E_\gamma) .
\label{eq_TXL}
\end{equation}
Average radiative widths, like neutron strength functions and the average spacing of resonances, are obtained from the analysis of parameter sets 
for resolved resonances \cite{Capote09,Mughabghab06}.

Data for $s$-wave average radiative width are available for about 228 nuclei \cite{Capote09} and have been used here to test PSF models. The predicted average 
radiative width remains however sensitive to the adopted NLD model, as discussed in Sec.~\ref{sec_comp_gamgam}.

\subsubsection{Maxwellian-averaged cross sections}
\label{sec_exp_macs}

The radiative neutron capture cross sections can also provide information on the PSF.  At keV neutron energies, the radiative neutron capture cross section 
is essentially proportional to the total photon transmission coefficient $\mathcal{T_{\gamma}}$ which in turn, like the average radiative width in 
Eq.~\ref{eq_gamgam}, is sensitive to the folding of PSF and NLD \cite{Goriely08a,Koning12}, as
\begin{eqnarray}
\mathcal{T_\gamma}=& & \sum_{J,\pi,X,L} \int_0^{S_n+E_n}   2\pi E_\gamma^{2L+1} f_{XL}(E_\gamma)\times \nonumber \\
& & \rho(S_n+E_n- E_\gamma,J,\pi) dE_\gamma~.
\label{eq_Tg}
\end{eqnarray}
A large compilation of about 240 experimental Maxwellian-averaged  neutron capture cross sections (MACS)  at 30~keV for nuclei with 
$20\le Z \le 83$ \cite{Dillmann06,Dillmann14}  is available and has been considered for testing the PSF models.

\section{Development of the experimental PSF database}
\label{sec_ass}

In this section, the assessment and selection criteria for the PSF data to be included in the library are discussed. Uncertainty analyses are elaborated in specific cases and re-analysed for two nuclei for which data from both the NRF and Oslo methods are available.

\subsection{Compilation of PSFs}
\label{sec_ass_compil}

The PSF data from the experimental methods described in Sec.\ref{sec_exp} were compiled in an experimental database.  The majority of PSF data were provided directly by the groups performing the measurements and include experimental results available as of January 2019. The photonuclear data were obtained from other databases and processed to obtain PSFs. In some cases, such as (p,$\gamma$) reactions, PSF data were extracted directly from tables in the respective publications, or when not available in table format, from the figures. 
It is important to emphasize that 
PSF data measured with the different methods is considered and included only from those publications where the original work extracted the PSF explicitly. Although valuable information can be obtained solely from capture cross section measurements, use of this data go beyond the scope of this work. 
Each set of data was assessed to verify the suitability for its inclusion as a data file in the library and accompanied by a README file, detailing key information to place the data into context. 

\subsubsection{PSFs extracted from NRF measurements}
\label{sec_ass_nrf}

The compilation comprises dipole strength functions $f_1$ that were deduced from absorption cross sections according to the prescription given in 
Sec.~\ref{sec_exp_nrf}. The data result from experiments covering the excitation-energy range from typically 4-5 MeV up to $S_n$, in which the 
absorption cross sections and the related dipole strength functions $\overrightarrow{f_1}(E_\gamma)$ were deduced as a smooth function of energy.

In the case of broad-band bremstrahlung measurements at $\gamma$ELBE \cite{Schwengner05}, $\gamma$-rays were measured with two shielded 
HPGe detectors placed at $90\degree$ to the beam and two at $127\degree$  to the beam. Spectra were response and efficiency corrected. 
The photon flux was determined by using known level widths in $^{11}$B. Background due to atomic processes in the target was determined in 
simulations and subtracted from the spectra. Subtracted spectra contain resolved peaks and nuclear quasicontinuum. These $\gamma$-ray spectra were corrected 
for feeding and branching intensities obtained from simulations of statistical $\gamma$ cascades. The absorption cross sections were obtained 
from scattering cross sections by using average branching ratios of ground-state transitions obtained from the simulations. Uncertainties of the 
absorption cross sections include statistical uncertainties, and 5\% uncertainties each for efficiency, photon flux and atomic background. The absorption cross sections are compiled in the EXFOR database.

Some experiments obtained data at energies above the neutron separation energy. These do not represent the total photoabsorption cross sections because of the 
opening of the competing ($\gamma$,n) channel and these PSF values are therefore not included in the data file.
Total dipole PSFs for 23 different nuclei for energies up to the neutron separation energies have been included in the PSF library. For 
3 nuclei measured at HI$\gamma$S \cite{Weller09} ($^{128}$Xe, $^{134}$Xe and $^{138}$Ba), both the $E1$ and $M1$ PSFs are available separately.
The assessment of data did not find grounds on which to exclude any of the available sets of data.

\subsubsection{PSFs extracted from the Oslo method}
\label{sec_ass_oslo}

The compilation includes total dipole PSFs $\overleftarrow{f_1}(E_\gamma)$ from the Oslo method analysis, as described in Sec.~\ref{sec_exp_oslo}. It also
includes data analyzed using the beta-Oslo method and data from inverse kinematics experiments (Inverse-Oslo) which have become available over the last few years. 
The data typically cover an energy range from about $E_\gamma \sim 1-2$~MeV up to a maximum energy $E_\gamma \sim S_n$. 

For data sets obtained prior to $\sim$ 2012, only statistical errors are included in the PSF data, while for newer data sets systematic errors are 
also considered, which is some cases also include uncertainties due to NLD models, $D_0$ and $\langle \Gamma_\gamma \rangle$. These are typically represented by upper and 
lower uncertainty bands. Where possible PSF data obtained from different NLD models are provided in separate data files. Where it was not possible 
to extract individual data sets for different normalizations then one data file is provided and the expected variations are provided in terms of 
error bars. 
In several cases, the published PSF data were re-analyzed, usually due to the availability of new data for NLDs and/or PSFs normalizations. In these cases the PSFs from both analyses are included in the library as they provide the user insight into the range of uncertainties due to model dependencies. 
 Similarly, if more experiments were performed in the same nucleus, the extracted PSFs from each unique experiment are included in the library since these sets of data are considered 
to be independent of each other (they may have different energy ranges, beam energy, detector arrangement, etc.).  
113 sets of total dipole PSFs $f_1$ for 72 different nuclei measured with the Oslo method for energies up to the neutron separation energies have been included in the PSF library.

\subsubsection{PSFs extracted from DRC/ARC}
 \label{sec_ass_arc}

Two different experimental techniques, ARC and DRC (see Sec.~\ref{sec_exp_arc}), were applied to obtain information on PSFs from resonance neutron capture experiments. The recent re-analysis of all available data from both types of experiments resulted in two separate databases, 
DRC-2018 and ARC-2019. The resulting PSF data files present the partial value averaged over measured resonances 
for each primary transition. For a detailed description of this work and processing of the data, we refer to 
Refs. ~\cite{Kopecky16,Kopecky17,Kopecky17b,Kopecky18}. The results were merged in the final DRC+ARC 2019 library, which includes information on PSFs for 88 nuclides with 
masses between $20 \le A \le 240$.
The list of nuclei available in the DRC+ARC-2019 library is shown in Table~\ref{tab_ass_arc}.

\begin{table*}
\begin{center}
\caption{Content of the DRC+ARC-2019 database. The symbol "x" corresponds to data that have been  included in the DRC (and DRC+ARC, if no ARC 
data is available) database, $\langle {\rm x} \rangle$ to DRC data for which binned results are available only, "0" to data not used due to insufficient averaging or missing transition rates,  and "xx" to  data that have been included in the ARC (hence ARC+DRC) database. The nucleus corresponds to the compound system. 
\label{tab_ass_arc} }
\begin{tabular}{lccc|lccc}
\hline\hline 
Nucleus	&	DRC	&	ARC	&	DRC+ARC	&	Nucleus	&	DRC	&	ARC	&	DRC+ARC	\\
\hline
F-20	&	x	&		&	x	&	Gd-156	&		&	xx	&	xx	\\
Mg-25	&	x	&		&	x	&	Gd-157	&	$\langle {\rm x} \rangle$	&	xx 	&	xx 	\\
Al-28	&	x	&		&	x	&	Gd-158	&		&	xx	&	xx	\\
Si-29	&	x	&		&	x	&	Gd-159	&	x	&	xx	&	xx	\\
Si-30	&	x	&		&	x	&	Dy-162	&		&	xx	&	xx	\\
S-33	&	x	&		&	x	&	Dy-163	&		&	xx	&	xx	\\
Cl-36	&	x	&		&	x	&	Dy-164	&		&	xx 	&	xx 	\\
Sc-46	&	x	&		&	x	&	Dy-165	&		&	xx	&	xx	\\
Cr-53	&	x	&		&	x	&	Ho-166	&		&	xx	&	xx	\\
Cr-54	&	x	&		&	x	&	Er-168	&	x	&	xx 	&	xx 	\\
Fe-57	&	x	&		&	x	&	Er-169	&	x	&		&	x	\\
Fe-59	&	x	&		&	x	&	Tm-170	&	x	&	xx 	&	xx 	\\
Co-60	&	x	&	0	&	x	&	Yb-172	&		&	xx	&	xx	\\
Cu-64	&	x	&	0	&	x	&	Yb-174	&	x	&	xx	&	xx	\\
Ge-74	&	x	&		&	x	&	Lu-176	&	x	&	xx 	&	xx 	\\
As-76	&		&	xx	&	xx	&	Lu-177	&	x	&		&	x	\\
Zr-92	&		&	xx	&	xx	&	Hf-178	&	x	&	xx 	&	xx	\\
Nb-94	&	x	&		&	x	&	Hf-180	&		&	xx 	&	xx	\\
Mo-93	&	x	&	0	&	x	&	Ta-182	&	x	&	xx 	&	xx 	\\
Mo-96	&		&	xx	&	xx	&	W-183	&	x	&		&	x	\\
Mo-98	&		&	xx 	&	xx 	&	W-184	&	x	&	xx	&	xx	\\
Mo-99	&	x	&	0	&	x	&	W-185	&		&	xx	&	xx	\\
Ru-100	&	x	&		&	x	&	W-187	&		&	xx 	&	xx	\\
Ru-102	&	x	&	xx 	&	xx 	&	Os-188	&		&	xx	&	xx	\\
Rh-104	&	x	&		&	x	&	Os-189	&		&	xx	&	xx	\\
Pd-106	&	x	&	xx 	&	xx 	&	Os-191	&		&	xx 	&	xx 	\\
Pd-109	&		&	xx	&	xx	&	Os-193	&		&	xx	&	xx	\\
Ag-108	&	$\langle {\rm x} \rangle$	&		&	$\langle {\rm x} \rangle$	&	Ir-192	&		&	xx	&	xx	\\
Cd-114	&		&	xx 	&	xx 	&	Ir-194	&		&	xx 	&	xx 	\\
In-116	&	x	&		&	x	&	Pt-195	&		&	xx	&	xx	\\
Sb-122	&	x	&		&	x	&	Pt-196	&	x	&	xx	&	xx	\\
Sb-124	&	x	&		&	x	&	Pt-197	&		&	xx 	&	xx 	\\
Te-124	&		&	xx	&	xx	&	Pt-199	&		&	xx	&	xx	\\
Te-126	&	x	&		&	x	&	Au-198	&	x	&	xx 	&	xx 	\\
I-128	&	x	&	xx 	&	xx 	&	Hg-199	&	x	&		&	x	\\
Ba-135	&		&	xx 	&	xx 	&	Hg-200	&	x	&		&	x	\\
Ba-136	&	x	&	xx 	&	xx 	&	Hg-202	&	x	&		&	x	\\
Nd-144	&	x	&		&	x	&	Th-233	&	x	&	xx 	&	xx 	\\
Nd-146	&	x	&	xx 	&	xx 	&	U-235	&	x	&		&	x	\\
Sm-148	&	x	&	xx	&	xx	&	U-236	&	x	&	xx 	&	xx 	\\
Sm-150	&	x	&	xx	&	xx	&	U-237	&	x	&		&	x	\\
Sm-155	&		&	xx 	&	xx 	&	U-239	&	x	&	xx 	&	xx 	\\
Eu-154	&		&	xx 	&	xx 	&	Np-238	&		&	xx	&	xx	\\
Gd-155	&	$\langle {\rm x} \rangle$	&	xx	&	xx	&	Pu-240	&	x	&	xx 	&	xx 	\\\hline
\hline
\end{tabular}
\end{center}
\end{table*}

\begin{figure}
\centering
\includegraphics[scale=0.36]{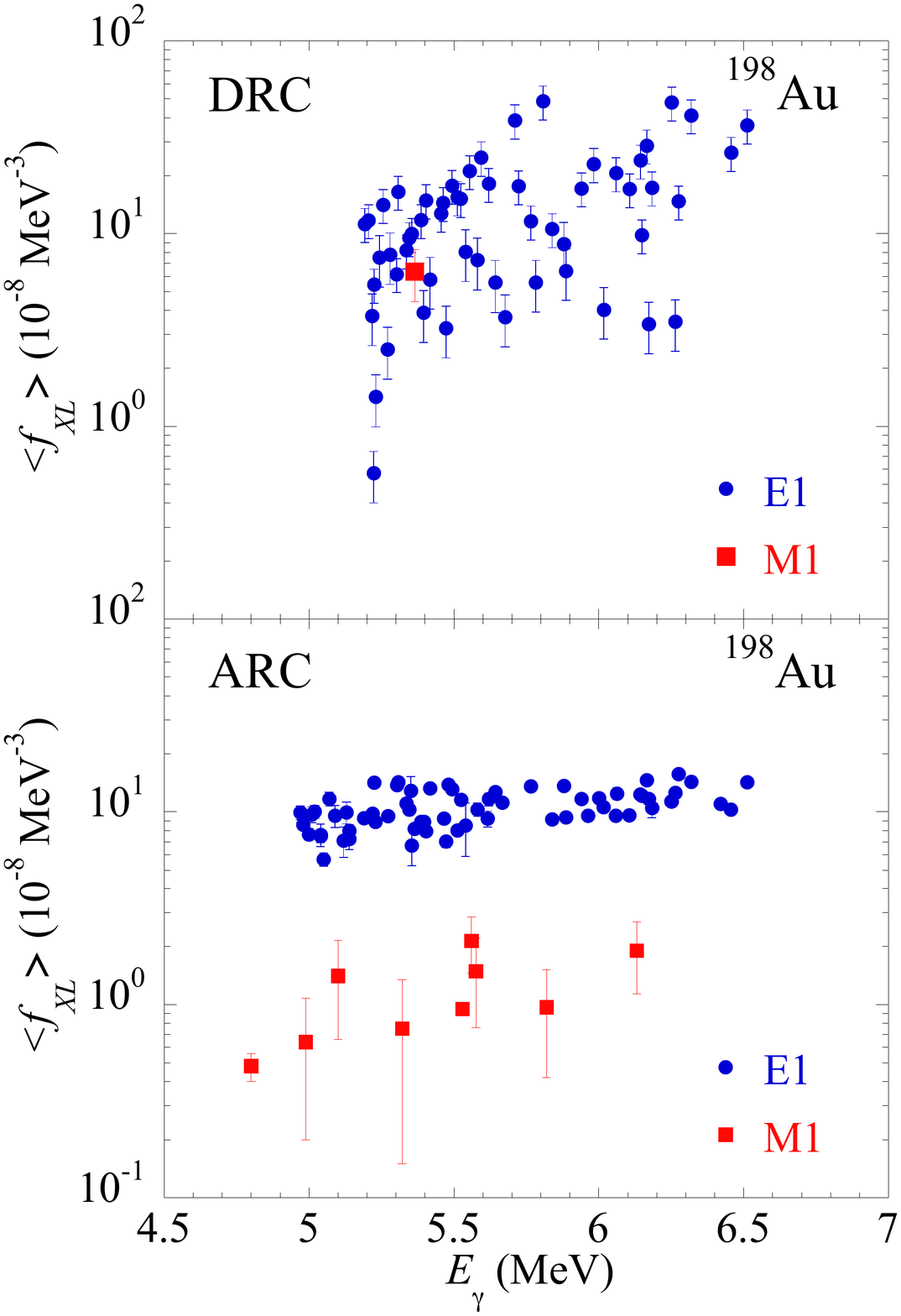}  
\vskip -0.5cm
\caption{(Color online) PSFs from DRC (upper panrel) and ARC (lower panel) measurements from the $^{197}$Au(n,$\gamma$)$^{198}$Au reaction. The large scatter of $E1$ transitions is for DRC data is from a  small number of 4 $s$-wave resonances (uncertainties due to the Porter-Thomas fluctuations are not included), while a very good averaging is obtained for ARC data where the averaging is made over $\approx$ 60 resonances.  The decreased detection sensitivity limit from lower TOF neutron fluence results in the detection of only one $M1$ transition with the rest undetected. Uncertainties are only statistical ones increased by 10\% due to estimated uncertainty in $\Gamma \gamma$ and $D$.}
\label{fig_arc_198Au}
\end{figure}

\begin{figure}
\centering
\includegraphics[scale=0.36]{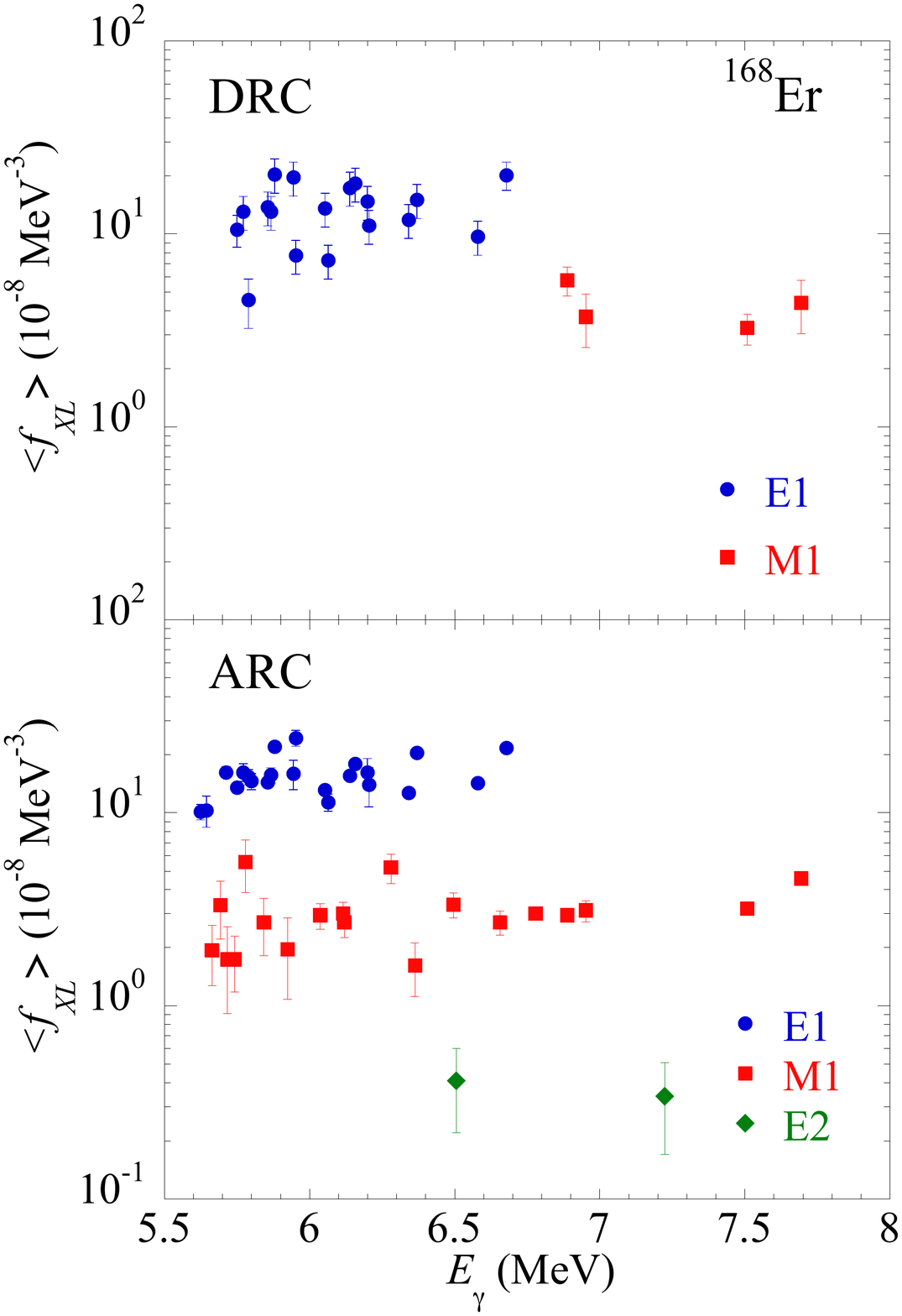}  
\vskip -0.5cm
\caption{(Color online) DRC (upper panrel) and ARC (lower panel) data from $^{167}$Er(n,$\gamma$)$^{168}$Er reaction. The comparable number of resonances, 81 in DRC and effectively about 250 from the boron filtered beam with a full width at half maximum of about 1 keV is averaged. The lower-detection sensitivity of DRC experiment prevents detection of low-energy $M1$ and all $E2$ transitions.}
\label{fig_arc_168Er}
\end{figure}

Recommended data were chosen from all extracted data sources and if data for both ARC and DRC experiments were available, the ARC filtered beam results 
were preferred because of better statistical accuracy due to averaging over a  much larger number of resonances compared to the DRC data. An example 
of this feature is given in Fig.~\ref{fig_arc_198Au} for DRC and ARC data for $^{198}$Au; the DRC values correspond to averaging over only 4 $s$-wave resonances. The uncertainty of the average value due to Porter-Thomas fluctuations is thus expected to be about 70\%. This uncertainty is not indicated in Fig.~\ref{fig_arc_198Au}. The error bars correspond only to uncertainties of measured transition intensities increased by $\Gamma_\gamma$ and $D$ uncertainty estimates of 10\%. When the number of resonances studied in a DRC experiment is 
large, the resulting data distribution is comparable to ARC measurements as shown in Fig.~\ref{fig_arc_168Er} for $^{168}$Er. 

In the absence of ARC data, DRC data, even those measured with a small number of resonances, were included in the final version of the DRC+ARC-2019 library. As long as DRC data is processed in a doubly average quasi-monoener\-getic format  of $\langle\langle f_{XL} \rangle\rangle$ (see Sec.~\ref{sec_exp_arc}), they give a satisfactory estimate of the absolute value of the PSF. 

The performance of this database was thoroughly validated against the previous evaluations \cite{Kopecky18} and in case of conflicting results, the reasons (such as difference in selected resonance parameters, different $E1$ or $M1$ assignments or applied spacing) are discussed in 
detail in Refs.~\cite{Kopecky16,Kopecky17b,Kopecky18}.

The PSF library includes DRC and ARC for 88 nuclei, out of which 34 are DRC data and 54 ARC data. Among the 34 DRC data sets, $E1$ strengths are 
available for 33 nuclei, $M1$ strengths for 29 nuclei and $E2$  for 8 nuclei only. The 54 ARC data sets include data on $E1$ for 52, 
$M1$ for 49 and $E2$ for 22 nuclei.

 \subsubsection{PSFs from the capture $\gamma$-ray library}
\label{sec_ass_egaf}

The intensities of primary transitions from thermal neutron captures, available in the EGAF library~\cite{EGAF} (see Sec.~\ref{sec_exp_egaf}), were exploited for the determination of the PSF for individual transition types. The EGAF library consists of approximately 32,000 prompt thermal neutron capture $\gamma$-ray cross sections for nearly all elements. For primary transitions these were converted to $f_{XL,f}$ as described in Sec. \ref{sec_exp_egaf}.

To suppress the influence of the Porter-Thomas fluctuations, several neighboring transitions were binned together and averaged. The average value needs to be in many cases corrected for weak, unobserved transitions within the energy bin. 

The expected number of transitions within each bin was obtained from the spin-dependent  level density based on a modified CT model \cite{Firestone17b} where the temperature is taken from the RIPL-3 library~\cite{Capote09} and the backshift energy is the yrast energy for each spin taken from ENSDF \cite{ENSDF}. The expectation value of the total unobserved transition intensity in a bin was estimated assuming the Porter-Thomas distribution \cite{Porter56} of individual intensities under the assumption that the observed transitions (after conversion to $f_{XL,f}$) are the strongest ones occurring in the bin. The PSFs values in the database were corrected for this estimate.

Given uncertainties correspond only to statistical uncertainties in the determination of the average value and an additional 20\% uncertainty in the estimate of the missing strength. Neither an uncertainty due to $D_0$ and $\langle \Gamma_\gamma \rangle$, nor the uncertainty in the determination of the average PSFs values (coming from the Porter-Thomas fluctuation) is included.  As the number of transitions in a bin is typically 3 to 10, the latter uncertainty is significant and it reaches values of about $40-80\%$. It can be deduced from the number of observed and expected transitions in the bin.

A total of 808 $\gamma$-ray binned PSF data have been extracted from thermal neutron capture data for a total of 209 nuclei, including $E1$ PSFs from 206 nuclei, $M1$ for 153 nuclei, $E2$ for 38 nuclei and $M2$ for 2 nuclei.

\subsubsection{PSFs from average resonance proton capture}
\label{sec_ass_arpc}

The data were published primarily in the pre-1990s with the technique described in Sec.~\ref{sec_exp_pg} and direct communication with the authors was not possible. Data has been extracted from the 
publications, either directly from the tables provided or by digitizing the graphs. 
No averaging of the PSF for a given nucleus over transitions to more low-lying levels has been performed.
The typical $E_\gamma$ range covered in this method is about $2-4$ MeV wide and is located between the proton and neutron separation energies 
of nuclei in the $A=46$ to $A=90$ mass region. More specifically, the measured excitation energy region typically starts at $\sim 2$~MeV above $S_p$, 
given by the minimum proton energy used in the experiment. Depending on the values of $S_p$ and $S_n$ the region investigated is generally 
between about 5 and 10 MeV although for $^{90}$Zr measurements have been performed beyond $S_n$ \cite{Szeflinski83} . 
Specific information, in particular details on the uncertainty analysis, may not be available at all or is only partially described in the original 
publications. A detailed understanding of the uncertainties assigned to most of the data is therefore lacking. In some publications, no mention on the origin of the uncertainties are made while estimates of statistical and/or systematic uncertainties are provided in other publications but generally without much detail on how these were obtained.  
Data on the total dipole PSF $\overleftarrow{f_1}$ from (p,$\gamma$) measurements are available for 22 nuclei and are included in the PSF library.

\subsubsection{PSFs from the Ratio method}
\label{sec_ass_rm}

The method (Sec.~\ref{sec_exp_ratio}) was developed recently and only relative values of the total dipole PSF $\overleftarrow{f_1}$ are obtained unless a normalization to GDR data is performed. 
Such a normalization has been performed for the case of $^{95}$Mo \cite{Krticka16} and the data is included in the database. The data covers a 
range from $E_\gamma \sim 1.5$~MeV to a few hundred keV below the neutron separation energy. For detailed discussions on the different sources of uncertainties see Refs. \cite{Wiedeking12,Krticka16}.

\subsubsection{PSFs from inelastic proton scattering}
\label{sec_ass_ips}

The compilation includes PSFs that were extracted from inelastic proton scattering reaction data using polarized proton beams. The 
measured intensities are converted to $E1$, $M1$ and total dipole PSFs and correspond to $\overrightarrow{f_1}$ as described in Sec.~\ref{sec_exp_pp}.
They are provided in separate files, covering the excitation-energy range from about 5 MeV up to approximately 20 MeV. 
Data is available for $^{96}$Mo, $^{120}$Sn and $^{208}$Pb nuclei. The uncertainties correspond to those published in the original papers \cite{Tamii11,Poltoratska12,Bassauer16,Martin17}.


\subsubsection{PSFs extracted from photonuclear data}
\label{sec_ass_photo}

Photoabsorption PSF data files have been compiled from photoneutron cross sections including the photofission cross section for fissioning nuclei and the photoproton cross sections as compiled in the EXFOR library \cite{EXFOR,Kawano19}. A full list of the corresponding photonuclear cross sections can be found in Ref.~\cite{Plujko18}. The spin-independent $E1$ PSF 
was extracted from the photoabsorption cross section as described in Refs.~\cite{Bartholomew73,Photo00,Capote09,Plujko18} by applying the more general  Eq.~\ref{eq_nrf1} to the special case of $E1$ photoabsorption, {\it i.e.}
\begin{equation}
\overrightarrow{f_{E1}} (E _{\gamma } )=\frac{\sigma _{E1} (E _{\gamma } )}{3E _{\gamma } \left(\pi \hbar c\right)^{2} } ,
\label{eq_photo1}
\end{equation}
\noindent where $\sigma _{E1} (E _{\gamma } )\equiv \sigma(\gamma ,abs)$ is the total photoabsorption cross section  of $E1$ $\gamma$-rays with energy $E _{\gamma } $ summed over final states with all possible spins. 

\begin{figure}
\centering
\noindent\includegraphics[width=1.\linewidth,clip]{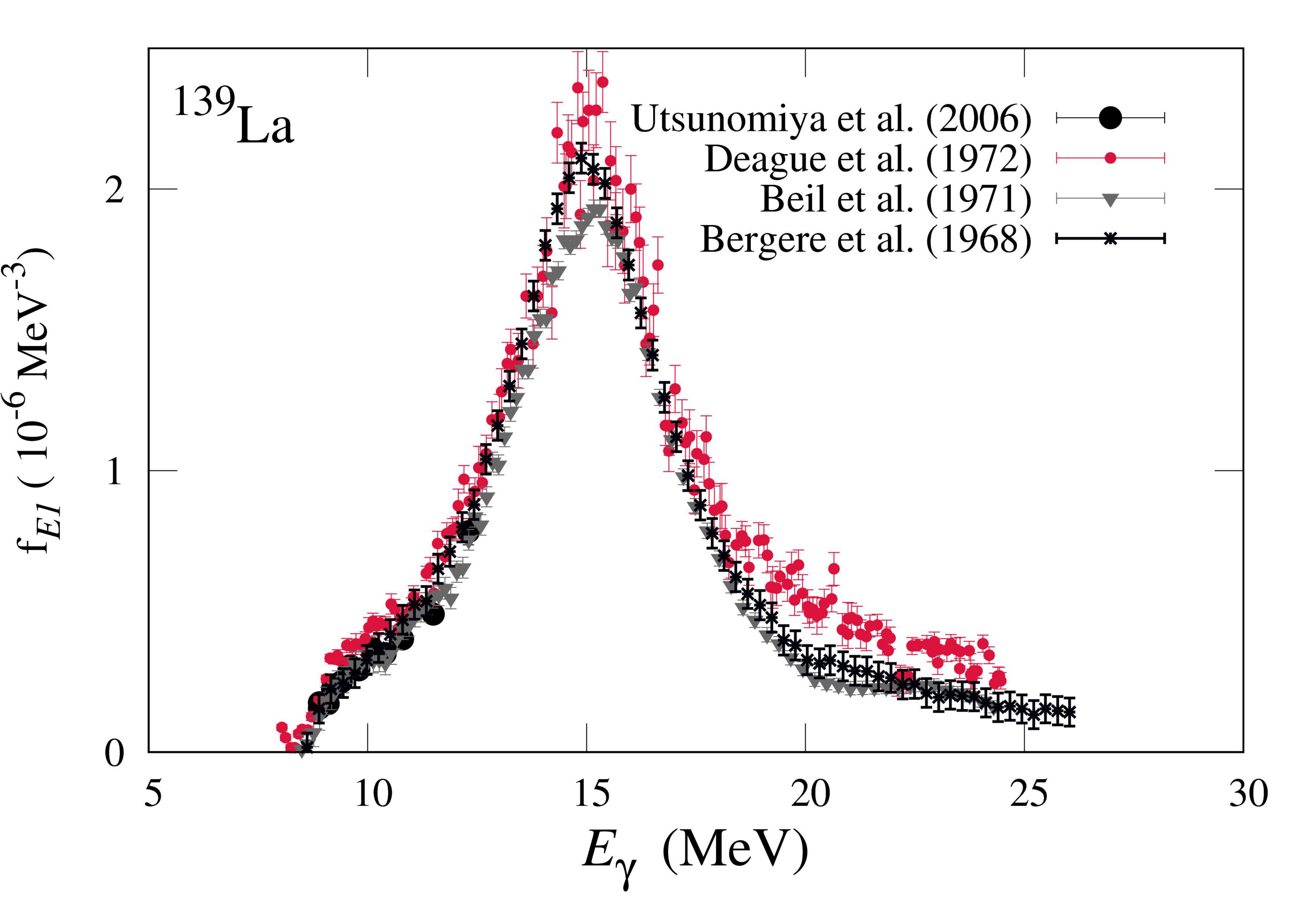}
\noindent\includegraphics[width=1.\linewidth,clip]{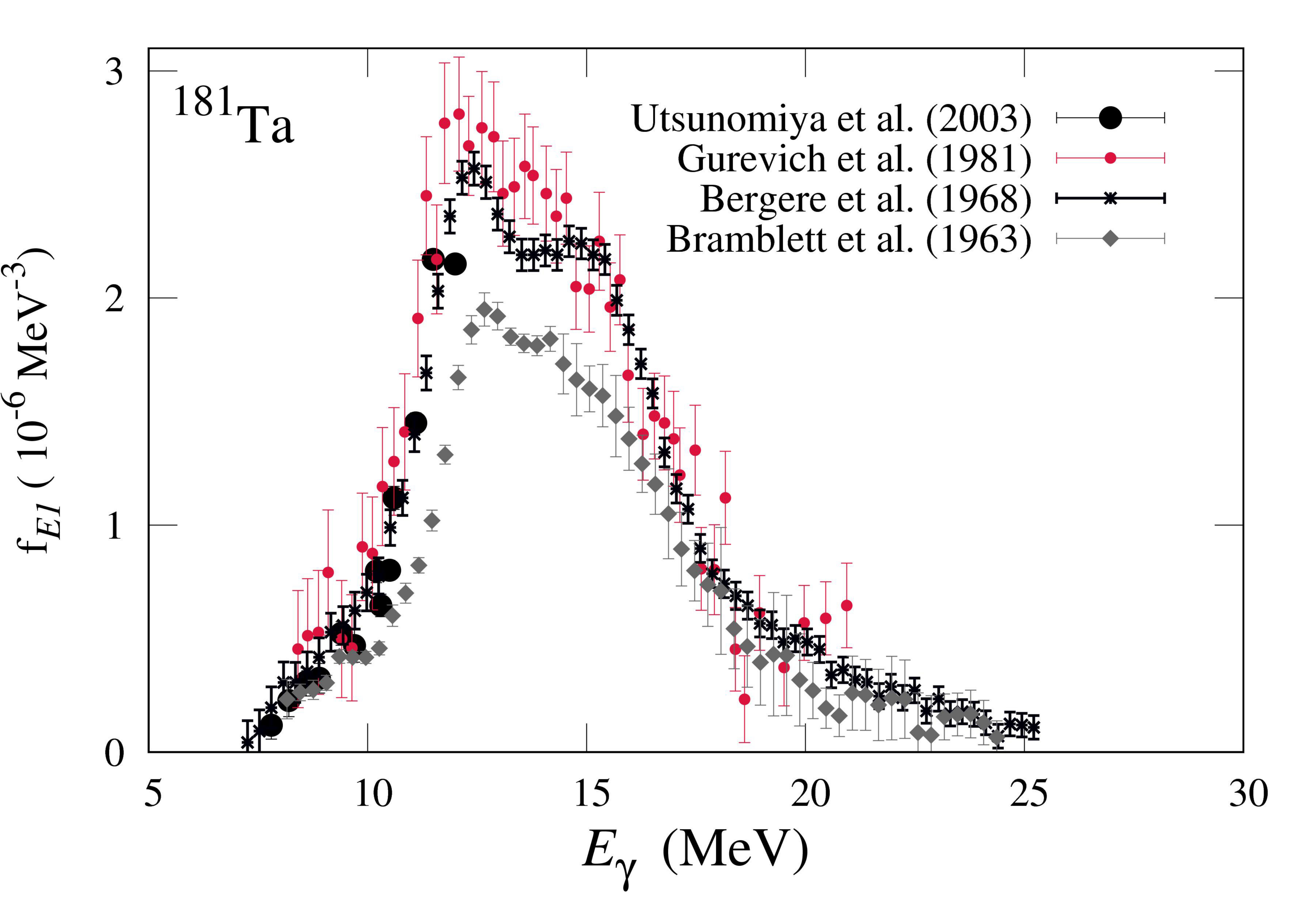}
\caption{(Color online) Photoabsorption  PSF for ${}^{139}$La and ${}^{181}$Ta extracted from experimental photo cross sections
\cite{Utsunomiya06,Deague72,Beil71,Bergere68,Utsunomiya03,Gurevich81,Bramblett63}.}
\label{fig_photo_La_Ta}       
\end{figure}

The $E1$ PSF uncertainties have been estimated with respect to the cross section uncertainties found in the EXFOR database \cite{EXFOR}. The mean values and uncertainties 
of the PSF extracted from the various experiments are different but results of recent experimental data are, as a rule, in agreement within experimental errors. Fig. \ref{fig_photo_La_Ta} shows representative examples of  PSFs extracted from different experiments. It can be seen that the relative uncertainties of the $E1$ PSFs in the vicinity of the GDR are of the order of 10 to 20 \% for recent experimental data. It should be mentioned that the experimental-theoretical re-evaluation method based on objective physical criteria of the data reliability  \cite{Varlamov14} could significantly decrease the relative uncertainties under discussion.

The $E1$ PSFs were extracted from photoneutron cross sections that include the emission of  particles, but do not include contributions from the 
($\gamma$,$\gamma$) channel.  Such a contribution dominates however just above the neutron separation energy. For this reason, when extracting 
PSF from photoabsorption cross sections (Eq.~\ref{eq_photo1}), only data lying sufficiently above the neutron threshold have been considered. 
More specifically, the present $E1$ PSF library only considers $\gamma$-ray energies for which the ($\gamma$,$\gamma$) cross section is expected to be at least ten times smaller than the photo-particle-emission cross section. The specific $\gamma$-ray energy interval ($\Delta \varepsilon$) for 
which the experimental cross section represents the total photoabsorption cross section was estimated using simulations of the 
photoabsorption cross section obtained using the nuclear reaction code TALYS \cite{Koning12} and typically $\Delta \varepsilon \lsimeq 1.5$~MeV. 
Below $S_n+\Delta \varepsilon$, the PSF obtained from Eq.~\ref{eq_photo1} has incorrectly small values if it is extracted from a ($\gamma$,n) 
cross section.

The procedure used to determine this low-energy cut-off ($\Delta \varepsilon$) requires a decomposition of the total photoabsorption cross 
section into partial cross sections through the following relations
\begin{equation}
\begin{array}{l}
\sigma (\gamma ,abs)=\sigma _{t} (\gamma ,\gamma )+\sigma '(\gamma ,abs), \\
\sigma ^\prime(\gamma ,abs)=\sigma (\gamma ,sn)+\sigma (\gamma ,cp)+\sigma (\gamma ,F). \\
\end{array}
\label{eq_photo3}
\end{equation}
 Here, $\sigma _{t} (\gamma ,\gamma )=\sigma (\gamma ,\gamma )+\sigma (\gamma ,\gamma ')$ is the total photon scattering cross section to excited states in the target nucleus, {\it i.e.} the sum of the cross sections of 
 elastic $\gamma$-ray scattering via different types of intermediate states (without shape-elastic component) and non-elastic $\gamma$-ray scattering;
 $\sigma ^\prime(\gamma ,abs)$ is the photoabsorption cross section with emission of the particles; 
 $\sigma (\gamma ,sn)$ is the total photoneutron reaction cross section; $\sigma (\gamma ,cp)$ is the photo-charged-particle-emission cross section 
 and $\sigma (\gamma ,F)$ the pho\-tofission  cross section. More details can be found in \cite{Plujko18,Kawano19,Wiedeking19}.

\begin{figure}
\centering
\noindent\includegraphics[width=1.\linewidth,clip]{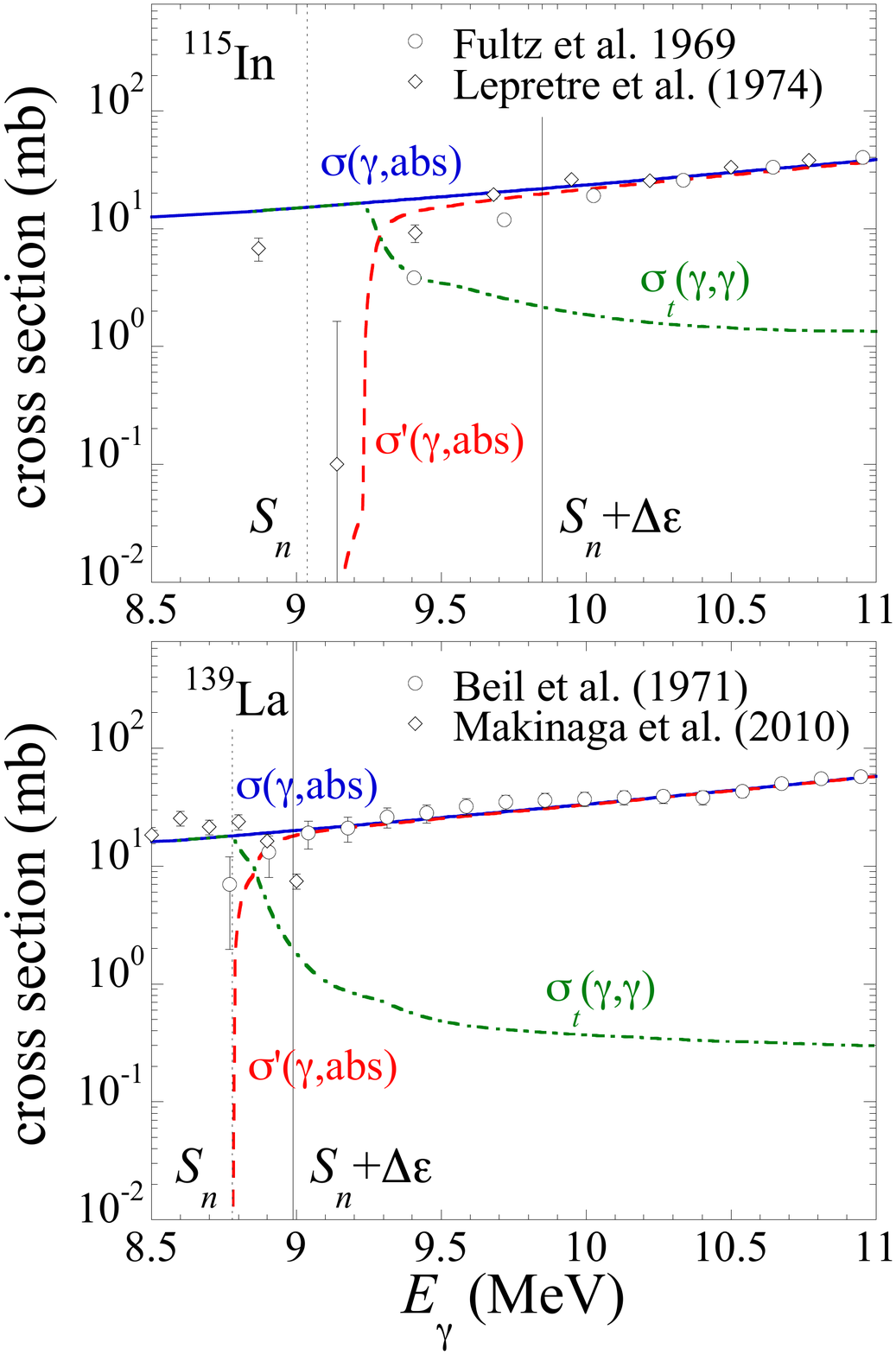}
\caption{(Color online) Comparison between theoretical and experimental \cite{EXFOR} photoabsportion cross sections for  $^{115}$In and $^{139}$La isotopes. Experimental cross sections are taken from Refs.~\cite{Beil71,Fultz69,Lepretre74,Makinaga10}. The vertical dashed lines correspond to $S_{n} $ and the vertical solid lines to the $\gamma$-ray energy $S_{n} $ + $\Delta \varepsilon $  where $\delta \sigma=0.1$. Also shown are the estimated SMLO total photoabsportion cross sections $\sigma(\gamma,abs)$ (blue solid line) and the partial cross sections $\sigma^\prime(\gamma,abs)$ (red-dashed line),  
and  $\sigma _{t} (\gamma ,\gamma )$ (green dot-dash line).}
\label{fig_photo_In_La}       
\end{figure}

For every nucleus in the photodata library, the specific energy interval $\Delta \varepsilon $ was estimated, so that the 
$\sigma _{t} (\gamma ,\gamma)$ cross section  from  ($\gamma,\gamma$) transitions does not exceed more than 10\% of the total photoabsorption cross section $\sigma (\gamma ,abs)$, {\it i.e.}
\begin{eqnarray}
\delta \sigma (E_{\gamma} =S_{n} +\Delta \varepsilon )&=&\frac{\sigma _{t} (\gamma ,\gamma )}{\sigma (\gamma ,abs)}  \nonumber \\
&=& \frac{\sigma (\gamma ,abs)-\sigma^\prime(\gamma ,abs)}{\sigma (\gamma ,abs)} = 0.1 .
\label{eq_photo4}
\end{eqnarray}
Fig. \ref{fig_photo_In_La} illustrates the experimental photoneutron cross sections for  $^{115}$In and $^{139}$La, together with the theoretical decomposition into various contributions.
The TALYS calculations were performed with the ``Simple Modified Lorentzian'' (SMLO) model of PSF (see Sec.~\ref{sec_th_smlo}),  the CT plus Fermi gas 
NLD model \cite{Koning08} and the default parameters for the additional  input quantities. 
It can be seen in Fig.~\ref{fig_photo_In_La} that neglecting the ($\gamma,\gamma$) contribution leads to a fast decrease of the 
photoabsorption cross section $\sigma^\prime(\gamma ,abs)$  for $\gamma$-ray energies approaching the neutron threshold. The experimental cross 
section for ${}^{139}$La denoted by diamonds  \cite{Makinaga10} corresponds to the ($\gamma,\gamma$) cross section measured in a NRF experiment and 
decreases just above $S_n$ due to the opening of the strong neutron emission channel.

\begin{figure}
\centering
\noindent\includegraphics[width=1.\linewidth,clip]{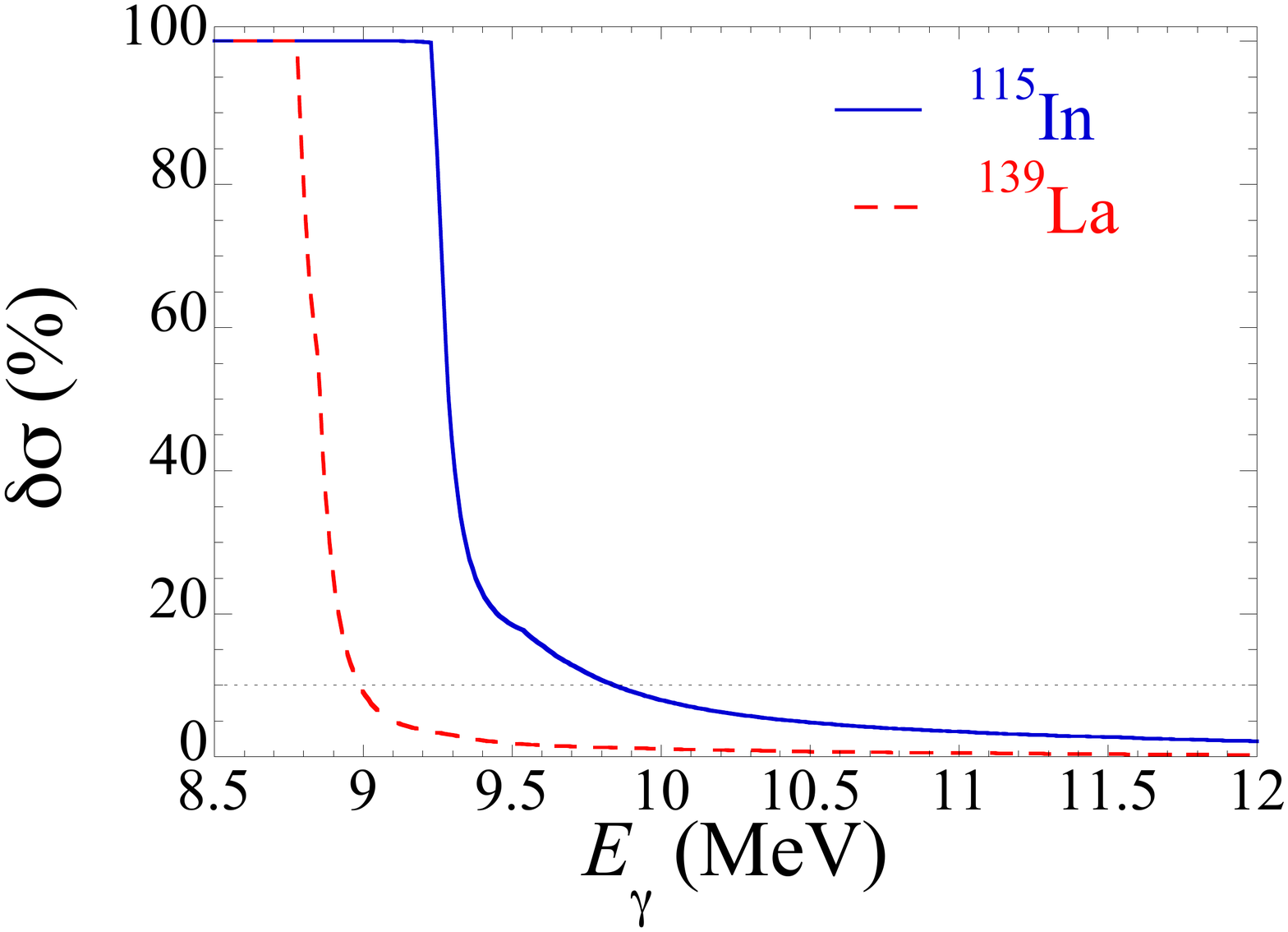}
\caption{(Color online) 
Contribution $\delta \sigma $ of  $\sigma_t (\gamma,\gamma)$ to $\sigma (\gamma,abs)$ as a function of the $\gamma$-ray energy for $^{115}$In (blue solid line) and $^{139}$La (dashed red line).}
\label{fig_photo_deltasigma}       
\end{figure}

\begin{figure}
\centering
\noindent\includegraphics[width=1.\linewidth,clip]{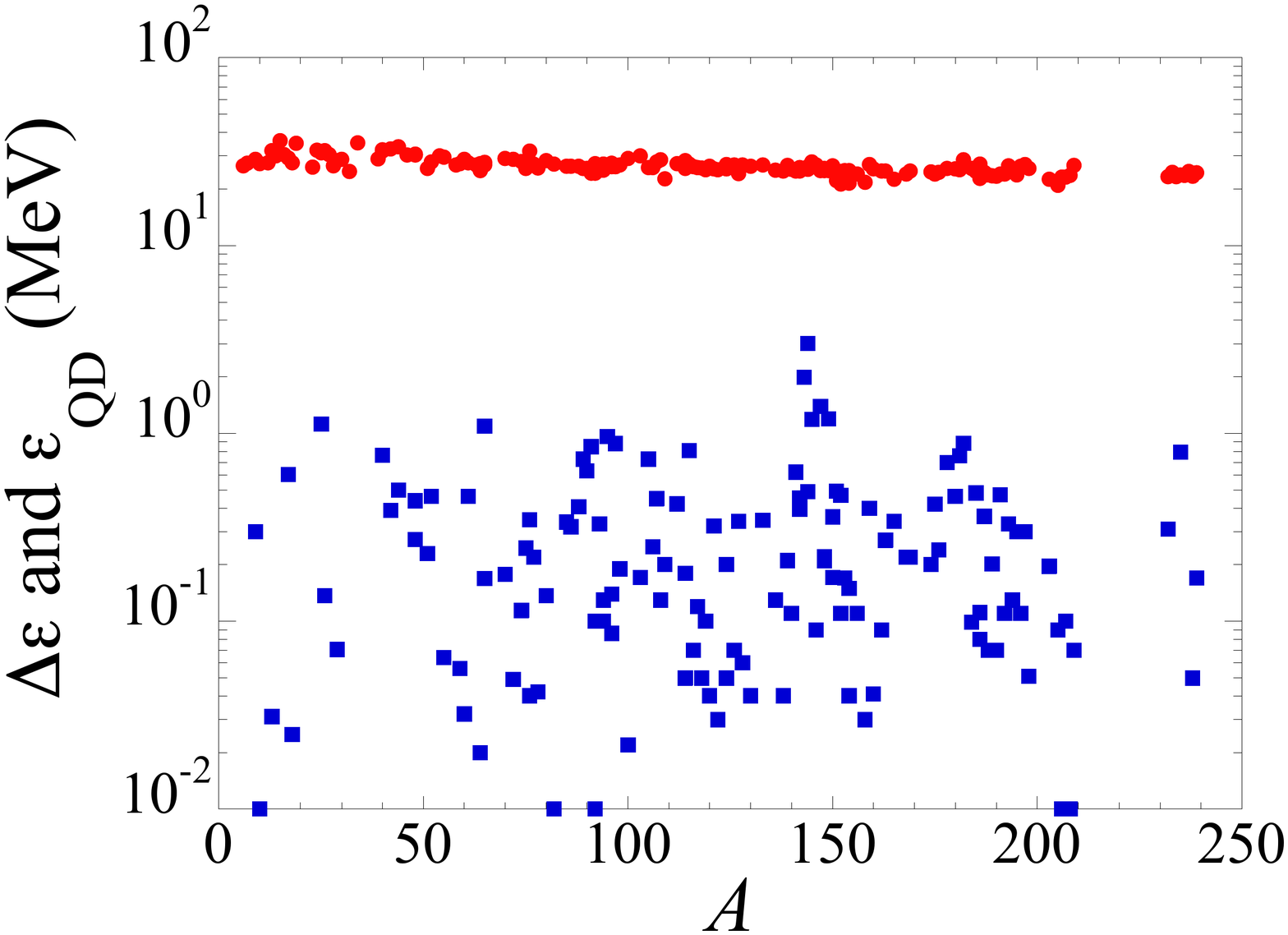}
\caption{(Color online) Low-energy cut-off $\Delta \varepsilon$ (blue squares) and high energy cut-off  $\varepsilon_{QD}$ (red circles)  as a function of  the  atomic mass number $A$. }
\label{fig_photo_borders}       
\end{figure}
Fig. \ref{fig_photo_deltasigma} illustrates the energy-dependence of the ratio of the cross section  from  ($\gamma,\gamma$) transitions to the total photoabsorption cross section (Eq.~\ref{eq_photo4}) for  ${}^{115}$In and ${}^{139}$La isotopes. Fig. \ref{fig_photo_borders} gives the resulting energy intervals $\Delta \epsilon$ for all nuclei for which the PSF has been extracted from the photonuclear library. 

Similarly to low energies in the vicinity of the neutron separation energy, at high energies the measured cross section may not reflect the 
total photoabsorption. In that case, many channels with various particle emissions are open and the quasi-deuteron breakup component dominates. 
For this reason, all PSFs from photoabsorption cross sections have been extracted only up to a maximum energy $\varepsilon_{QD}$ at which the 
quasi-deuteron component is expected to become higher than 10\% of the total photoabsorption cross section. All cross sections have been estimated on the basis of the 
SMLO PSFs (Sec.~\ref{sec_th_smlo}) and the quasi-deuteron component from the standard model of Ref.~\cite{Chadwick91}. The resulting values of $\varepsilon_{QD}$ are shown in Fig.~\ref{fig_photo_borders}.

The $E1$ PSFs were extracted from all available experimental data on photoreaction cross sections from the EXFOR database \cite{EXFOR}
and the recent update of the photonuclear library \cite{Kawano19}. 
In total, the $E1$ PSFs are given for 159 isotopes between $^{6}$Li and $^{239}$Pu including 19 elements of natural isotopic composition corresponding all together to 465 different entries.
The PSF values for $\gamma$-ray energies below $S_{n}+\Delta \varepsilon $ and above $\varepsilon_{QD}$ were discarded from the data files but are included in the README files (see Sec.~\ref{sec_data}). 


\subsection{Uncertainty analysis on test cases}
\label{sec_ass_unc}

PSFs from different experimental techniques are often not consistent \cite{Dimitriou13}. The inconsistencies can be substantial especially between results from Oslo and NRF data. The PSFs deduced from the Oslo and NRF techniques are based on several assumptions and depend, for instance, on the NLD model used during the data processing procedure. Below we describe different sources of uncertainty in these two techniques and perform an enhanced uncertainty and NLD model analysis, in two nuclei for which Oslo and NRF results exist. 

\subsubsection{Uncertainties in the NRF method}
\label{sec_ass_unc_nrf}

\begin{figure}
\epsfig{file=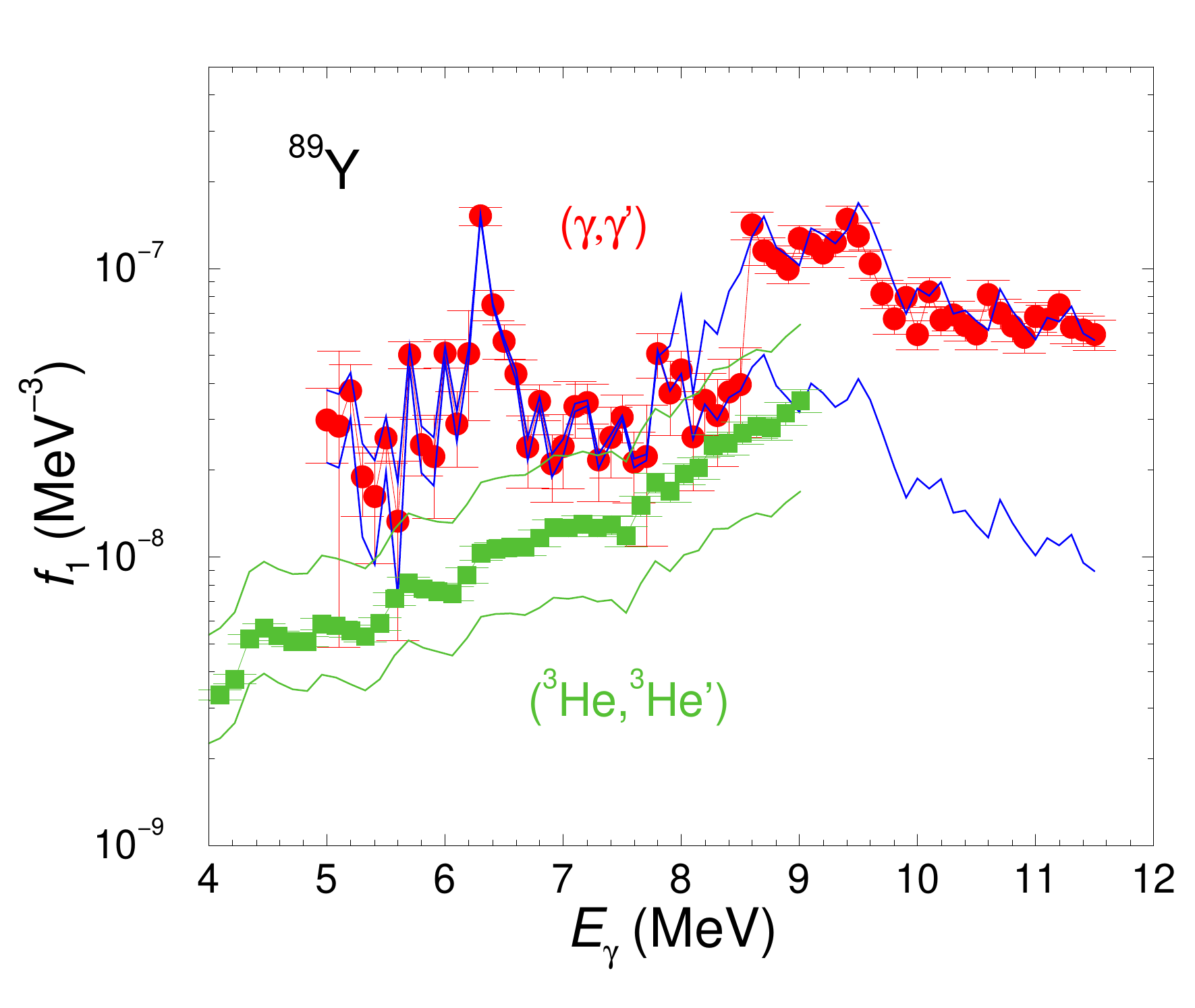,width=8.0cm}
\caption{\label{fig_89Yf1}(Color online) The PSFs deduced from ($\gamma$,$\gamma^\prime$) data of $^{89}$Y \cite{Benouaret09} (red circles). Maximum uncertainties obtained from applying extreme limits of level densities in the simulations of $\gamma$ cascades are shown by blue solid lines. 
The data were re-processed using the code $\gamma$DEX \cite{Massarczyk12} for the cascade simulations.
Oslo data from the ($^3$He,$^3$He$^\prime$) reaction \cite{Larsen16}  (green squares) are shown for comparison, together with extreme uncertainty limits (green solid lines).}
\end{figure}

\begin{figure}
\epsfig{file=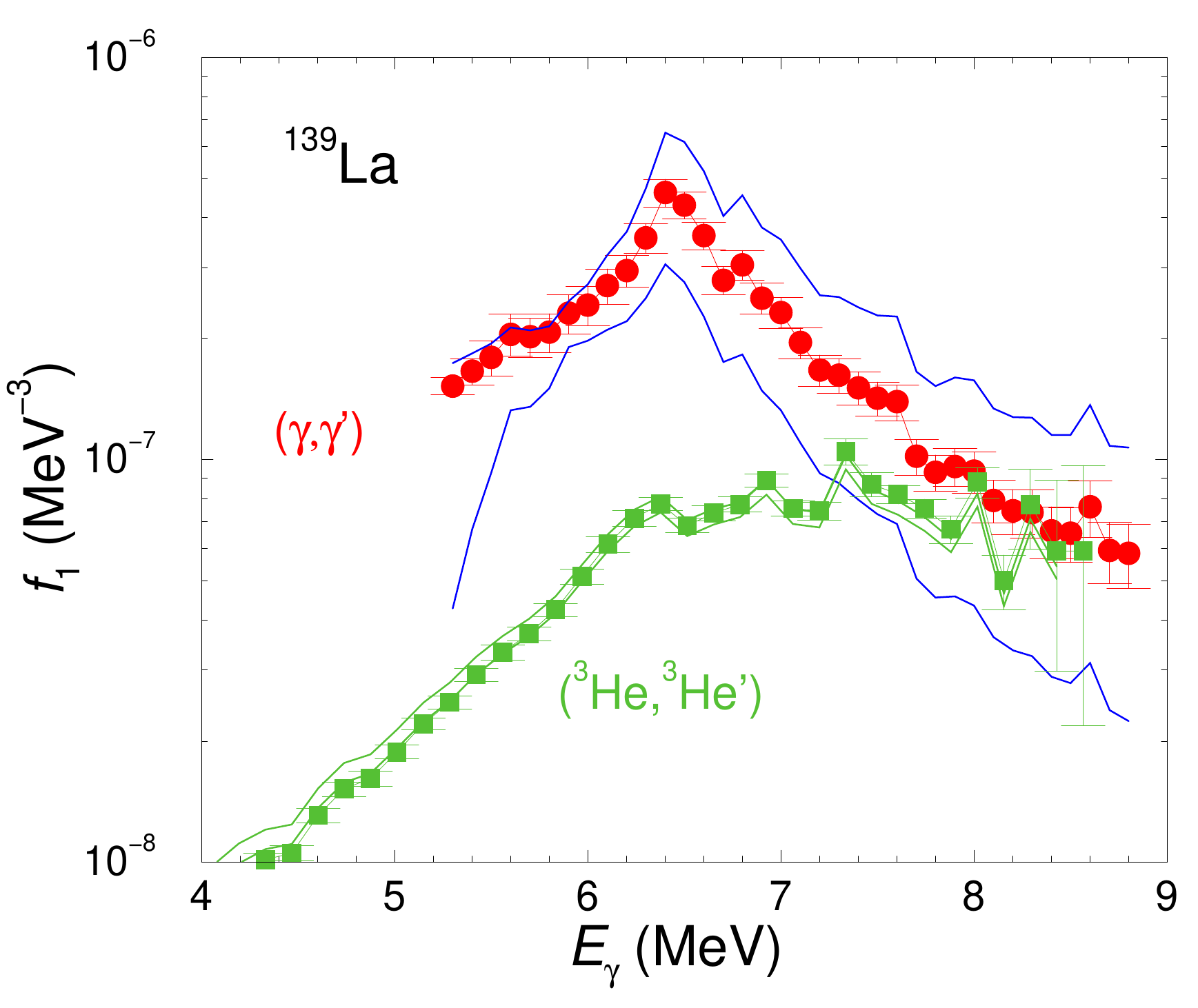,width=8.0cm}
\caption{\label{fig_139Laf1}(Color online) The PSFs deduced from  ($\gamma$,$\gamma^\prime$) data of $^{139}$La \cite{Makinaga10} (red circles). Maximum uncertainties obtained from applying extreme limits of level densities in the simulations of $\gamma$ cascades are shown by blue lines. Oslo data from the ($^3$He,$^3$He$^\prime$) reaction \cite{Kheswa17} (green squares) are shown for comparison, together with extreme uncertainty limits (green lines).}
\end{figure}

In NRF experiments, photoabsorption cross sections are deduced from intensity distributions that include resolved peaks as well as a quasicontinuum, determined as the intensity after subtracting the atomic background (see Sec.~\ref{sec_exp_nrf}). For the determination of the photoabsorption cross section the intensities of inelastic transitions have to be subtracted from the total intensity distribution. Furthermore, the remaining ground-state transitions have to be corrected for their branching ratios (see Sec.~\ref{sec_exp_nrf}). The relative intensities of elastic and inelastic transitions can be estimated by simulations of statistical $\gamma$ cascades. The initial values of the PSFs and NLDs are input data in these simulations. The initial strength functions for $E1$, $M1$, and $E2$ radiation are Lorentzian-shaped using parameters taken from the RIPL database \cite{Capote09}. Absorption cross sections are determined with an iterative technique \cite{Massarczyk12}, in which the $E1$ input PSF is taken from the output of the preceding step. Level density parameters are taken from the compilation \cite{Egidy09}. The given uncertainties are taken into account in the simulations for the CT plus Fermi gas model as well as the back-shifted Fermi gas (BSFG) model. The extreme limits of the resulting strength functions can be determined by combining PSFs obtained using the limits of the uncertainties given in Ref.~\cite{Egidy09}. This has been done for the cases of $^{89}$Y \cite{Benouaret09}
and $^{139}$La \cite{Makinaga10} in the present uncertainty analysis. Error bars include statistical uncertainties and uncertainties of detector efficiencies, of photon flux as well as a 1$\sigma$ deviation from the mean values in the individual simulations. In the present analysis, all combinations of upper and lower limits of the level-density parameters were applied. To determine the extreme
lower and upper limits of the strength functions, the values with the greatest deviations from the mean were combined. The results are shown in Figs.~\ref{fig_89Yf1}- \ref{fig_139Laf1} and are compared with the data obtained in experiments based on the Oslo method \cite{Oslo}. 

\subsubsection{Uncertainties in the Oslo method}
\label{sec_ass_unc_oslo}

For most data, uncertainties given for the PSF are only the statistical uncertainties, which are propagated through the unfolding procedure \cite{Guttormsen96}, the first generation iteration \cite{Guttormsen87} and finally the $\chi^2$ minimization procedure \cite{Schiller00} of the Oslo method.
In recent years, systematic errors have been included as well and these contributions are represented by upper and lower limits. No standard procedure was applied though. A detailed review of possible uncertainties in the Oslo method can be found in Ref.~\cite{Larsen11}.
In the Oslo method analysis, the most significant source of systematic uncertainties originates from the absolute normalization of the NLD and PSF. The slopes of the NLD and PSF are interdependent, {\it i.e.} when the slope of one is known, the slope of the other is fixed. As discussed in Sec.~\ref{sec_exp_oslo}, the NLD is typically normalized by comparison to the known discrete levels at low-excitation energy and to the NLD at $S_n$ which is determined from the average neutron resonance spacing $D_0$ and the spin cutoff parameter $\sigma_c$ in a process detailed in Ref. \cite{Schiller00}.
The number of discrete levels at low-excitation energy is usually well known and does not contribute significantly to the uncertainties. Instead, one of the main contributor to the uncertainties arises from determining the NLD at $S_n$, which includes the experimental error bars on the measured $D_0$ value and the assumptions made on the spin distribution at $S_n$. 
In some cases the libraries \cite{Capote09} and \cite{Mughabghab18} disagree and provide different recommended $D_0$ values. Such different values will contribute to the upper and lower limits of the systematic uncertainties.
 
For some nuclei, $D_0$ and/or $\langle \Gamma_\gamma \rangle$ have not been measured experimentally and the normalization procedure relies on systematics from neighbouring nuclei or on theoretical input to estimate these values. In such situations the systematic uncertainties become more significant. In most cases the measured transmission coefficients are normalized to $\langle \Gamma_\gamma \rangle$, as detailed in \cite{Voinov01}, and converted to the PSF by using Eq.~\ref{eq_dipole_psf}.

Since the presence of strong yrast transitions can result in large uncertainties in the primary $\gamma$-ray matrix, only $E_\gamma$ values above typically 1-2 MeV are used in the analysis. This results in the NLD data being available to an excitation energy 1-2 MeV below $S_n$. The absolute normalization therefore relies on an interpolation between the 
highest NLD data points measured and NLD at $S_n$, which is typically made using the CT formula \cite{Gilbert65}. The larger the gap between the last measured data point and $S_n$, the more sensitive the normalization is to the choice of the NLD model used for the interpolation.

\begin{figure}
    \centering
    \includegraphics[width=\columnwidth, angle = 0]{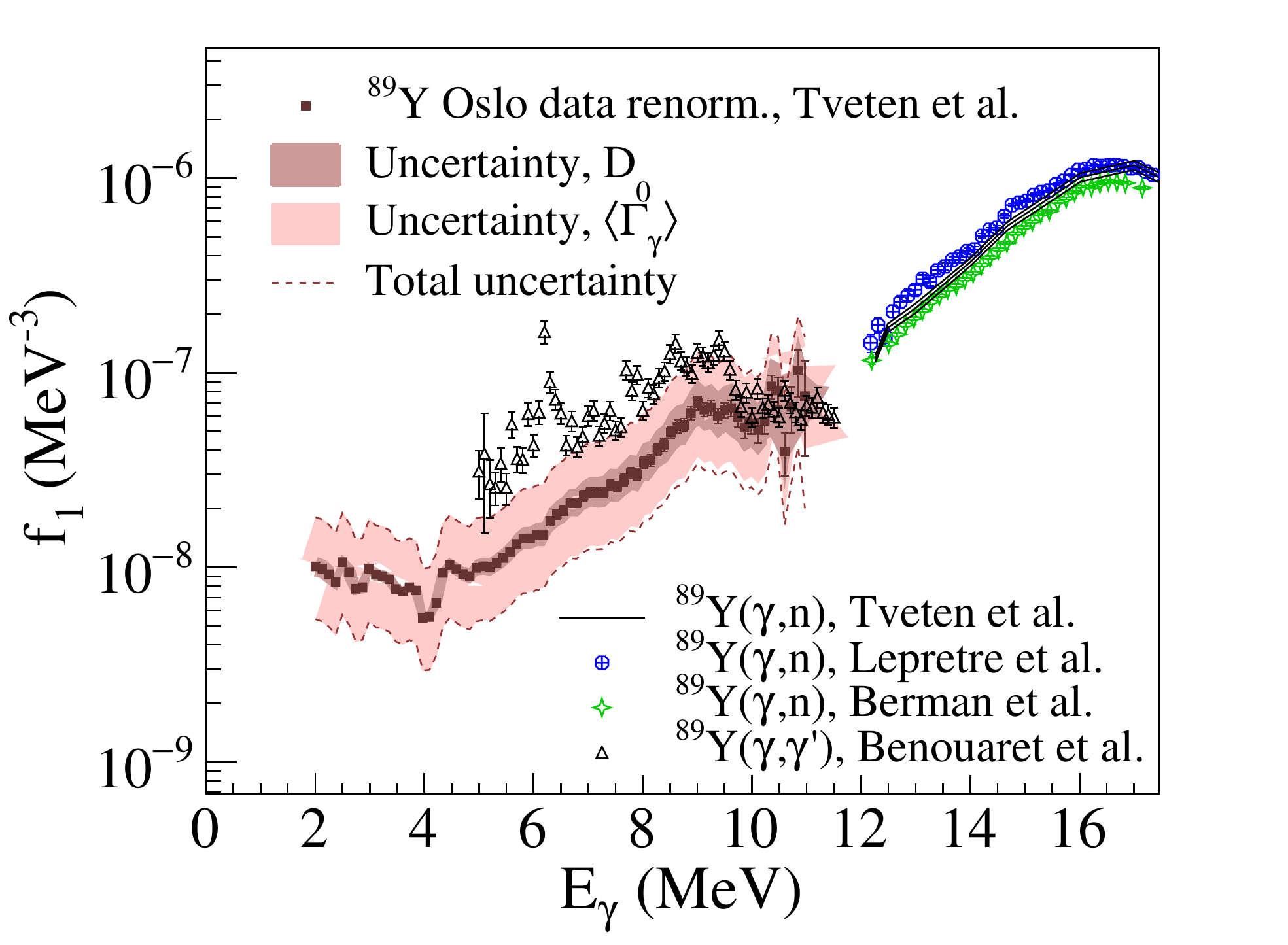}
    \caption{(Color online) The uncertainty in the PSF of $^{89}$Y broken down into the different origins. For $^{89}$Y, $\langle \Gamma_\gamma \rangle$ is not available and it had to be estimated for the data from the Oslo method (black squares). This was accomplished from systematics of neighbouring nuclei, which exhibited no clear trend but could be fitted in the extreme limit by a constant value or by an increasing slope. The value used for $^{89}$Y is the average of these two extrapolations, with the different extrapolations yielding upper and lower limits, resulting in the large systematic uncertainties seen as the light-shaded area.  Oslo data (squares) and ($\gamma$,n) (solid line) data are taken from Ref. \cite{Tveten19}. NRF data (triangles) are taken from \cite{Benouaret09} while photodata (circles, crosses) are from \cite{Berman67,Lepretre71}.}
\label{fig_unc_89Y}
\end{figure}

Different procedures have been applied to estimate upper and lower limits for systematic uncertainties. For $^{89}$Y, shown in Fig.~\ref{fig_unc_89Y}, the different components contributing to the total uncertainty are decomposed. The statistical uncertainties are relatively small and are given as error bars on the data points. The dark shaded area represents the systematic uncertainties due to the $D_0$ value. In the case of $^{89}$Y the largest contribution to the systematic uncertainties is due to the unavailability of the average total radiative width $\langle \Gamma_\gamma \rangle$ needed to normalize the PSF. This is shown by the light-shaded area of Fig.~\ref{fig_unc_89Y}. This  uncertainty analysis was published in Ref. \cite{Tveten19}.

Finally, note that in the case of heavy isotopes, especially in the actinide region, a small proportion of the available spins may be populated by charged-particle-induced reactions. A similar situation is found in the beta-Oslo method \cite{Spyrou14,Spyrou17,Larsen18} for which the nucleus of interest is populated by $\beta$-decay. In these cases, a good knowledge of the populated spin distribution is crucial for a proper determination of the PSF by the Oslo method, as being recently studied in Ref.~\cite{Zeiser19}. 

\subsection{Oslo method versus NRF experimental PSFs}
\label{sec_ass_rec}

So far, for nine nuclides the PSFs below the neutron separation energy have been studied in NRF experiments at the 
bremsstrahlung facility $\gamma$ELBE \cite{Schwengner05} as well as in charged-particle-induced reactions at OCL. 
{\bf
There are several nuclei showing considerable differences in the shape and magnitude of the PSFs from these experiments. 
}
We have compared the results of uncertainty analyses for both methods. For the NRF method, this was the first time such an analysis was performed. In the case of the Oslo method, we use 
the analyses performed in Ref.~\cite{Kheswa17} and Ref.~\cite{Larsen16} for $^{139}$La and $^{89}$Y, respectively.

For $^{89}$Y and $^{139}$La, considerable differences are found in the shape and magnitude of the PSFs despite the uncertainty analysis which includes statistical and systematic uncertainties.
The NRF data have been processed as described in Sec.\ref{sec_exp_nrf}. The intensities of ground-state transitions were corrected for their branching ratios to deduce the absorption cross section and the error bars of these cross section values include statistical uncertainties and uncertainties of detector efficiencies, of photon flux as well as a $1\sigma$ deviation from the mean values in the individual simulations. In the cascade simulations, described in Sec.~\ref{sec_ass_nrf} and \ref{sec_ass_unc_nrf}, the BSFG level density was used with the parameters given in Ref. \cite{Egidy09}. The parameters were varied within their uncertainties in the individual realizations of level schemes. 
In the present analysis, combinations of extreme upper and lower limits of the level-density parameters were applied. In addition to the BSFG level densities, the CT plus Fermi gas model was also tested with the uncertainties as given in Ref. \cite{Egidy09}. On average, the BSFG NLDs result in greater cross sections compared with those resulting from CT plus Fermi gas NLDs. To determine the extreme lower and upper limits of the cross sections, values with the greatest deviation were combined. 
Figs.~\ref{fig_89Yf1} and ~\ref{fig_139Laf1} illustrate the effect of the above mentioned uncertainties on the NRF data.  

In the case of the Oslo method, the analytical methodology follows the prescription of Sec.\ref{sec_exp_oslo} and the procedure yields a functional form
for $\rho(E_f)$ and $T_\gamma(E_\gamma)$ which must be normalized to known experimental data to obtain physical solutions. The statistical uncertainties are carried through the Oslo method (Sec.~\ref{sec_ass_unc_oslo}). For $^{139}$La, the systematic model dependencies have been explored \cite{Kheswa17}, where two theoretical models were used to obtain different values of $\rho(S_n)$. These are the parity-dependent Hartree-Fock Bogoliubov (HFB) plus combinatorial model \cite{Goriely08b} and the CT plus Fermi gas model where both parities are assumed to have equal contributions \cite{Koning08}. In the latter
case, two spin cut-off parameter prescriptions were considered. Thus we explored a total of three different normalizations. The first normalization with the CT plus Fermi gas model is based on the spin cut-off parameter of Ref. \cite{Egidy05} and in the second approach, $\rho(E_x,J)$ was calculated with the spin cut-off parameter equation as implemented in the TALYS code \cite{Koning12}.

For $^{139}$La, the $D_0$ and $\langle \Gamma_\gamma \rangle$ values are averages
of experimental values taken from \cite{Mughabghab18, Capote09}. The three different normalizations are included in Fig.~\ref{fig_139Laf1} in the form of the upper and lower uncertainty bands. The agreement of PSFs from NRF and Oslo data is not very satisfying. Significant deviations are apparent for $E_\gamma < 7.5$ MeV. This raises the question if certain structures effects are enhanced/reduced depending on the reaction used. While the Oslo method includes both $E1/M1$ isoscalar and isovector components, the NRF method probes exclusively the isovector component. Alternative experimental approaches, using (p,p$^\prime$) and ($\alpha$,$\alpha^\prime$) reactions at zero degree relative to the beam, using magnetic spectrometers to investigate the $E1$ isoscalar and isovector components specifically, may be necessary to fully understand the discrepancy and to disentangle the different components of the PSF in $^{139}$La. 
In this context, we also note that recent shell model calculations \cite{Sieja18} for $^{108,134}$Xe show structural differences between $M1$ strengths for photoabsorption from the ground state and photoemission from a number of high-lying states. This suggests that the photoabsorption approach is more sensitive to nuclear structure effects for energies below 5 MeV. Similar effects may play a role for $E1$ excitations at higher energies and may provide a possible explanation for the discrepancies observed between NRF and Oslo data. This may limit the ability to compare Oslo and NRF data principally because of the different excitation mechanisms.

In the case of $^{89}$Y, the uncertainty analysis of NRF data is performed as outlined above for $^{139}$La. For the Oslo data \cite{Larsen16}, the spin distribution up to the neutron separation energy is determined from three different approaches; {\it i)} a phenomenological BSFG spin-cutoff parameter \cite{Egidy09}, {\it ii)} the BSFG spin-cutoff parameter of Ref. \cite{Egidy06} where a rigid-body moment of inertia is assumed, and {\it iii)} microscopic calculations within the HFB plus combinatorial model \cite{Goriely08b}. The absolute values and slope of the  NLD are obtained by normalizing to discrete levels at low-excitation energy and to average $s$-wave neutron resonance spacing at $S_n$ when this information is available. However, in this nucleus the NLD at $S_n$ is not available and instead, systematics of $s$-wave resonance spacings for this mass region are considered by using  the RIPL-3 database \cite{Capote09}. More specifically, the average value for $^{88,89}$Sr, $^{90}$Y and $^{91}$Zr are used to estimate $D_0$. In addition, systematic errors due to the spin distribution at $S_n$ are taken into account \cite{Larsen11}. The uncertainty bands including these dependencies are shown in Fig.~\ref{fig_89Yf1}.

Within the uncertainties, there is an overlap between the NRF and Oslo PSF data for most parts of the energy region for $^{89}$Y. The main discrepancy appears from structures observed in the NRF data around 6.5 MeV which are not seen in the Oslo method data.

\begin{figure}
\epsfig{file=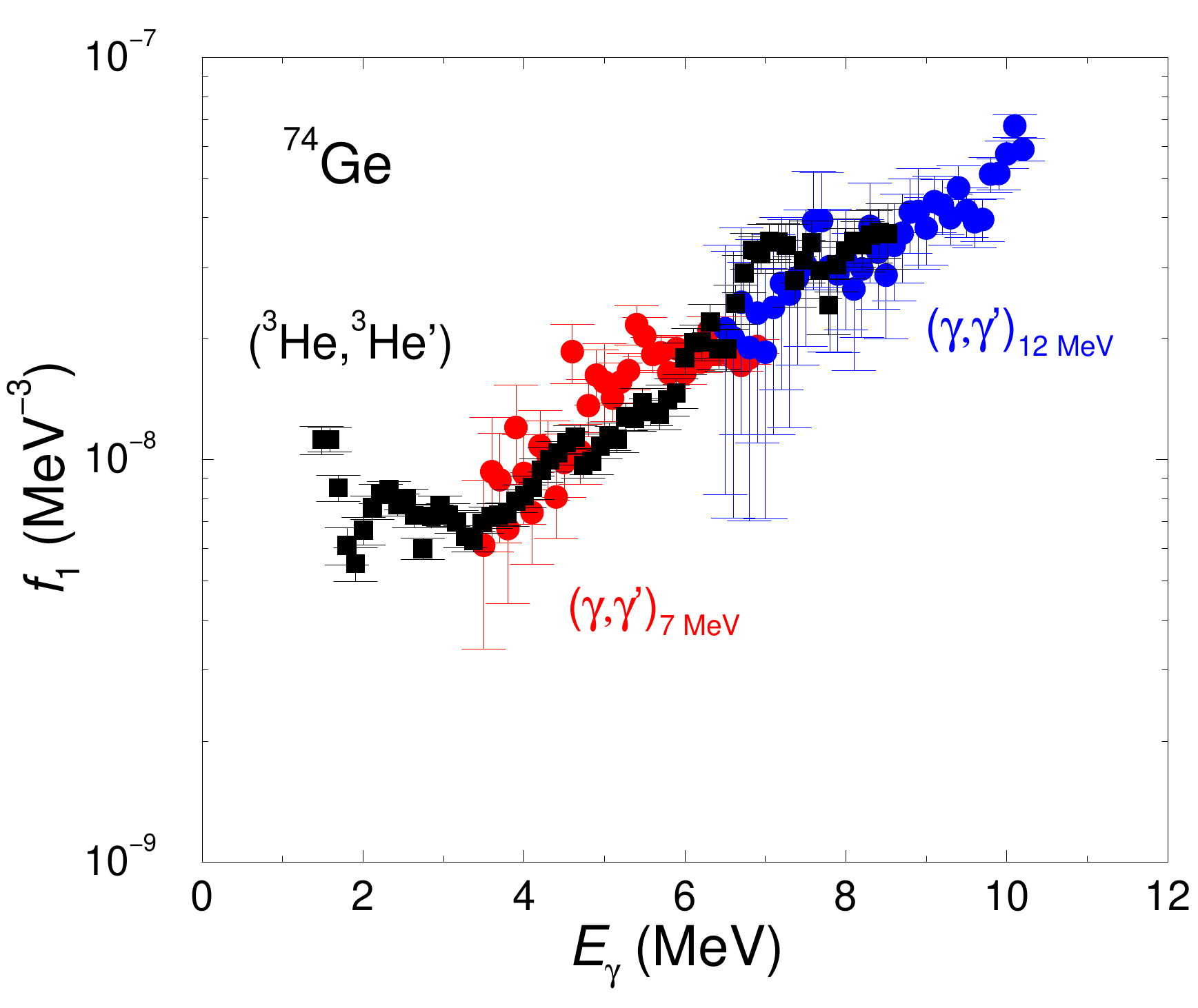,width=8.0cm}
\caption{\label{fig_74Ge}(Color online) $^{74}$Ge PSFs deduced from $(\gamma,\gamma')$ data (red and blue circles corresponding to two different maximum $\gamma$ beam energies) and from the Oslo method using the ($^3$He,$^3$He$^\prime$) reaction (black squares). Only statistical uncertainties are shown.}
\end{figure}

The best agreement between NRF and Oslo methods is found in the case of $^{74}$Ge \cite{Renstrom16, Massarczyk15}, as shown in Fig.~\ref{fig_74Ge}, 
despite the fact that an error analysis similar to that of $^{89}$Y and $^{139}$La has not been performed. So far, $^{74}$Ge is  the only case where both PSF data sets can be recommended. 
 One may speculate that the extra strength which is observed in several nuclei around 6 to 9 MeV in NRF measurements (see Figs.~\ref{fig_89Yf1}, \ref{fig_139Laf1} as well as 
Fig.~\ref{fig_comp_nrf} in Sec.~\ref{sec_comp_nrf}) is most pronounced in nuclei near shell closures. It results from prominent $1^- \rightarrow  0^+$ transitions, which in the case of 
ion-induced reactions are mixed with many other transitions (if they are excited in the same way at all). No such prominent peaks are found in the NRF PSF of deformed nuclei, such as $^{74}$Ge. In this case, the strength is fragmented and is found to be compatible with the Oslo PSF in contrast to spherical nuclei.

\section{PSF models}
\label{sec_th}

\subsection{Introduction}
\label{sec_th_intro}

The total photon transmission coefficient $\mathcal{T_{\gamma}}$ (Eq.~\ref{eq_Tg}) from an excited state is normally dominated by the dipole $E1$ and $M1$ transitions. 
Simple semi-classical models assume a Lorentzian shape for the photoabsorption cross section that is dominated by a giant resonance, at least for medium- and heavy-mass nuclei.  Experimental photoabsorption data confirms the simple semi-classical picture of a Lorentzian shape at energies around the $E1$ resonance energy \cite{Axel62,Berman75}.  The photonuclear data near the peaks of the GDR can be fitted by Lorentz, Breit-Wigner or Gaussian functions equally well, but their low- and high-energy tails differ significantly \cite{Bartholomew73,Gardner84}. Lorentz and Breit-Wigner shapes of the photoabsorption cross sections can be derived from different theoretical approaches \cite{Brink55,Plujko08b}. The Lorent\-zian shape can be transformed to a  Breit-Wigner form but with a shape width that depends on the photon energy, resonance energy and width of the Lorentzian \cite{Plujko18}. A Lorentz shape is more suitable for fitting the photonuclear data because the standard Breit-Wigner expression is obtained without taking account of time reversal invariance and is adequate for describing a strong resonance state when the width is small with respect to the resonance energy \cite{Dover72}. However, both line shapes correspond to a nuclear response to an electromagnetic field which proceeds through the excitation of one strong collective state that exhausts the energy-weighted sum rule. The photoabsorption and $\gamma$-decay processes that occur on the wings of the GDR are governed by the excitation of states of a different nature and therefore, the fitting of experimental data by Lorentzian functions should be limited to small energy ranges around the GDR peak in order to obtain reliable values for the GDR parameters.
For axially deformed nuclei, the GDR is found experimentally to be well represented by the sum of two Lorentzian-like components. This double peak structure is interpreted as the collective vibrations along and perpendicular to the axis of symmetry \cite{Axel62,Berman75}. Extension to the superposition of three Lorentzian functions to describe triaxility has also been considered \cite{Alhassid88,Alhassid90,Grosse17}.

Significant deviations from a standard Lorentzian (SLO) have been observed in both theoretical and experimental studies below the neutron threshold \cite{McCullagh81,Allen82,Kopecky90}. In particular, low-energy $E1$ DRC data were explained by introducing an energy-dependent width of the SLO functional to reduce the $E1$ strength around the neutron separation energy \cite{McCullagh81}. A generalization of the Lo\-rentzian shape, to account for data below and above the neutron threshold, was achieved through the inclusion of an energy- and temperature-dependent width. This family of Lorentzian models is based on the theory of Fermi liquids \cite{Kadmenskii83,Kamerdzhiev04} and has shown to improve significantly the calculations of the experimental radiative widths and $\gamma$-ray spectra \cite{Kopecky90,Goriely98,Capote09,Plujko11,Plujko18}. Until recently, this generalized Lorentzian (GLO) functional has been the only $E1$ model used in practical applications, and more specifically in global calculations for large sets of nuclei. While the $E1$ mode has been widely studied, less effort has been devoted to the parametrisation of the $M1$ PSF. For the $M1$ PSF, the most commonly used formula is an SLO expression describing the spin-flip mode only \cite{Kopecky90,Capote09} that neglects  the low-energy $M1$ mode for deformed nuclei (the so-called scissors mode). Only a few works \cite{Schramm12,Mumpower17,Grosse17,Goriely18b} proposed a systematic phenomenological description of the low-energy scissors mode. 

The Lorentzian GDR approach, however, even in the generalized form, suffers from shortcomings of various sorts. On the one hand, it is unable to predict the enhancement of the $E1$ strength at energies around the neutron separation energy (such as the pygmy resonance) as demonstrated by different experiments (see {\it e.g.} \cite{Kneissl06,Tonchev10,Savran11,Savran13}). On the other hand, even if a Lorentzian function provides a suitable representation of the $E1$ strength, the location of its maximum and its width remain to be predicted from a model for each nucleus or from systematics. For nuclear applications, these properties have often been obtained from a droplet-type model \cite{Myers77}. This approach clearly lacks reliability when dealing with exotic nuclei. The situation is even less satisfactory for the $M1$ component, where systematics are limited or scarce due to the limited amount of experimental information available. 

In view of this situation, and considering that the GDR properties and low-energy resonances may influence substantially the predictions of radiative capture cross sections, it is clearly of great interest to develop PSF models of the microscopic type which are expected to be more reliable and have some predictive power. Since the early 70', different mean-field approaches, such as the QRPA, the quasiparticle-phonon model and some of their improved variants, have been developed and successfully applied to the description of giant multipole resonances in both the non-relativistic  \cite{Bertsch73,Liu76,Blaizot77,Khan04,Kamerdzhiev04,Peru08,Schwengner08,Peru11,Martini16,Arsenyev17,Deloncle17,Colo01,Sarchi04,Tsoneva08,Papakonstantinou09,Papakonstantinou12,Papakonstantinou15,Avdeenkov11,Achakovskiy15,Gambacurta18} 
and relativistic 
\cite{Vretenar01,Ring01,Ma02,Ring07,Pena08,Daoutidis12a,Litvinova13a,Litvinova13b,Egorova16}
mean field frameworks. The nuclear shell model has also been extensively used to describe electromagnetic excitations \cite{Sober85,vonNeumann98,Caurier05,Heyde10,Loens12,Schwengner13,Schwengner17,Brown14,Sieja17,Sieja17b,Sieja18,Midtbo18,Karampagia17}, but restricted to light nuclei.
Despite such a huge effort in developing microscopic models, only a few attempts have been made to provide systematic large-scale calculations that can compete with more phenomenological Lorentz-type models.  The QRPA $E1$ PSFs obtained within the Hartree-Fock+BCS \cite{Goriely02} as well as HFB \cite{Goriely04,Martini16,Goriely18a} frameworks have been shown to reproduce satisfactorily the location and width of the GDR and the ARC data at low energies for the bulk of existing data. The aforementioned QRPA calculations have been performed for all the $8\le Z\le 110$ nuclei lying between the two drip lines. In the neutron-deficient region as well as along the valley of $\beta$-stability, the QRPA distributions are very close to a Lorentzian profile. Similar attempts have been performed within the relativistic mean-field framework but in a less systematic way \cite{Daoutidis12a,Egorova16}.

Due to the applicability to a large set of nuclei and the fact that they are inherently different approaches, both the phenomenological Lorentzian-type and the mean field plus QRPA models are proposed for developing global models of the dipole PSF. The recommended theoretical PSFs are detailed below and their ability to reproduce experimental data is discussed in Sec.~\ref{sec_comp}.

\subsection{Phenomenological  $E1$ \& $M1$ SMLO model}
\label{sec_th_smlo}

In this section, we describe the phenomenological SMLO model that was developed to estimate the $E1$ and $M1$  
PSFs \cite{Plujko18,Goriely18b} for all nuclei with $8\le Z\le 124$. Both $E1$ and $M1$ SMLO formulas have been adjusted on a large number of experimental data, as shown in Sec~\ref{sec_comp}.

\subsubsection{The $E1$ SMLO model}
\label{sec_th_smlo_e1}

For the $E1$ PSF for cold and heated nuclei, we consider  a rather simple expression given by \cite{Plujko18,Goriely18b}
\begin{eqnarray}
\overleftarrow{f_{E1}}(E_{\gamma},T)=&& \frac{1}{3\pi^2\hbar^2c^2} \frac{1}{1-\exp(-E_{\gamma}/T)} \sigma_{TRK} \times \nonumber \\
&&  \frac{2}{\pi} \sum_{j=1}^{j_m} s_{r,j} \frac{E_\gamma ~\Gamma_j(E_{\gamma},T)}{(E_\gamma^2-E_{r,j}^2)^2+E_\gamma^2 \Gamma_j(E_{\gamma},T)^2},
\label{eq_smlo}
\end{eqnarray}
where $T$ denotes the temperature of the heated nucleus, $j_m$ the number of normal vibration modes of the GDR excitation ($j_m=1$ for spherical nuclei and 2 for axially deformed ones) and $\sigma_{TRK}$ the $E1$ Thomas-Reiche-Kuhn (TRK) sum rule given by 
\begin{equation}
\sigma_{TRK}=60 \frac{NZ}{A}=15A(1-I^2) ~{\rm (mb~MeV)},
\label{eq_trk}
\end{equation}
where $I=(N-Z)/(N+Z)$ is the neutron-proton asymmetry factor. The Lorentzian function in Eq.~\ref{eq_smlo} is characterized by GDR parameters corresponding to the peak energy $E_{r,j}$, the width at half maximum $\Gamma_j$ and the possible deviation of the integrated cross section from the TRK sum rule $s_{r,j}$. 

The SMLO width, $\Gamma_j$, related to the relaxation mechanism of the giant vibration $j$-mode, is taken to be energy- and tempe\-rature-dependent as follows
\begin{equation}
\Gamma_j(E_\gamma,T)=\frac{\Gamma_{r,j}}{E_{r,j}} \left(E_{\gamma}+\frac{4\pi^2}{E_{r,j}}T^2 \right),
\label{eq_gamj}
\end{equation}
where the linear dependence on the energy $E_{\gamma}$ arises from the inverse  $E_{\gamma}$-dependence of the average squared matrix element in the transitions of the 1~particle -- 1~hole states to 2~particles -- 2~holes states \cite{Capote09}. The quadratic temperature dependence in Eq.~\ref{eq_gamj} originates from the Fermi liquid theory \cite{Kadmenskii83}.

The GDR resonance energies of the $j$-mode are taken such that $E_{r,j=1}<E_{r,j=2}$, for deformed nuclei while for spherical nuclei, $E_{r,j=1}=E_{r,j=2}$. These energies are connected to the energies $E_a$ and $E_b$ of the vibrations along and perpendicular to the symmetry axis (note that for prolate nuclei, we take $E_{r,1}=E_a$ and $E_{r,2}=E_b$, while for oblate ones, $E_{r,1}=E_b$ and $E_{r,2}=E_a$). Finally, the $s_{r,j}$ factor gives the weight of the $j$-mode with respect to the TRK sum rule.

Whenever experimental photoabsorption data are available in the vicinity of the GDR, the GDR parameters $E_{r,j}$, $\Gamma_{r,j}$ and $s_{r,j}$ are adjusted to the data. A compilation of such data can be found in Ref.~\cite{Plujko18}. However, when no data exists, systematics of the GDR parameters is used. Such a systematics was obtained by performing a least-square fit to the recommended experimental GDR parameters in spherical nuclei as well as deformed nuclei in the $150<A<190$ and $220<A<253$ ranges, where to a good approximation deformed nuclei can be considered as axially deformed. The following expression was adopted for the centroid energy $E_r$ of the GDR: 
\begin{equation}
E_{r}=e_1 (1-I^2)^{1/2} \frac{A^{-1/3}}{(1+e_2 A^{-1/3})^{1/2}}  ,
\label{eq_er}
\end{equation}
where $e_1=128.0\pm 0.9$~MeV and $e_2=8.5\pm 0.2$. Eq.~\ref{eq_er} corresponds approximately to the eigenenergy of the GDR vibration within the hydrodynamical liquid drop model \cite{Myers77}, and agrees with sum rule prescriptions \cite{Lipparini89,Bohigas79,Gleissl90}. For deformed nuclei, we assume the equiprobability of the normal mode excitations and the twofold degeneracy of the giant collective vibration perpendicular to the axis of symmetry.  In this case, the centroid  energy  is expressed as $E_r=(E_a+2E_b)/3$ and the energies $E_a$ and $E_b$ along the two ellipsoid semi-axes are approximated as 
\begin{eqnarray}
E_a & = & \frac{3E_r}{1+2D_{ab}}\\
E_b &=& D_{ab} E_a  ,
\label{eq_ab}
\end{eqnarray}
where  $D_{ab}=0.911a/b+0.089$ is determined from the ratio of the ellipsoid semi-axis lengths $a/b=(1+\alpha_2)/(1-\alpha_2/2)$, which in turn is a function of the quadrupole deformation parameter $\beta_2$ since $\alpha_2=\sqrt{5/4\pi}\beta_2$.

Slight deviations from the TRK sum rule are known to exist from experimental photoabsorption data. For this reason, the possible deviation $s_{r,j}$ of the $j$-mode is estimated assuming that $s_\Sigma=\sum_j s_j=1.2$, {\it i.e.} $s_1=s_\Sigma/3$, $s_2=2s_\Sigma/3$ for prolate nuclei and $s_1=2s_\Sigma/3$, $s_2=s_\Sigma/3$ for oblate nuclei. 

The GDR width is estimated from a simple power-law expression $\Gamma_{r,j}=c E_{r,j}^d$ with $c=0.42\pm0.05$~MeV and $d=0.90\pm0.04$. More details on the model and the adjustment can be found in Refs.~\cite{Plujko08,Plujko11,Plujko18,Goriely18b}. The prescription for calculating the high-energy quasi-deuteron contribution, {\it i.e.}, the photoabsorption cross section leading to the production of a neutron -- proton pair can be found in Ref.~\cite{Plujko18}.

Finally, the temperature $T$ is derived from the excitation energy $U$ using a simple Fermi gas expression. Since the temperature entering Eq.~\ref{eq_smlo} corresponds to the temperature of the final state, it reads $T=\sqrt{(U-E_\gamma)/\tilde{a}}$ where the level density parameter $\tilde{a}=A/10$~MeV$^{-1}$ is adopted. More details can be found in \cite{Plujko18,Goriely18b}. Note that the temperature dependence gives rise to a non-zero limit of the $E1$ PSF at zero $E_\gamma$ energy, as shown in Fig.~1 of Ref.~\cite{Goriely18b}. The photoabsorption PSF for cold nuclei can directly be deduced from the deexcitation PSF as $\overrightarrow{f_{E1}}(E_\gamma)=\overleftarrow{f_{E1}}(E_\gamma,T=0)$.

\subsubsection{The $M1$ SMLO model}
\label{sec_th_smlo_m1}

An empirical $M1$ PSF \cite{Goriely18b} was built, inspired by the HFB+ QRPA strength \cite{Goriely18a} obtained with the D1M Gogny force \cite{Goriely09a} (hereafter D1M+QRPA, see Sec. \ref{sec_th_qrpa})  considering an SLO-type function for both the low-energy scissors ($sc$) mode and the spin-flip ($sf$) components, {\it i.e.}
\begin{eqnarray}
\overrightarrow{f_{M1}^{SMLO}}(E_\gamma)= & & \frac{1}{3\pi^2\hbar^2c^2} \sigma_{sc} \frac{E_\gamma ~\Gamma_{sc}^2}{(E_\gamma^2-E_{sc}^2)^2+E_\gamma^2 \Gamma_{sc}^2}  \nonumber \\
&&  + \frac{1}{3\pi^2\hbar^2c^2}\sigma_{sf} \frac{E_\gamma ~\Gamma_{sf}^2}{(E_\gamma^2-E_{sf}^2)^2+E_\gamma^2 \Gamma_{sf}^2} ,
\label{eq_smlo_m1}
\end{eqnarray}
where $\sigma_i$ is the peak cross section, $E_i$ the energy at the peak and $\Gamma_i$ the width at half maximum for both the spin-flip ($i=sf$) or the scissors mode ($i=sc$). 

Following the $A$ and deformation dependences of both the spin-flip and scissors mode resonances predicted within the D1M+QRPA approach, the energy, width and strength of the Lo\-rentzian-type function have been determined in a simple manner.  The D1M+QRPA calculations predict that, globally, the spin-flip peak energy scales like $A^{-1/6}$ and the peak strength increases linearly with $A$, so that the peak cross section scales like $A^{5/6}$. For the scissors mode, present only in deformed nuclei, the centroid energy remains rather constant between 4~MeV for light nuclei and 3~MeV for heavier ones. It can be rather well described by the simple $A^{-1/10}$ decreasing function \cite{Goriely18b},
while the peak strength $f_{sc}=\sigma_{sc}/E_{sc}$ is found to be globally proportional to $A$ and to the quadrupole deformation parameter $\beta_2$. The amplitude of the strength is then determined by comparing the $M1$ approximation (Eq.~\ref{eq_smlo_m1}) with existing data, namely ARC data \cite{Kopecky17} for the spin-flip model and NRF data \cite{Pietralla98} in the rare-earth region for the scissors mode. 

When considering the deexcitation PSF, deviations from the photoabsorption strength can be expected, especially for $\gamma$-ray energies approaching the zero limit. In particular, shell-model calculations \cite{Schwengner13,Brown14,Sieja17,Sieja17b,Sieja18,Karampagia17,Schwengner17,Midtbo18} predict an exponential increase of the $M1$ deexcitation PSF at decreasing $\gamma$-ray energies approaching zero. Such an upbend of the PSF observed experimentally (see {\it e.g.} \cite{Voinov04,Guttormsen05}) has therefore been assumed to be of $M1$ nature, though no strong experimental evidence for this assignement exists at the moment \cite{Jones18}. For the deexcitation $M1$ PSF, the zero-$E_\gamma$ limit determined in Ref.~\cite{Goriely18a} can be added to the photoabsorption expression, leading to 
\begin{equation}
\overleftarrow{f_{M1}^{SMLO}}(E_\gamma)=\overrightarrow{f_{M1}^{SMLO}}(E_\gamma)+ C \exp(-\eta E_\gamma)  ,
\label{eq_smlo_m2}
\end{equation}
where the parameters $C$ and $\eta$ can be tuned on available shell-model and low-energy experimental results \cite{Goriely18a,Goriely18b}.

The final parameters of the three $M1$ modes read 
\begin{itemize}
\item the spin-flip resonance: $\sigma_{sf}=0.03A^{5/6}$~mb, $E_{sf}=18~A^{-1/6}$ MeV and $\Gamma_{sf}=4$~MeV;
\item the scissors mode: $\sigma_{sc}=10^{-2}|\beta_2| A^{9/10}$~mb, $E_{sc}=5~A^{-1/10}$ MeV and  $\Gamma_{sc}=1.5$~MeV;
\item the upbend: $\eta=0.8$ and $C =3.5 \times 10^{-8}~e^{-6\beta_2}$~MeV$^{-3}$,
\end{itemize}
where the final amplitude of the spin-flip and scissors mode strength has been globally tuned on ARC and NRF data. The quadrupole deformation parameter $\beta_2$ can be extracted from mean field calculations, such as HFB \cite{Goriely09a}. More details can be found in Ref.~\cite{Goriely18b}.

\subsection{Mean-Field + QRPA model}
\label{sec_th_qrpa}

With respect to phenomenological approaches that have just been described, the reliability of the PSF predictions can be greatly improved by the use of microscopic or semi-microscopic models. Such an effort can be found in Refs. \cite{Goriely02,Goriely04,Daoutidis12a,Martini16,Goriely16b,Goriely18a} where a complete set of $E1$ and $M1$  PSFs was derived from mean field plus QRPA calculations. When compared with experimental data and considered for practical applications, all mean field plus QRPA calculations need however some phenomenological corrections. These include a broadening of the QRPA strength  to take the neglected damping of  collective motions into account as well as a shift of the strength to lower energies due to the contribution beyond the 1 particle - 1 hole excitations and the interaction between the single-particle and low-lying collective phonon degrees of freedom  \cite{Colo01,Sarchi04,Tsoneva08,Papakonstantinou09,Papakonstantinou12,Papakonstantinou15,Avdeenkov11,Achakovskiy15,Gambacurta18,Litvinova13b,Egorova16}.
In addition, most of the mean field plus QRPA calculations assume spherical symmetry, so that phenomenological corrections need to be included in order to properly describe the splitting of the giant dipole resonance in deformed nuclei. State-of-the-art calculations including effects beyond the 1 particle - 1 hole excitations and phonon coupling are now available \cite{Colo01,Sarchi04,Tsoneva08,Papakonstantinou09,Papakonstantinou12,Papakonstantinou15,Avdeenkov11,Achakovskiy15,Gambacurta18,Litvinova13b,Egorova16} but they remain computer-wise intractable for large-scale applications.

Axially-symmetric-deformed QRPA calculations based on HFB calculations using the finite-range Gogny interaction have been shown to provide rather satisfactory predictions of the $E1$ and $M1$ strengths \cite{Peru08,Peru11,Peru14,Martini16,Goriely16b,Deloncle17,Goriely18a}.  The effects beyond the 1 particle - 1 hole QRPA are empirically included by considering an energy shift that increases with energy. More specifically, both the $E1$ and $M1$  QRPA strengths are shifted by an energy of $\Delta=0.5$~MeV for $E_\gamma \le 0.5$~MeV,  $\Delta=2.5$~MeV for $E_\gamma= 18$~MeV and $\Delta=5$~MeV for $E_\gamma \ge 21$~MeV. For energies in the  $0.5\le E_\gamma\le 21$~MeV range, the energy shift $\Delta$ is interpolated linearly between the anchor values at 0.5, 18 and 21~MeV.
Similarly, an empirical damping of the collective motions is introduced in the QRPA strength by folding each $E1$ strength by a SLO function of width $\Gamma$ that has been adjusted on photoabsorption data and is assumed to differ for both possible projections $K$ of the angular momentum, but also to be dependent on the atomic mass $A$ and the quadrupole deformation $\beta_2$. More precisely, for the $E1$ strength, the width is expressed as $\Gamma(K=0^-)=\Gamma_0/(1+\beta_2)$ and $\Gamma(K=1^-)=\Gamma_0\times(1+\beta_2)$, where  $\Gamma_0~{\rm (MeV)}=7-A/45$ for $A\le 200$, and 2.5~MeV otherwise. For the $M1$ strength, a constant value of $\Gamma=0.5$~MeV is adopted \cite{Goriely16b}. Note that these QRPA calculations are applied to even-even nuclei, the PSF for odd nuclei being derived by interpolation \cite{Martini16}.The resulting model is referred to as  D1M+QRPA.

When considering the deexcitation PSF, deviations from the photoabsorption strength are expected, especially for $\gamma$-ray energies approaching the zero limit. As proposed in Ref.~\cite{Goriely18a}, a constant $E1$ strength and an $M1$ upbend, both inspired by shell-model calculations can be assumed for $\gamma$-ray energies approaching zero \cite{Schwengner13,Brown14,Sieja17,Sieja17b,Sieja18,Karampagia17,Schwengner17,Midtbo18}. The $E1$ and $M1$ PSFs, including the low-energy contributions and hereafter denoted as D1M+QRPA+0lim, finally read
\begin{eqnarray}
\overleftarrow{f_{E1}^{QRPA}}(E_\gamma) & = &  f_{E1}^{QRPA}(E_\gamma) +f_0 U / [1+e^{(E_\gamma-E_0)}] \label{eq_qrpa1}\\ 
\overleftarrow{f_{M1}^{QRPA}}(E_\gamma)  &=&  f_{M1}^{QRPA}(E_\gamma) +  C ~ e^{-\eta E_\gamma} , \label{eq_qrpa2}
\end{eqnarray}
where $f^{QRPA}_{X1}$ is the D1M+QRPA dipole strength at the photon energy $E_\gamma$, $U$ is the excitation energy of the initial deexciting state and $f_0$, $E_0$, $C$, and $\eta$  are free parameters that have been adjusted on shell-model results and available low-energy experimental data such as those obtained with the Oslo method (see {\it e.g.} \cite{Voinov04,Guttormsen05,Algin08}), the average radiative widths \cite{Capote09} or MSC and MD spectra \cite{Krticka19}. Such a study \cite{Goriely18a,Krticka19} led to $f_0=10^{-10}$~MeV$^{-4}$, $E_0=3$~MeV, $\eta=0.8$~MeV$^{-1}$, and $C= 1\times 10^{-8}$~MeV$^{-3}$ for all nuclei with $A \ge 105$ and $C=3 \times 10^{-8} \exp(-4\beta_{2})$ MeV$^{-3}$ for lighter nuclei \cite{Krticka19} (where the qua\-drupole deformation parameter $\beta_2$ is taken consistently from HFB calculations based on the D1M Gogny force \cite{Goriely09a}).
The deformation dependence of the zero-$E_\gamma$ limit $C$ is mainly based on shell-model calculations \cite{Sieja17b,Karampagia17,Schwengner17,Midtbo18},  the $M1$ upbend being less pronounced for deformed nuclei for which part of the strength is transferred into the scissors mode region. 

\section{Comparison between experiments and models}
\label{sec_comp}

In addition to capturing the underlying physics, the recommended theoretical models of dipole PSF should also be able to reproduce the existing experimental data.
The PSF data described in  Secs.~\ref{sec_exp} and \ref{sec_ass} are available for a relatively large number of nuclei and span a broad region of $\gamma$-ray  energies of $\sim 1-25$~MeV, allowing us to probe either the $E1$ and $M1$ components separately or the total dipole PSF. In the present section, the theoretical models described in Sec.~\ref{sec_th} are compared with the following data:
\begin{itemize}
\item NRF data  for the total dipole strength below $S_n$ for 23 nuclei; the extraction of PSF is sensitive to the adopted NLD model; NRF methods also provide integrated $M1$ strength for  about 47 nuclei in the $2-4$~MeV region;
\item Oslo data for the total dipole PSF below $S_n$ for 72 nuclei; the extraction of PSF is sensitive to the adopted normalisation for both the PSF and NLD;
\item ARC and DRC data in the $5-8$~MeV region for 88 nuclei which are available separately for the $E1$ and $M1$ modes; extracted PSF values are sensitive to the systematics adopted within the normalisation procedure;
\item PSFs determined from intensities of primary transition following the thermal neutron capture, including $E1$ data for 206 nuclei and $M1$ data for 153 nuclei;
\item (p,$\gamma$) data for 22 nuclei with $50 \lsimeq A \lsimeq 90$ in the energy range of $1-4$~MeV below the proton separation energy;
\item (p,p$^\prime$) data for $^{96}$Mo, $^{120}$Sn and $^{208}$Pb available separately for the $E1$ and $M1$ modes;
\item Photoneutron and photoabsorption data for 159 nuclei which are sensitive to the dominant $E1$ PSF in the GDR region;
\item Singles $\gamma$-ray spectra from thermal neutron capture yielding information on the PSF  below $S_n$ for 5 nuclei; comparative simulations are sensitive to the adopted NLD model;
\item MSC and MD spectra from individual neutron resonances  yielding information on the PSF below $S_n$. Simulations are sensitive to the adopted NLD model. The comparison was performed for 15 nuclei, in some of them for spectra from resonances of different spin and parity;
\item TSC spectra following thermal neutron capture for 2 nuclei; predictions are sensitive to the adopted NLD model 
\item Average radiative width $\langle \Gamma_\gamma \rangle$ yielding information on the integrated PSF below $S_n$ for 228 nuclei; estimates of the average radiative width are sensitive to the adopted NLD model;
\item MACS at 30~keV yielding information on the integrated PSF below $S_n$ for 240 nuclei; estimates of the MACS are sensitive to the adopted NLD model.
\end{itemize}
Comparisons between the above-mentioned data and the two theoretical models, namely SMLO and D1M+QRPA+0lim, are illustrated in the following subsections.

\subsection{Comparison with NRF data}
\label{sec_comp_nrf}

\begin{figure*}
\includegraphics[scale=0.6]{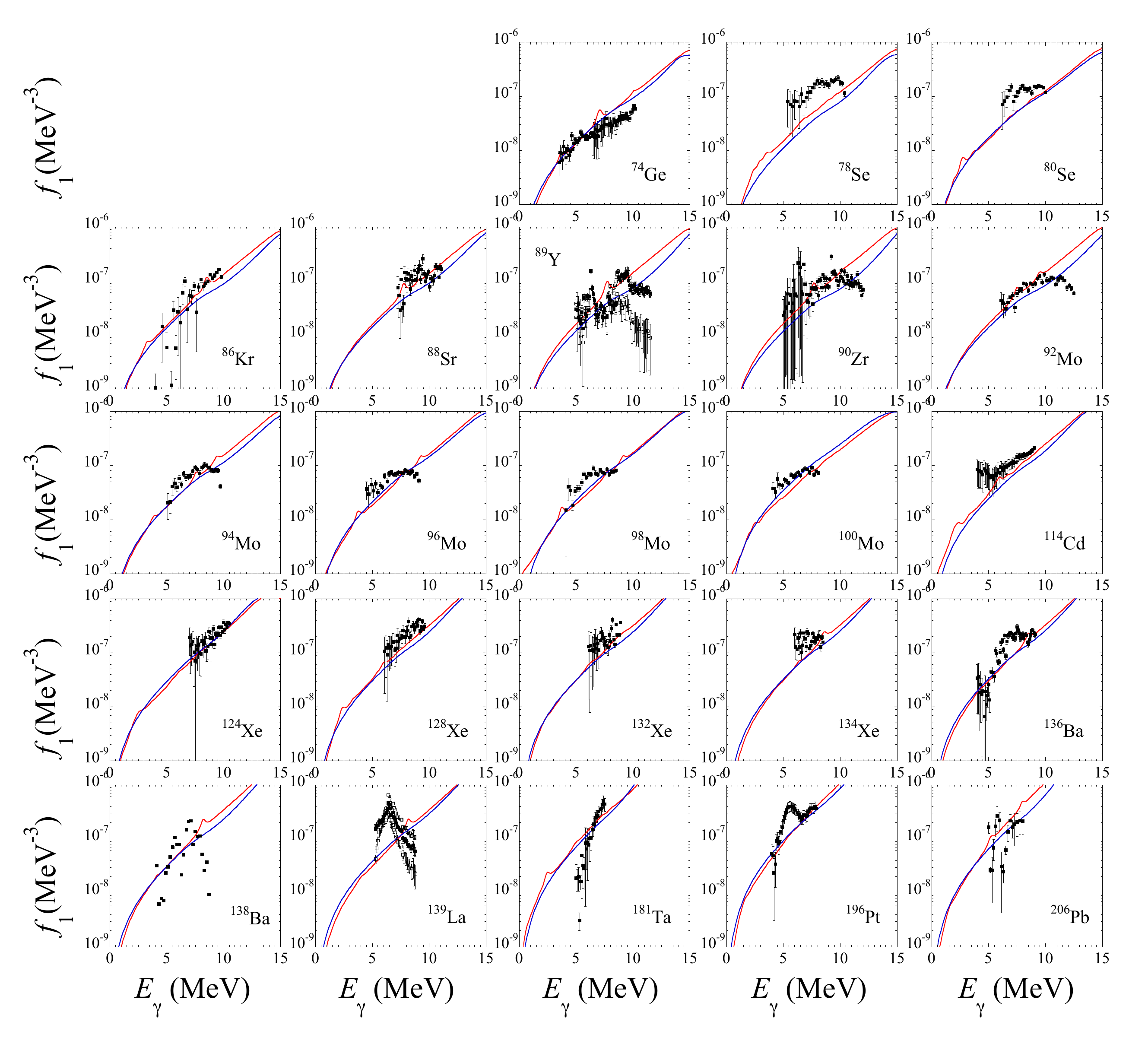}  
\caption{(Color online)  Comparison between experimental total dipole PSFs extracted from NRF data (Sec.~\ref{sec_ass_nrf}) (black squares) and theoretical $T=0$ SMLO (blue lines) and D1M+QRPA (red lines) models (without the zero energy limit). Systematic uncertainties due to model dependencies are only included for $^{89}$Y and $^{139}$La (open squares).}
\label{fig_comp_nrf}
\end{figure*}

The total dipole PSFs extracted from NRF measurements are compared with the SMLO and D1M+QRPA models in Fig.~\ref{fig_comp_nrf} for all the 23 different nuclei for which data exist. Data extend from the neutron separation energy down to around $4-5$~MeV. Note that in this case, the PSF data correspond to photoexcitation strengths, so that no temperature dependence or additional low-energy contribution is included in the calculations. A rather fair agreement is obtained, except in some cases like Se isotopes, $^{139}$La, $^{181}$Ta or $^{196}$Pt for which the energy dependence obtained experimentally does not agree with the model description. It should however be stressed that the NRF data presented in Fig.~\ref{fig_comp_nrf} do not include systematic uncertainties due to model dependencies, except for $^{89}$Y and $^{139}$La, as discussed in  Sec.~\ref{sec_ass_unc_nrf}. 

Some extra strength around 6 to 9 MeV is observed in NRF data, especially for nuclei near closed shells. This strength stems 
from prominent $1^-$ to $0^+$ transitions that in the case of ion-induced reactions are mixed with many other transitions. It could consequently be attributed to the experimental method itself, as discussed in Sec.~\ref{sec_ass_unc_nrf}.
Without a full uncertainty analysis to reveal the impact of  input NLD models on the shape and magnitude of the PSF, we are unable to draw any definite conclusions about the agreement between experiment and model calculations. In addition, discrepancies between PSFs obtained with different methods are also observed for some of these nuclei, as discussed in Sec.~\ref{sec_database_exp} (see in particular Fig.~\ref{fig_allexp_other}). For this reason, the presence of a low-lying dipole strength, known as the pygmy resonance, is not discussed here. The readers are referred to Ref.~\cite{Savran13}.

\begin{figure}
\begin{center}
\includegraphics[scale=0.33]{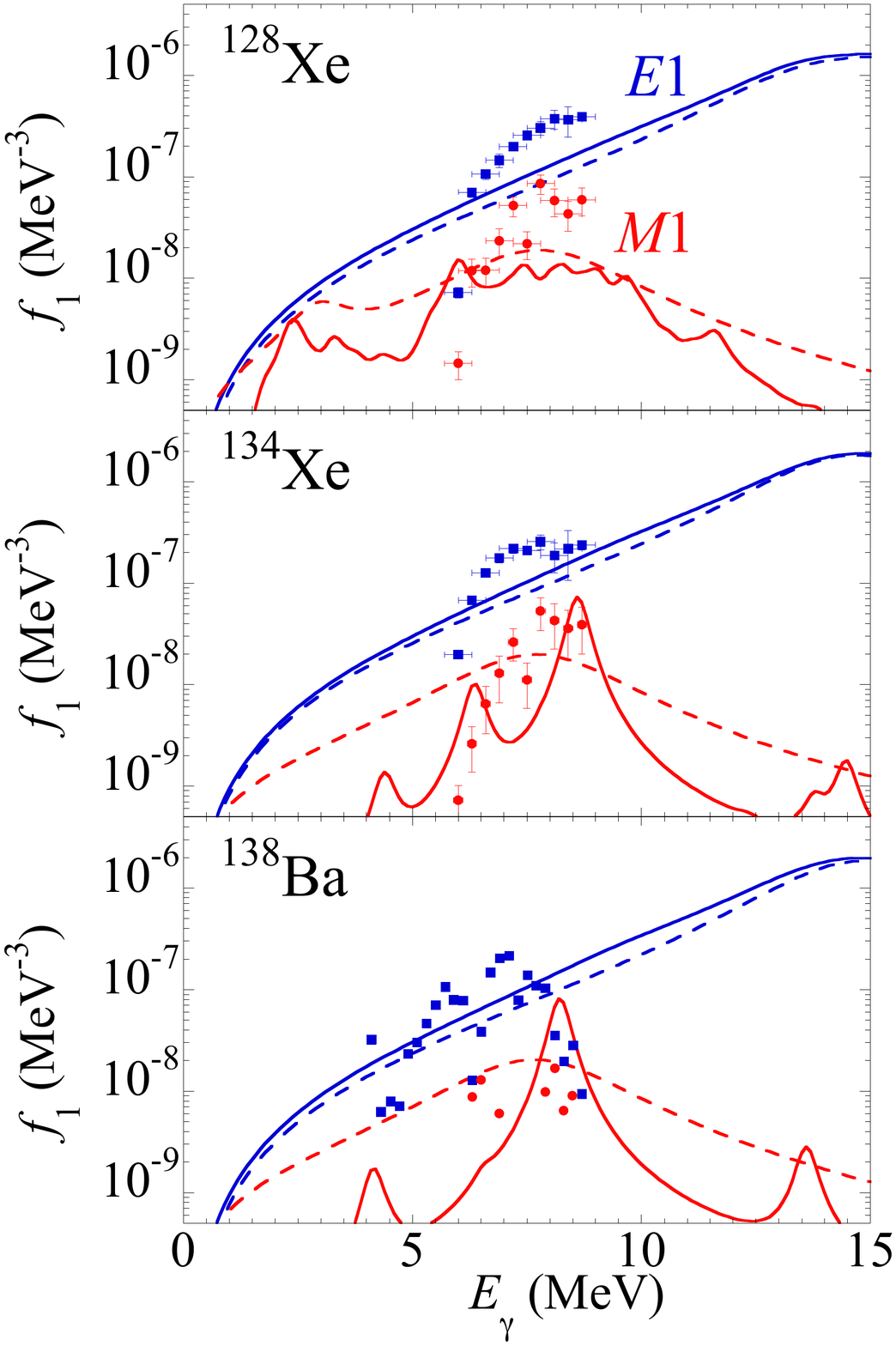}  
\caption{(Color online) Comparison between experimental $E1$ (blue squares) and $M1$ (red dots) PSFs extracted from NRF data for $^{128}$Xe, $^{134}$Xe \cite{Massarczyk14} and 
$^{138}$Ba \cite{Tonchev10} and theoretical $T=0$ SMLO (dashed lines) and D1M+QRPA (solid lines) models (without the zero energy limit). The $E1$ PSFs are shown in blue and the $M1$ in red.}
\label{fig_comp_nrf_E1M1}
\end{center}
\end{figure}

In Fig.~\ref{fig_comp_nrf_E1M1}, we compare the experimental $E1$ and $M1$ PSFs of the slightly deformed $^{128}$Xe and spherical  $^{134}$Xe and $^{138}$Ba nuclei obtained with  quasi-mono\-energetic and linearly polarized $\gamma$-ray beams art HI$\gamma$S \cite{Tonchev10,Massarczyk14} with the D1M+QRPA and SMLO models. Like in Fig.~\ref{fig_comp_nrf}, the dominant $E1$  strength obtained with NRF differs from the Lorentzian-like theoretical PSFs in the 6--9~MeV region. As can be seen in Fig.~\ref{fig_comp_nrf_E1M1}, the SMLO width for the $M1$ spin-flip mode is assumed to  be constant ({\it i.e.} $A$- and $E_\gamma$-independent), in contrast to what is obtained with the D1M+QRPA calculation and observed experimentally.

\begin{figure}
\includegraphics[scale=0.3]{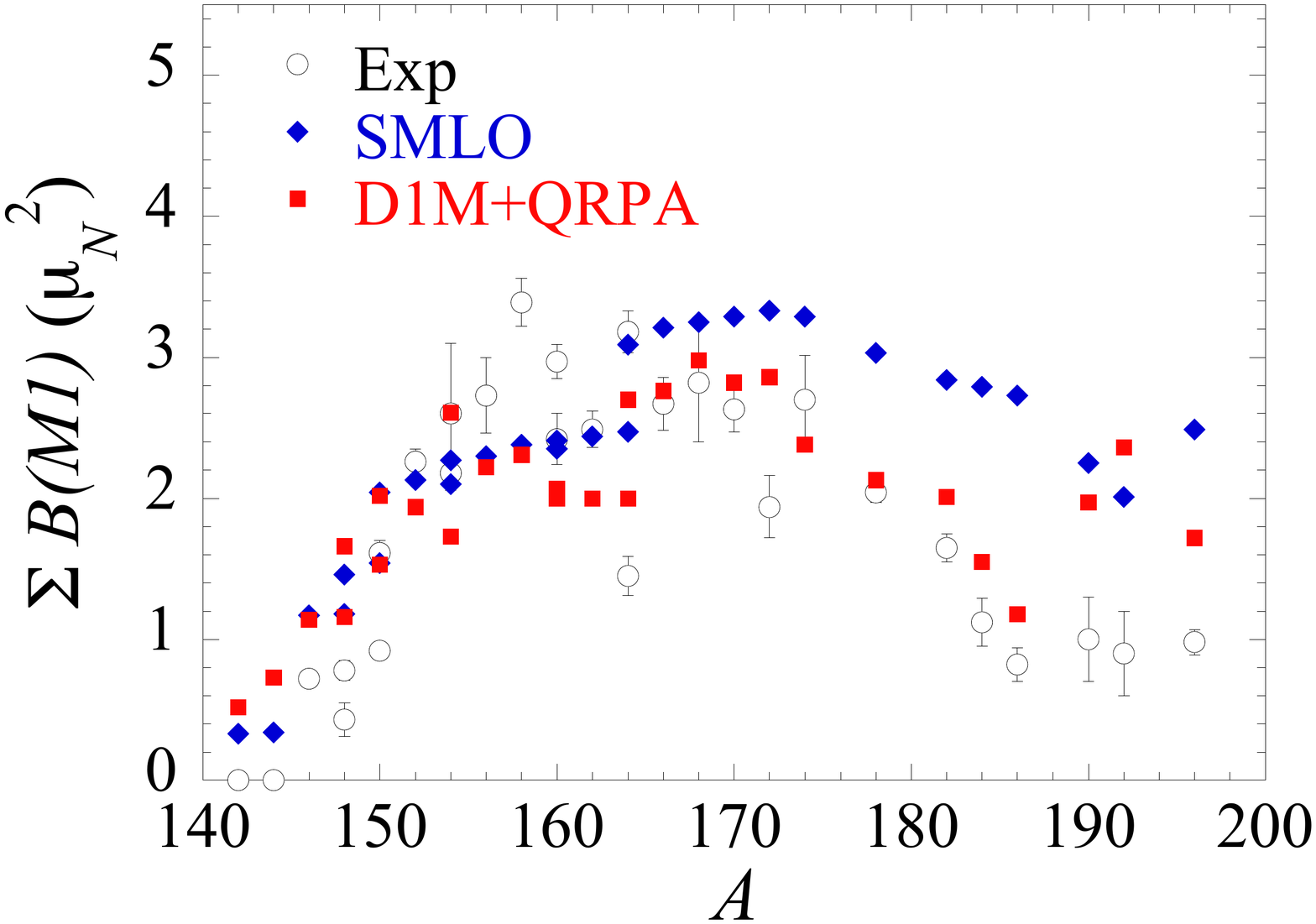}  
\caption{(Color online)  Comparison between experimental \cite{Pietralla95,Fransen99,Linnemann03} (open circles), SMLO (blue diamonds) and D1M+QRPA (red squares) values of the integrated 
strength $\sum B(M1)(\mu^2_N)$ for rare-earth nuclei in the well-defined energy range of 2--4~MeV corresponding to the measurements.}
\label{fig_sumM1_exp_th}
\end{figure}

The NRF method also allows to estimate the integrated strength that can provide valuable information on the low-energy PSF, especially for the $M1$ mode 
\cite{Laszewski86,Laszewski87,Laszewski88,Alarcon89,Richter90,Frekers90,Pietralla95,Luttge96,Fransen99,Linnemann03,Adekola11,Mettner87}. Fig.~\ref{fig_sumM1_exp_th} compares the integrated  experimental $M1$ strength obtained from ($\gamma$,$\gamma^\prime$) NRF experiments  with the one predicted by the SMLO and D1M+QRPA models for the rare-earth nuclei with $140 \lsimeq A \lsimeq 195$. The data integrated over energies ranging from 2 to 4~MeV typically correspond to the scissors mode (circles in Fig.~\ref{fig_sumM1_exp_th}). Overall agreement is observed between experimental and calculated integrated strength. 
In the deformed-spherical transition region around $A=180-190$, the SMLO model tends to overestimate the experimental data due to uncertainties in the determination of the quadrupole deformation parameter \cite{Goriely18b}. A better description is found with the D1M+QRPA model.

The excitation energy of the scissors mode is found to be systematically located around 3~MeV for all rare-earth nuclei for which experimental data is available \cite{Heyde10,Pietralla98}. Both the D1M+QRPA and SMLO mean energies are compatible with the measured data \cite{Goriely16b,Goriely18b}.

\subsection{Comparison with Oslo data}
\label{sec_comp_oslo}

\begin{figure*}
\includegraphics[scale=0.6]{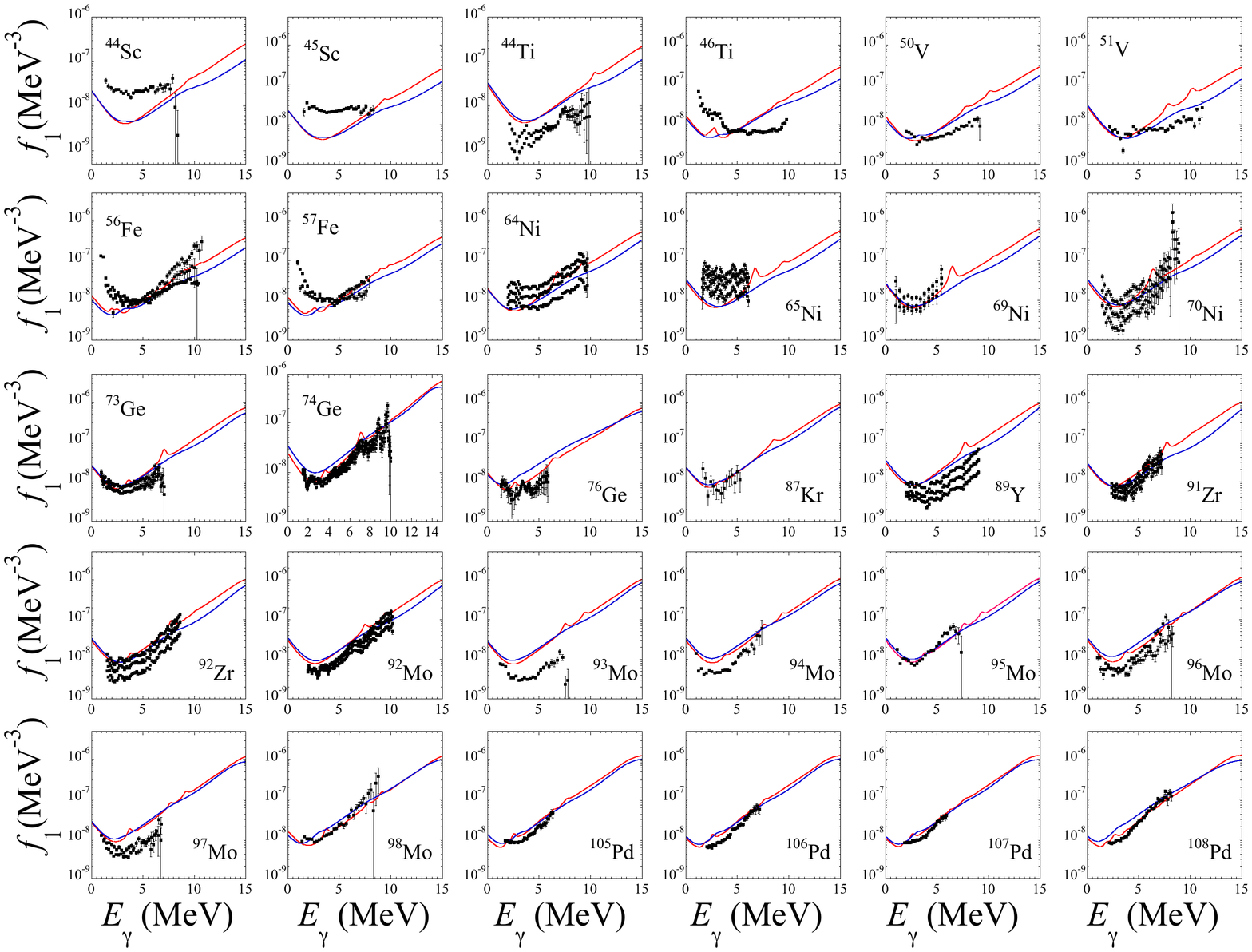}  
\caption{(Color online)  Comparison between experimental total dipole PSFs extracted from the Oslo method (Sec.~\ref{sec_ass_oslo}) (black squares) and theoretical SMLO (blue lines) and D1M+QRPA+0lim (red lines) models for nuclei between Sc and Pd. }
\label{fig_comp_oslo1}
\end{figure*}

\begin{figure*}
\includegraphics[scale=0.6]{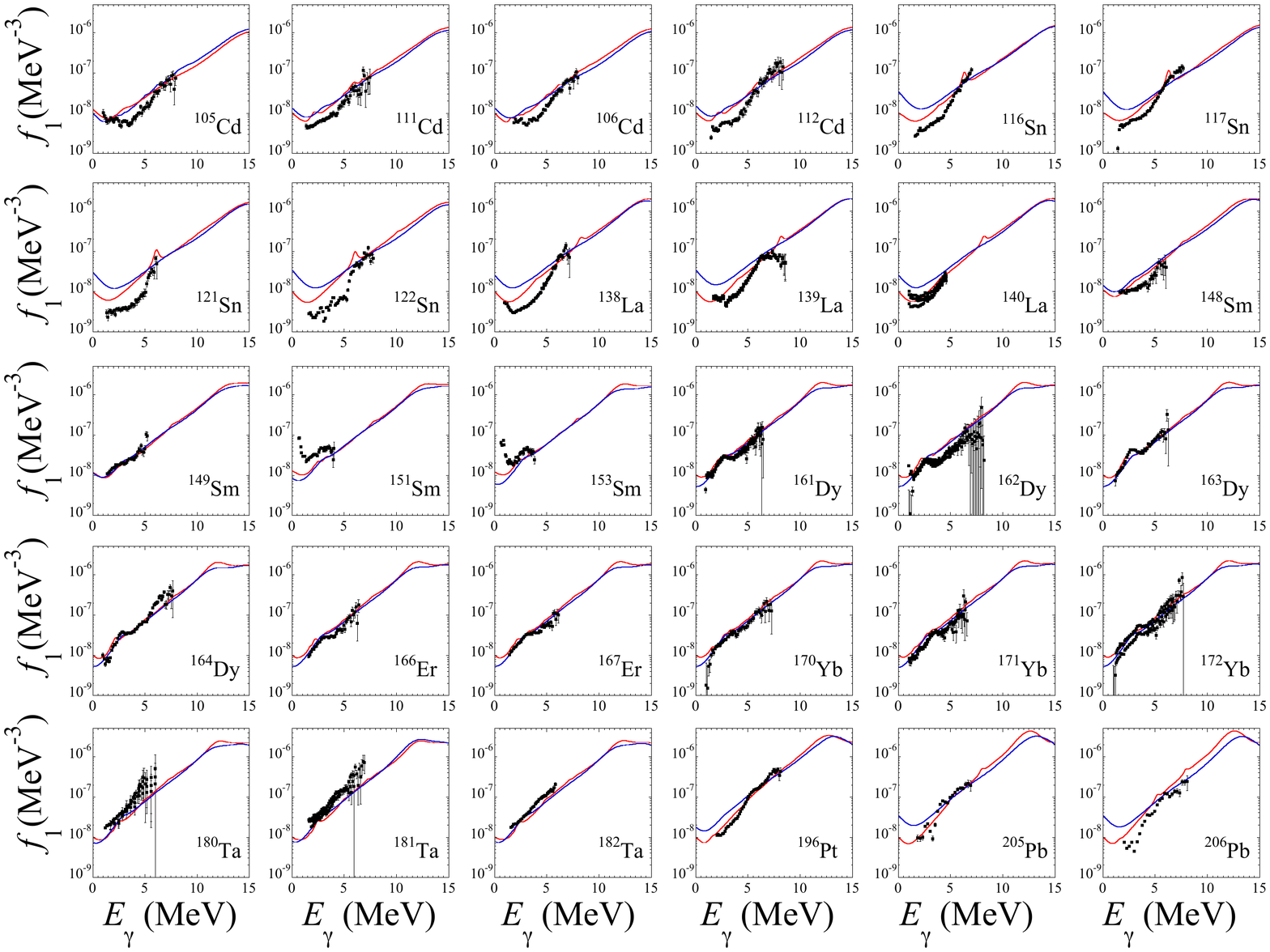}  
\caption{(Color online) Same as Fig.~\ref{fig_comp_oslo1} for nuclei between Cd and Pb.}
\label{fig_comp_oslo2}
\end{figure*}

\begin{figure*}
\vskip -7cm
\includegraphics[scale=0.6]{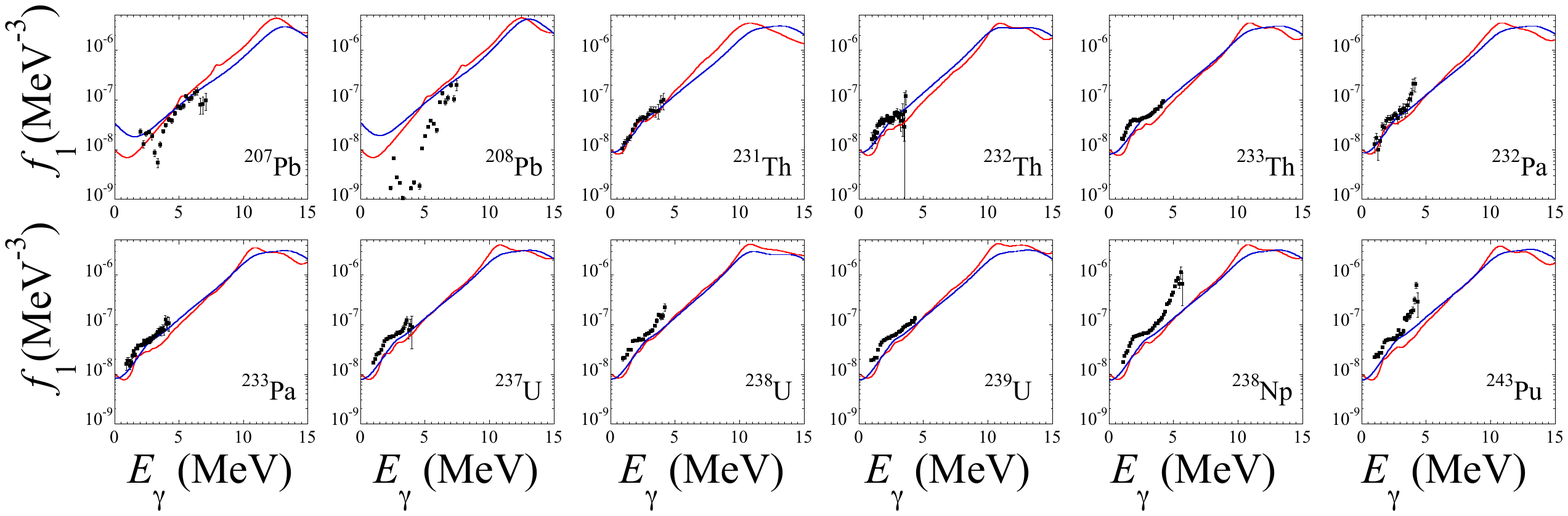}  
\caption{(Color online) Same as Fig.~\ref{fig_comp_oslo1} for nuclei between Pb and Pu.}
\label{fig_comp_oslo3}
\end{figure*}

The total dipole PSFs obtained by the Oslo method are compared with the SMLO and D1M+QRPA models in Figs.~\ref{fig_comp_oslo1}--\ref{fig_comp_oslo3} for all the 72 nuclei for which data exist. An uncertainty analysis including model dependencies, as discussed in Sec.~\ref{sec_ass_unc_oslo}, has only been performed in the case of $^{64-65}$Ni, $^{69}$Ni, $^{89}$Y, $^{92}$Zr, $^{138-140}$La and $^{180-182}$Ta. This could explain why global models may deviate for some nuclei, such as the Mo or Sn isotopes, since in these cases the impact of using different NLD models in the extraction of the experimental PSF has not been investigated and is therefore, not reflected in the plotted uncertainties. 
A comprehensive uncertainty analysis of the Oslo data (Sec.~\ref{sec_ass_unc_oslo}) could consequently help to reconcile experiment and theory. However, for some nuclei, like the Sc isotopes, a strong modification of the Oslo PSF slope, {\it i.e.} of the NLD normalization, would be required. 

It is thus not possible to draw definite conclusions on the agreement between experimental and model PSFs, as was also mentioned in Sect.~\ref{sec_comp_nrf}. In the few cases mentioned above, where an uncertainty analysis has been performed, the calculations lie within or at the limits of the range of uncertainties. It should also be noted that the upper and lower limits in these cases do not correspond to a full uncertainty analysis, but to a partial analysis including one or two different NLD models and the uncertainties of the normalisation and total average radiative width, as described in Sec.~\ref{sec_ass_unc_oslo}. Since the Oslo data describe a deexcitation strength, the model calculations include here the low-energy contributions described in Sec.~\ref{sec_th}.


\subsection{Comparison with ARC/DRC data}
\label{sec_comp_arc}

\begin{figure}
\includegraphics[scale=0.35]{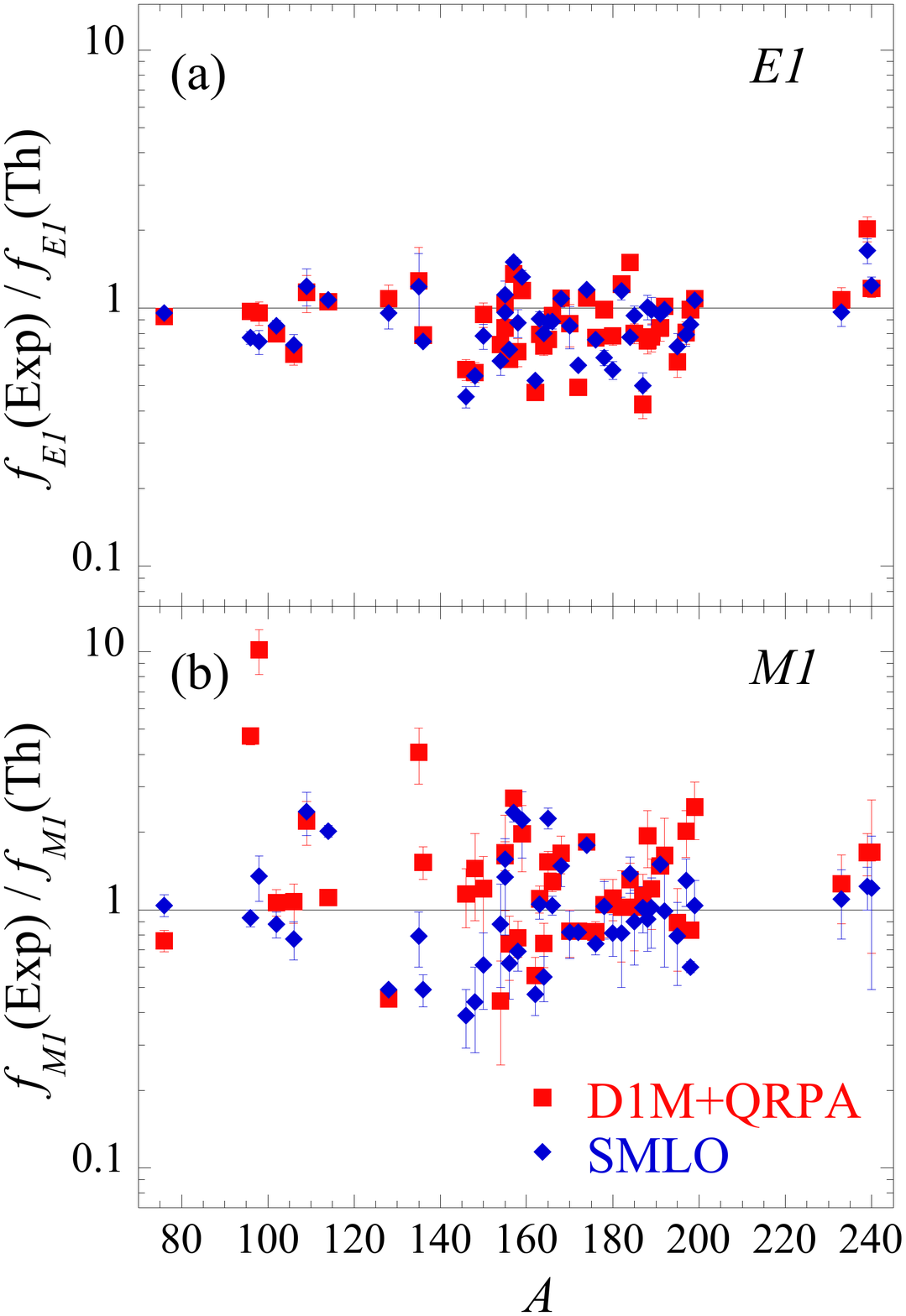}  
\caption{(Color online)  Ratio between PSFs from ARC experiments in the energy range of $6.5\pm0.5$~MeV and theoretical predictions, SMLO (blue diamonds) and D1M+QRPA+0lim (red squares) . (a) The upper panel correspond to the $E1$ PSFs and (b) the lower panel to the $M1$ PSFs.}
\label{fig_comp_ARC_E1M1}
\end{figure}

\begin{figure*}
\begin{center}
\includegraphics[scale=0.6]{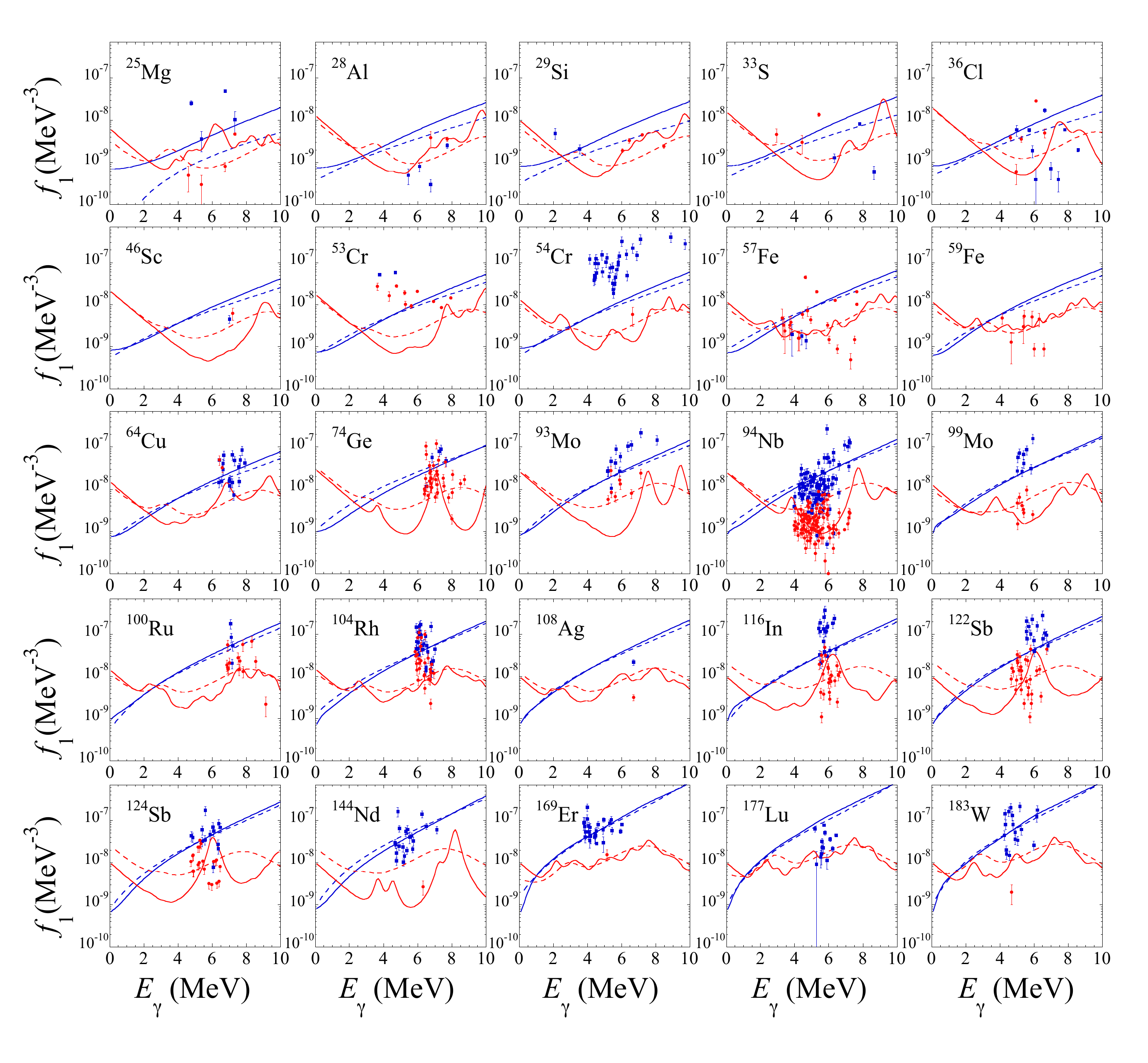}  
\caption{(Color online)  Comparison between experimental PSFs extracted from DRC data  and theoretical predictions, SMLO (dashed lines) and D1M+QRPA+0lim (solid lines). Blue squares and lines correspond to the $E1$ PSFs and red circles and lines to the $M1$ PSFs.}
\label{fig_comp_drc_all}
\end{center}
\end{figure*}

\begin{figure*}
\begin{center}
\includegraphics[scale=0.6]{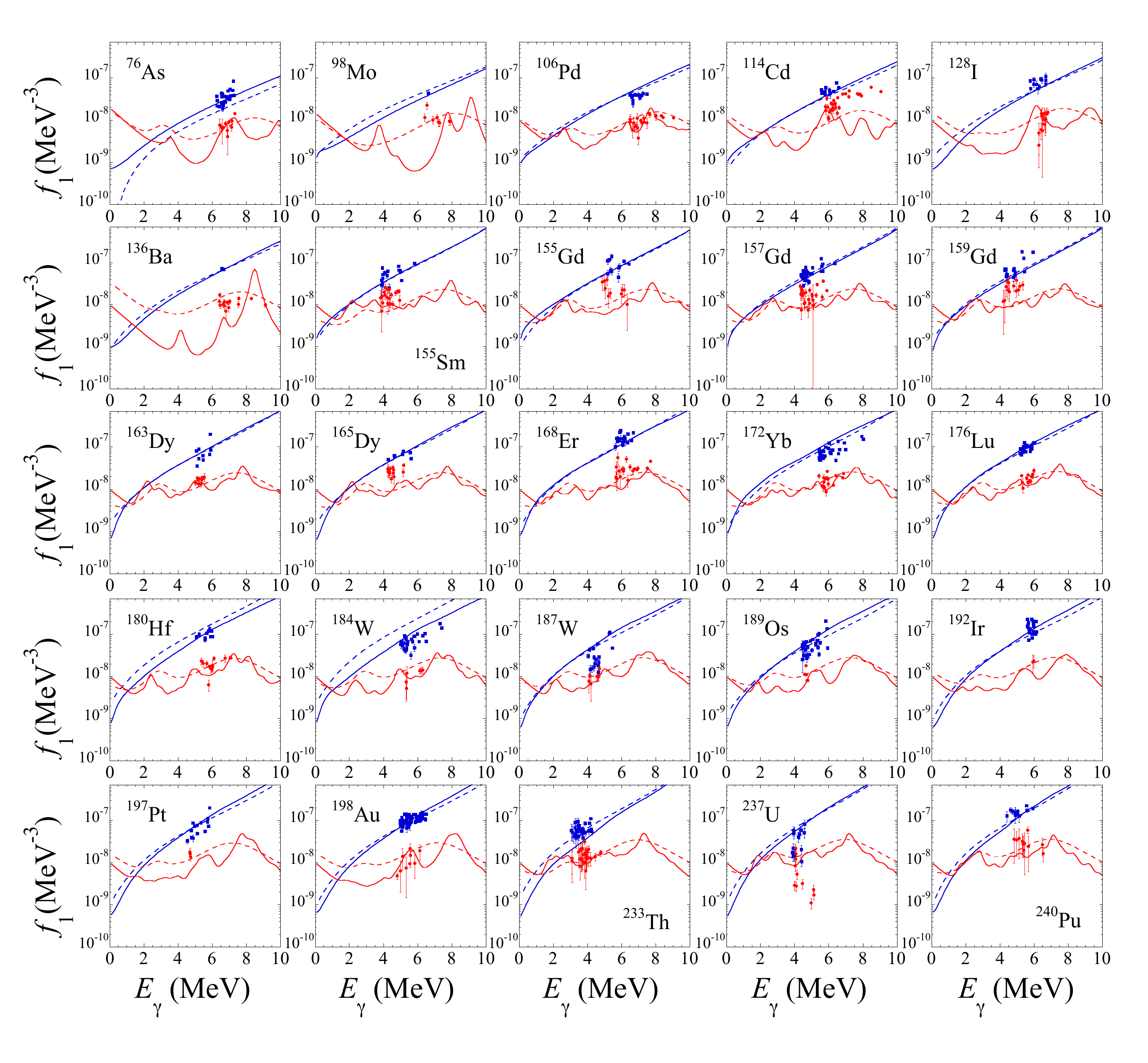}  
\caption{(Color online)  Comparison between experimental PSFs extracted from ARC data  and theoretical predictions, SMLO (dashed lines) and D1M+QRPA+0lim (solid lines) for a sample of 25 nuclei. Blue squares and lines correspond to the $E1$ strength and red circles and lines to the $M1$ strength.}
\label{fig_comp_arc_all}
\end{center}
\end{figure*}

The quasi-monoenergetic representation of PSF values in the energy window $E_\gamma= 6.5 \pm 0.5$~MeV extracted from the differential DRC and ARC data (see Sec.~\ref{sec_exp_arc}) is  compared in Fig.~\ref{fig_comp_ARC_E1M1} with the model predictions averaged within the same energy interval.  The comparison includes both  the $E1$ and $M1$ transitions. It is seen that both the D1M+QRPA+0lim and SMLO models, on average, reproduce the whole set of ARC data, from the lightest to the heaviest species. More scatter is obtained for the $M1$ PSF, and in particular with D1M+QRPA+0lim,  due to the peaked spin-flip resonance found mainly in spherical nuclei (see also Figs.~\ref{fig_comp_drc_all}-\ref{fig_comp_arc_all}). 

A more detailed comparison of DRC and ARC data, now for individual values of $f_{XL}$, can be found in Figs.~\ref{fig_comp_drc_all} and \ref{fig_comp_arc_all}, respectively. As discussed in Sec.~\ref{sec_ass_arc}, DRC data are included in the library when no equivalent ARC measurement is available, even if the DRC PSF has been extracted out of a small number of resonances. In this case, a large scatter  can be expected, as seen in Fig.~\ref{fig_comp_drc_all}, especially for the lightest nuclei where the non-statistical pattern of the few resonances DRC data can clearly be  observed. In such cases, a detailed comparison with models is meaningless. 

Whenever available, the ARC filtered beam results are preferred over the DRC data because of the higher statistical accuracy and averaging over a much larger number of resonances that is  involved compared to the DRC data. This is illustrated in Fig. \ref{fig_comp_arc_all} for a sample of 25 nuclei out of the 54 known cases (see Table~\ref{tab_ass_arc}). In this case, it can be seen that the agreement between ARC and model PSFs is relatively good for both models and both transition types, from the lightest to the heaviest nuclei. 

\subsection{Comparison with thermal neutron data}
\label{sec_comp_egaf}

\begin{figure}
\includegraphics[scale=0.35]{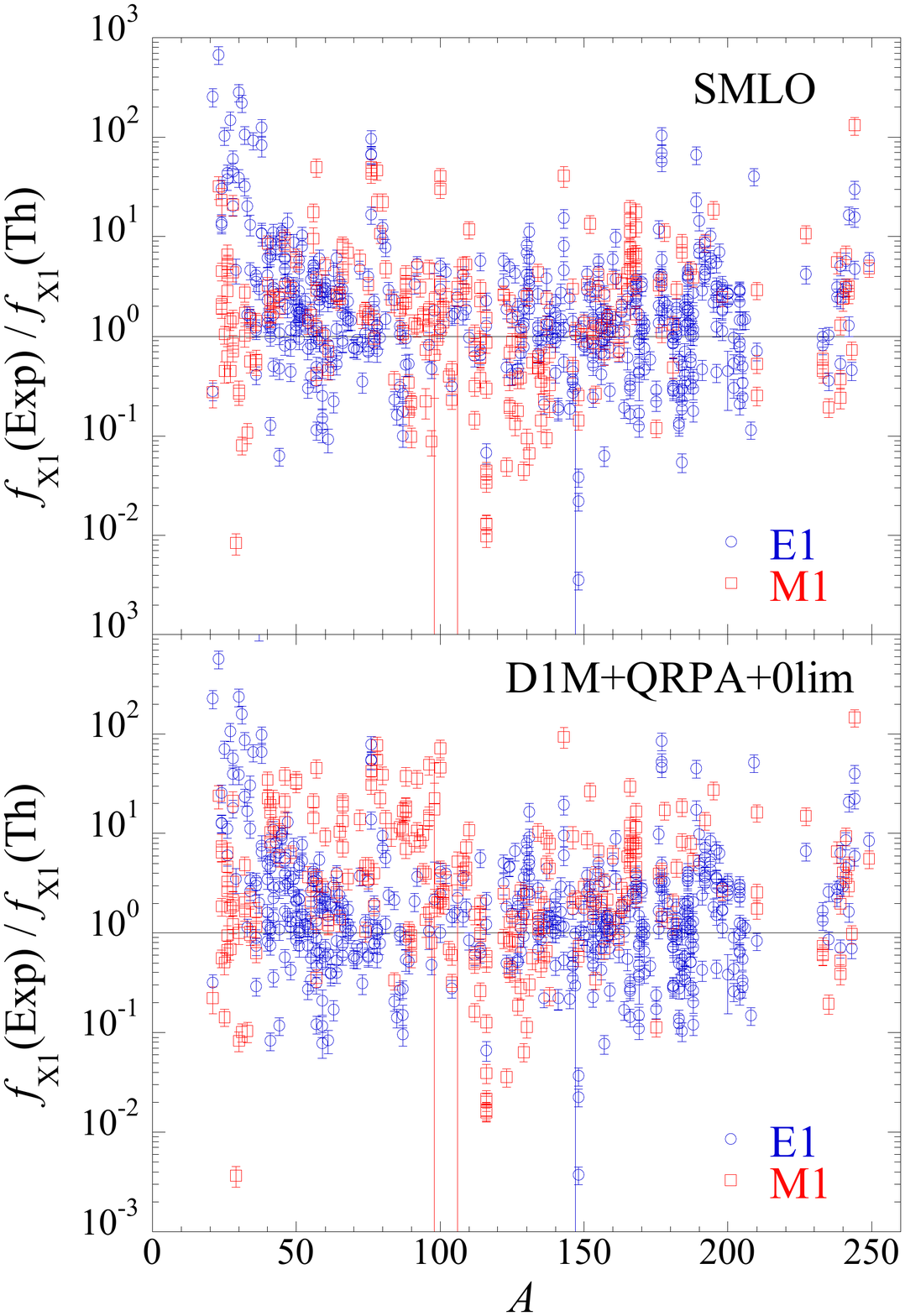}  
\caption{(Color online)  Ratio between experimental $E1$ (blue circles) and $M1$ (red squares) PSFs from primary transitions from thermal neutron capture data  and theoretical predictions, SMLO (upper panel) and 
D1M+QRPA+0lim (lower panel), as a function of the atomic mass.}
\label{fig_comp_egaf_A}
\end{figure}

\begin{figure}
\includegraphics[scale=0.35]{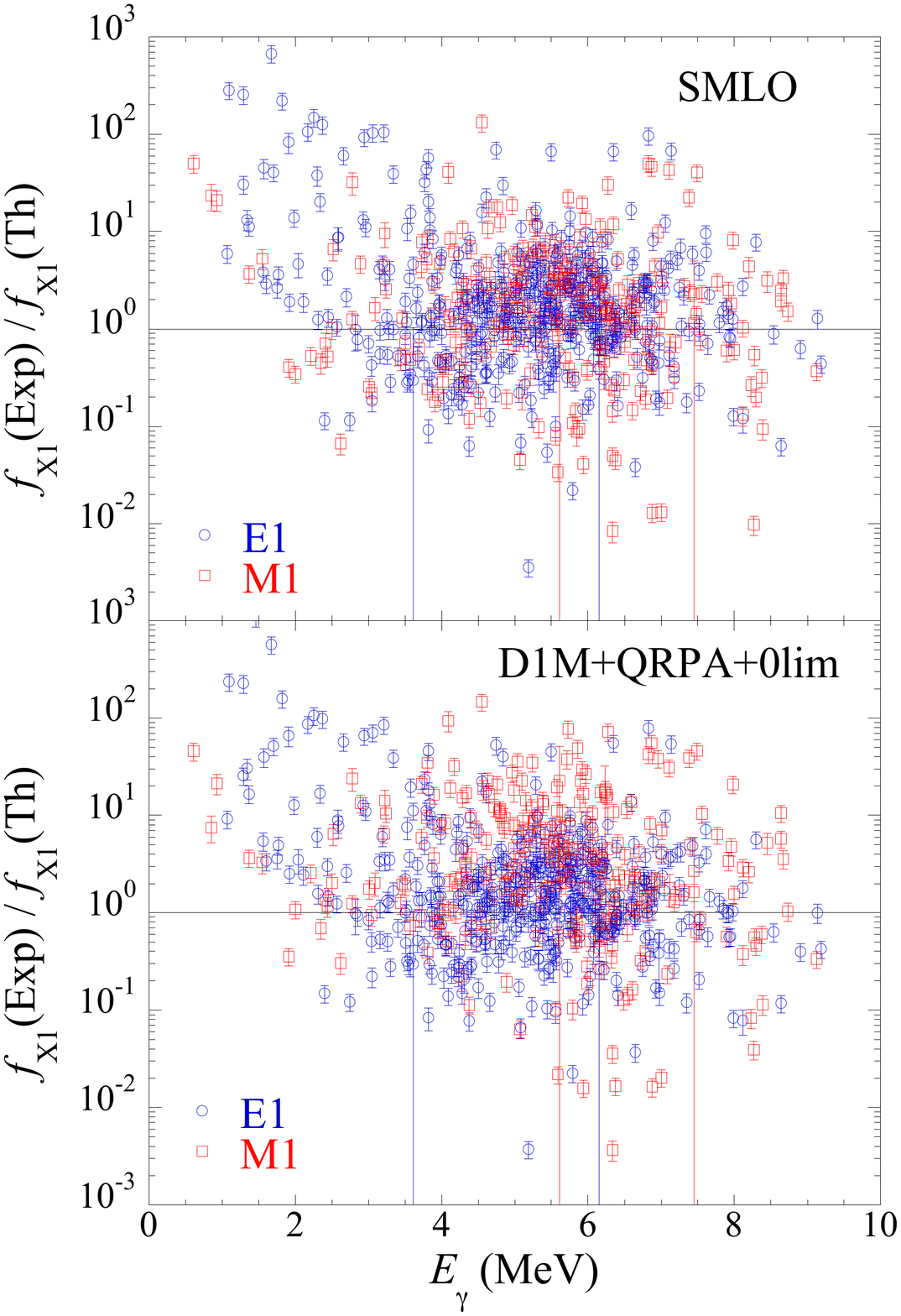}  
\caption{(Color online)  Ratio between experimental $E1$ (blue circles) and $M1$ (red squares)  PSFs from primary transitions from thermal neutron capture data and theoretical predictions, SMLO (upper panel) and D1M+QRPA+0lim (lower panel), as a function of the $\gamma$-ray energy.}
\label{fig_comp_egaf_Eg}
\end{figure}

PSFs from the thermal neutron capture data have been extracted as described in Sec.~\ref{sec_ass_egaf}. The  808 $\gamma$-ray binned PSF data for $E1$ and $M1$ components are compared with 
model predictions in Fig.~\ref{fig_comp_egaf_A} as a function of the atomic mass $A$ and in  Fig.~\ref{fig_comp_egaf_Eg} as a function of the $\gamma$-ray energy. Model predictions are seen to globally underpredict the data within a factor of 10. However, much larger deviations can be observed for both 
model predictions, especially for light nuclei. It should, however, be recalled here (see Sec.~\ref{sec_ass_egaf}) that extracting average strengths from this kind of data is challenging as the 
resulting PSFs are affected significantly by 
Porter-Thomas fluctuations. These data are therefore, expected to have large associated uncertainties.

\subsection{Comparison with (p,$\gamma$) data}
\label{sec_comp_pg}

\begin{figure*}
\begin{center}
\includegraphics[scale=0.6]{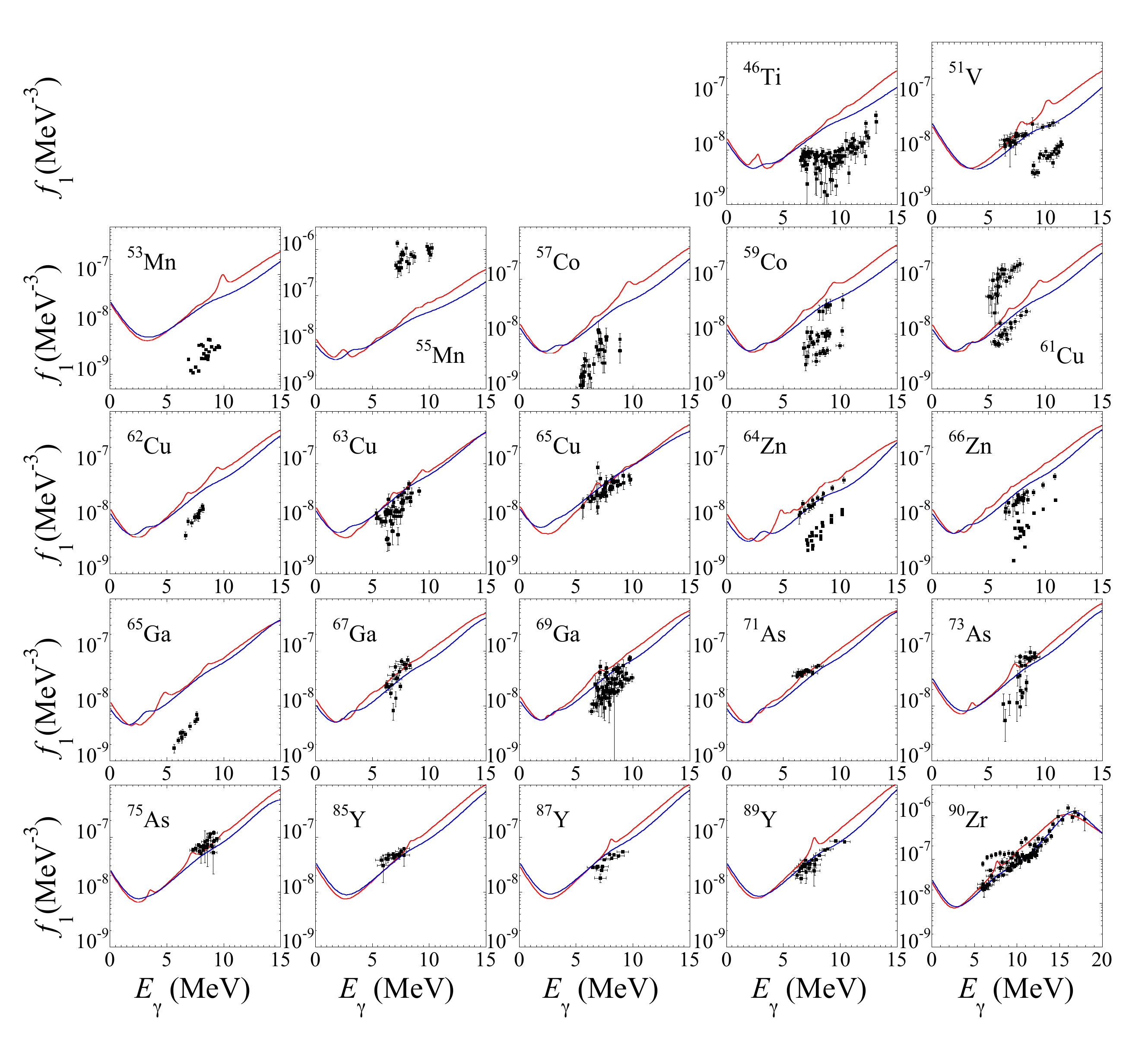}  
\caption{(Color online) Comparison between experimental PSFs extracted from (p,$\gamma$) data  and  SMLO (blue lines) and D1M+QRPA+0lim (red lines) models for the 22 nuclei for which data exist.}
\label{fig_comp_pg}
\end{center}
\end{figure*}

In Fig.~\ref{fig_comp_pg}, a comprehensive comparison is shown between PSFs extracted from (p,$\gamma$) data  and theoretical SMLO and D1M+QRPA+0lim predictions for the 22 nuclei for which data exist. In some cases, in particular for the Mn, Co, Cu and Zn isotopes, the agreement is rather poor. New proton capture measurements, including a detailed  analysis of both experimental and theoretical uncertainties, are recommended to confirm the determination of the PSF. For heavier nuclei, including $^{90}$Zr for which data up to the GDR region is available (see also Fig.~\ref{fig_allexp_other}), the agreement between experiment and models is quite satisfactory and shows the relevance of this method for extracting PSFs.

\subsection{Comparison with (p,p$^\prime$) data}
\label{sec_comp_pp}

\begin{figure}
\includegraphics[scale=0.35]{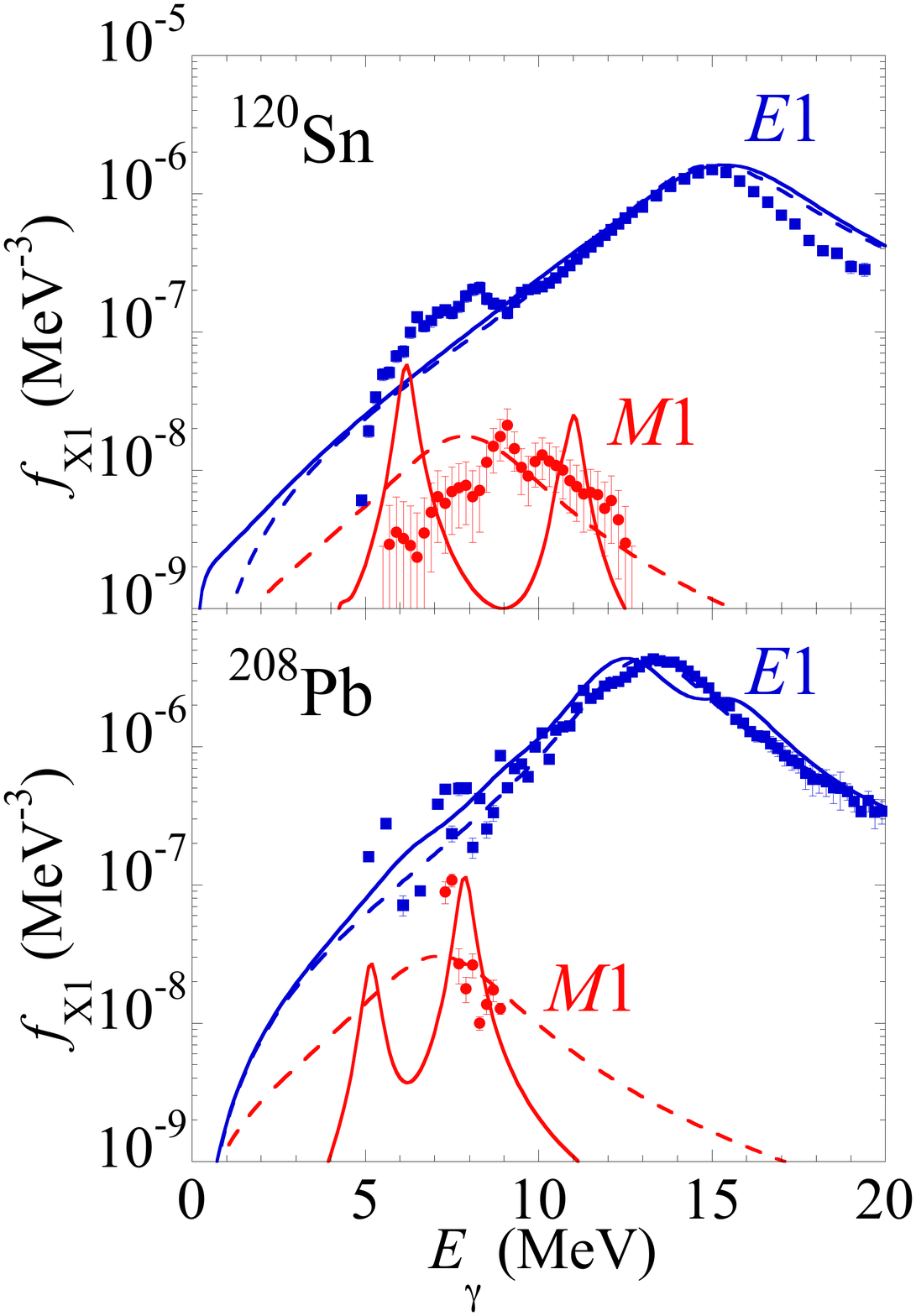}  
\caption{(Color online) $^{120}$Sn and $^{208}$Pb  $E1$ and $M1$ PSFs extracted from (p,p$^\prime$) experiments \cite{Tamii11,Poltoratska12,Bassauer16} and compared with D1M+QRPA (solid lines) and 
SMLO (dashed lines) predictions. $E1$ data (squares) and model predictions are shown in blue colors and $M1$ data (circles) and model predictions in red.}
\label{fig_comp_Darmstadt}
\end{figure}

 We show in Fig.~\ref{fig_comp_Darmstadt} the $E1$ and $M1$ PSFs extracted from (p,p$^\prime$) experiments \cite{Tamii11,Poltoratska12,Bassauer16} for $^{120}$Sn and $^{208}$Pb. As pointed out in Sec.~\ref{sec_th_smlo}, SMLO GDR parameters have been adjusted on photoneutron data, so that the location and width of the $E1$ strength reproduce rather well the (p,p$^\prime$) data in the GDR region, as expected. In contrast, the D1M+QRPA model is a global approach and while it reproduces rather well the  $^{120}$Sn  data in the GDR region, discrepancies are seen for $^{208}$Pb. As far as the $M1$ PSF is concerned, the experimental PSF is in rather good agreement with the SMLO spin-flip resonance for $^{120}$Sn (though the centroid energy is about 1.5~MeV too low), but not for $^{208}$Pb. D1M+QRPA calculation gives a fair description of $^{208}$Pb $M1$ data, but cannot reproduce the $^{120}$Sn data. Note that the $^{96}$Mo   (p,p$^\prime$) data \cite{Martin17} are not discussed separately here, but will be compared with data obtained from other methods in Sec.~\ref{sec_database_exp} (see in particular Fig.~\ref{fig_allexp_Mo}).

\subsection{Comparison with photodata}
\label{sec_comp_photo}

\begin{figure*}
\includegraphics[scale=0.62]{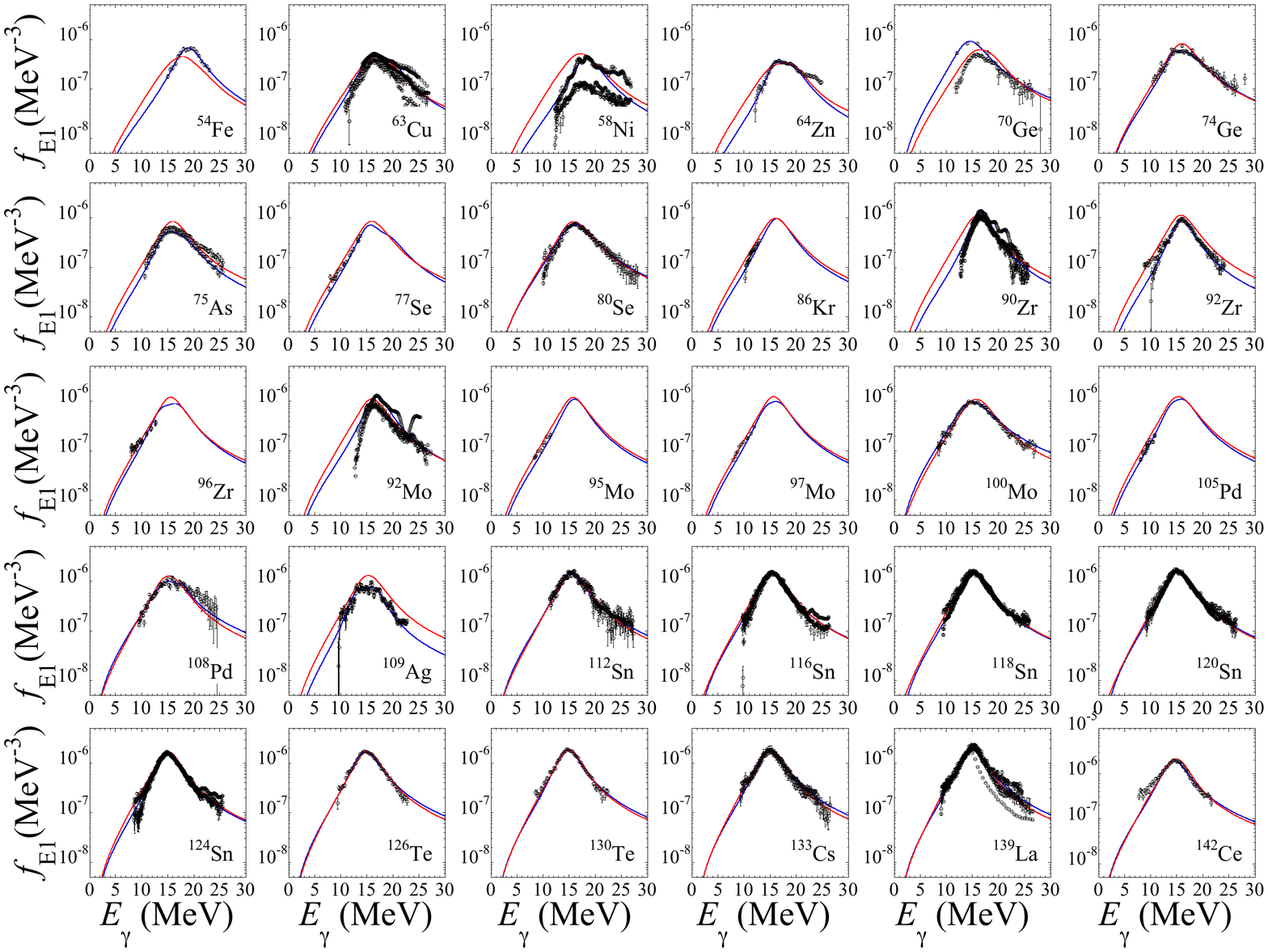}  
\caption{(Color online)  Comparison between experimental $E1$ PSFs extracted from photonuclear cross sections (open circles) and D1M+QRPA (red lines) and SMLO (blue lines)  strengths for a sample of 30 nuclei between Fe and Ce.}
\label{fig_comp_photo1}
\end{figure*}

\begin{figure*}
\includegraphics[scale=0.62]{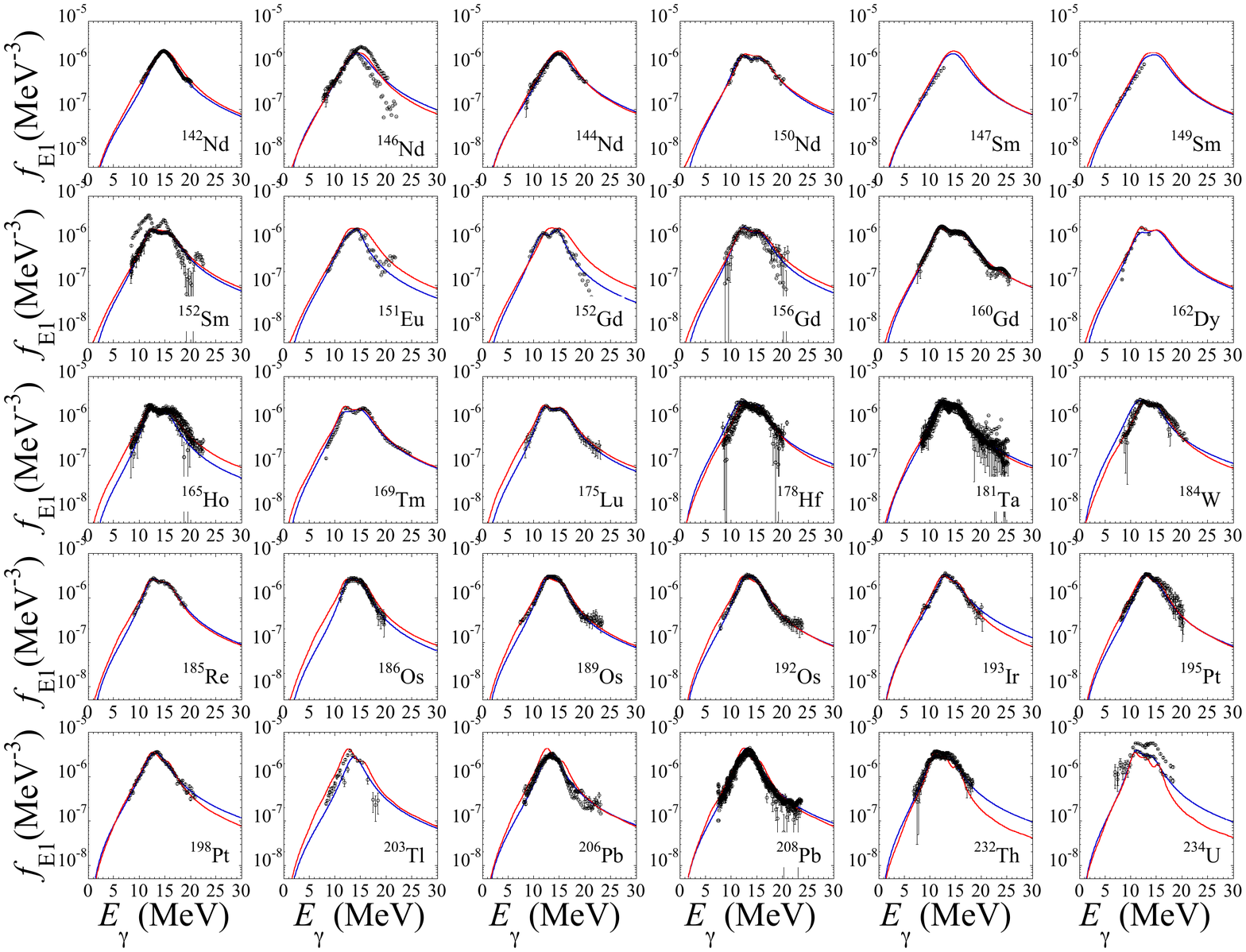}  
\caption{(Color online)  Same as Fig.~\ref{fig_comp_photo1} for a sample of 30 nuclei between Nd and U.}
\label{fig_comp_photo2}
\end{figure*}

Figs.~\ref{fig_comp_photo1}-\ref{fig_comp_photo2} compare the PSFs extracted from experimental photoabsorption and/or photoneutron cross sections with the SMLO and D1M+QRPA PSFs for a sample of 60 nuclei out of the 159 for which photodata is available. We recall that the SMLO calculations and the corresponding $E1$ Lorentzian GDR parameters have been directly adjusted to experimental photonuclear cross sections \cite{Plujko18,Kawano19} and therefore, the global comparison in Figs. ~\ref{fig_comp_photo1}-\ref{fig_comp_photo2} illustrates the quality of the fitting procedure as well as the adequacy of the Lorentzian phenomenological approach to describing the $E1$ PSF. In contrast, the D1M+QRPA model 
is a global model that has not been tuned on individual cross sections, but globally renormalized, as described in Sec.~\ref{sec_th_qrpa}, to account for missing effects, such as contributions 
beyond the 1 particle -- 1 hole excitations and the interaction between the single-particle and low-lying collective phonon degrees of freedom. This global renormalisation is shown in 
Figs.~\ref{fig_comp_photo1}-\ref{fig_comp_photo2} to lead to a rather satisfactory description of the $E1$ PSF in the GDR region, though for light nuclei, the centroid energy is usually found at 
lower energies than observed. The $K$-dependence introduced in the $E1$ spreading width $\Gamma(K)$ (see Sec.~\ref{sec_th_qrpa}) is found to give the correct hierarchy between the two GDR peaks in 
the case of deformed nuclei. 

\subsection{Comparison with singles spectra from cold neutron capture}
\label{sec_comp_ncap}

Unfolded singles spectra obtained from cold neutron capture can provide important constraints on the PSF and 
NLD models. Due to the low neutron energy, the angular distribution of the $\gamma$-rays is isotropic, simplifying the determination of the experimental $\gamma$-ray branching ratios.
Experimentally it is possible to provide $\gamma$-ray multiplicity $M_\gamma$ based on the unfolded spectrum using the internal cross 
section calibration and the total radiative capture cross section.
The multiplicity is a ratio of the sum of partial $\gamma$-ray production cross sections in the spectrum and total cross section. 
Once the spectrum is well described by the model, the observed multiplicity is correctly predicted. However, the inverse is not true, reproducing the experimental multiplicity does not imply that the model is correct.
In the current version of the BITS code (see Sec.~\ref{sec_exp_ncap2}), only the lowest multipolarity is taken into the account to generate the average partial widths from the PSFs used to calculate the branching ratios above the critical energy. 
The discrete level part of the decay scheme is built up from the ENSDF library data \cite{ENSDF}. The experimental branching ratios below the critical energy are adopted from  intensities measured in the present study and the ENSDF conversion coefficients are included to estimate the electromagnetic decay probability. When no experimental conversion coefficient is available, it is calculated for the lowest multipolarity and the corresponding type using the BrICCs code \cite{Kibedi08}. Above the critical energy, the electromagnetic decay probability does not include the mixing ratio and is calculated for the lowest multipolarity. The mixing ratio can significantly influence the simulation results which is practically limited to the case of $M1$ plus $E2$ mixing through the PSF choice in the determination of the transition width. Because it is generally not known for higher lying levels, we neglect this possibility by considering the lowest multipolarity. The electron conversion however may significantly influence the modelling results, since the low-energy transitions in high mass nuclei will not appear in the decay spectrum but appears as feeding of the final levels; this requires double administration of the transition matrix elements which is  implemented in the BITS model. This effect is expected in the case of an $E2$ transition below 300~keV and $M1$ below 1~MeV, but is negligible above 0.1~MeV for $E1$ transitions in heavy nuclei, such as actinides.

\begin{figure}
\includegraphics[width=\columnwidth, angle = 0]{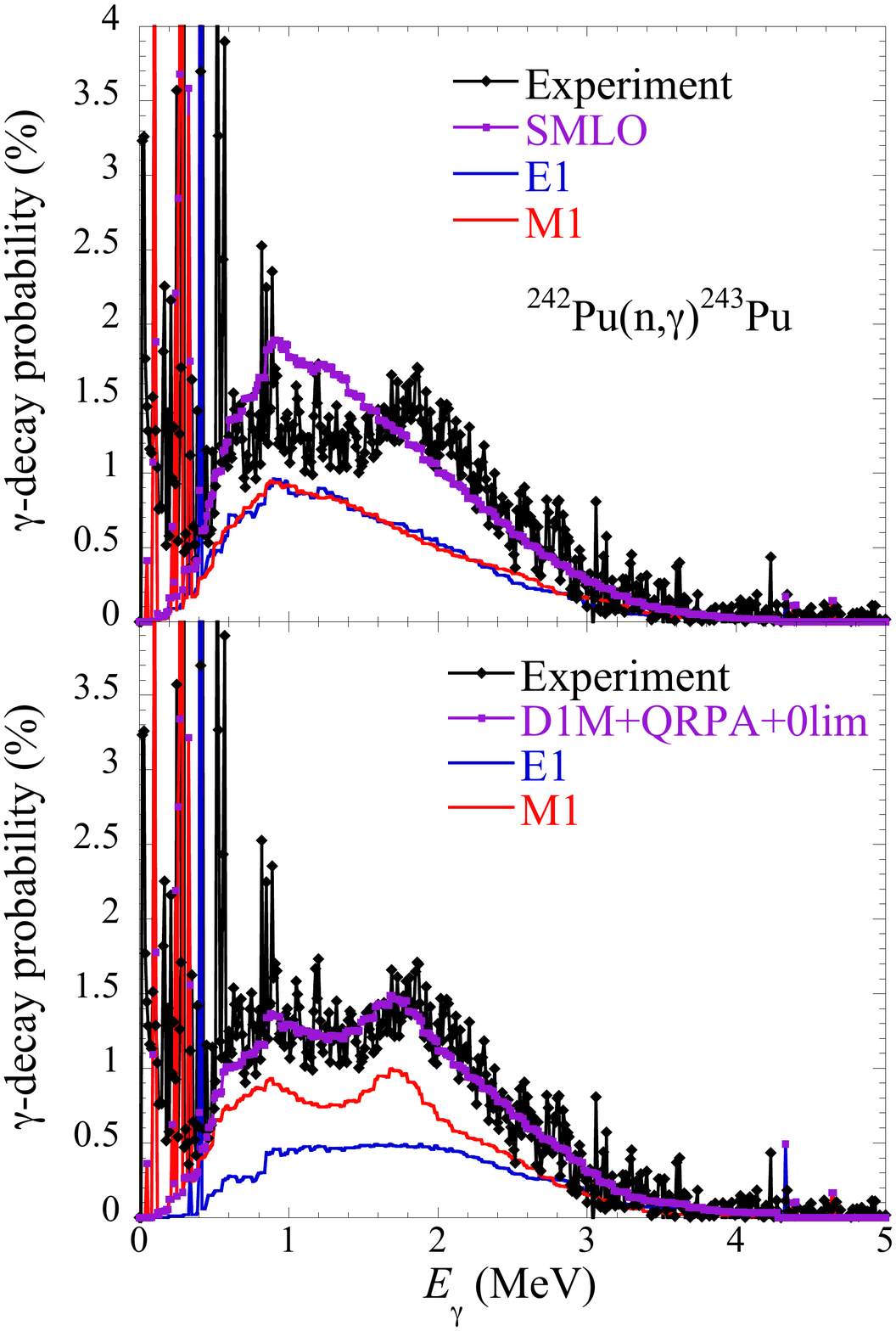}
\vskip -0.5cm
\caption{(Color online) Comparison of singles $\gamma$-ray spectrum from the $^{242}$Pu(n,$\gamma$)$^{243}$Pu reaction with the BITS simulations based on the SMLO (upper panel) and D1M+QRPA+0lim (lower panel) PSFs.  The  $M1$ and $E1$ contributions to the calculated statistical spectrum are indicated by blue and red lines, respectively. The $E2$ contribution is found to be negligible.}
\label{fig_comp_nth_243Pua}
\end{figure}

\begin{figure}
\centering
\includegraphics[width=\columnwidth, angle = 0]{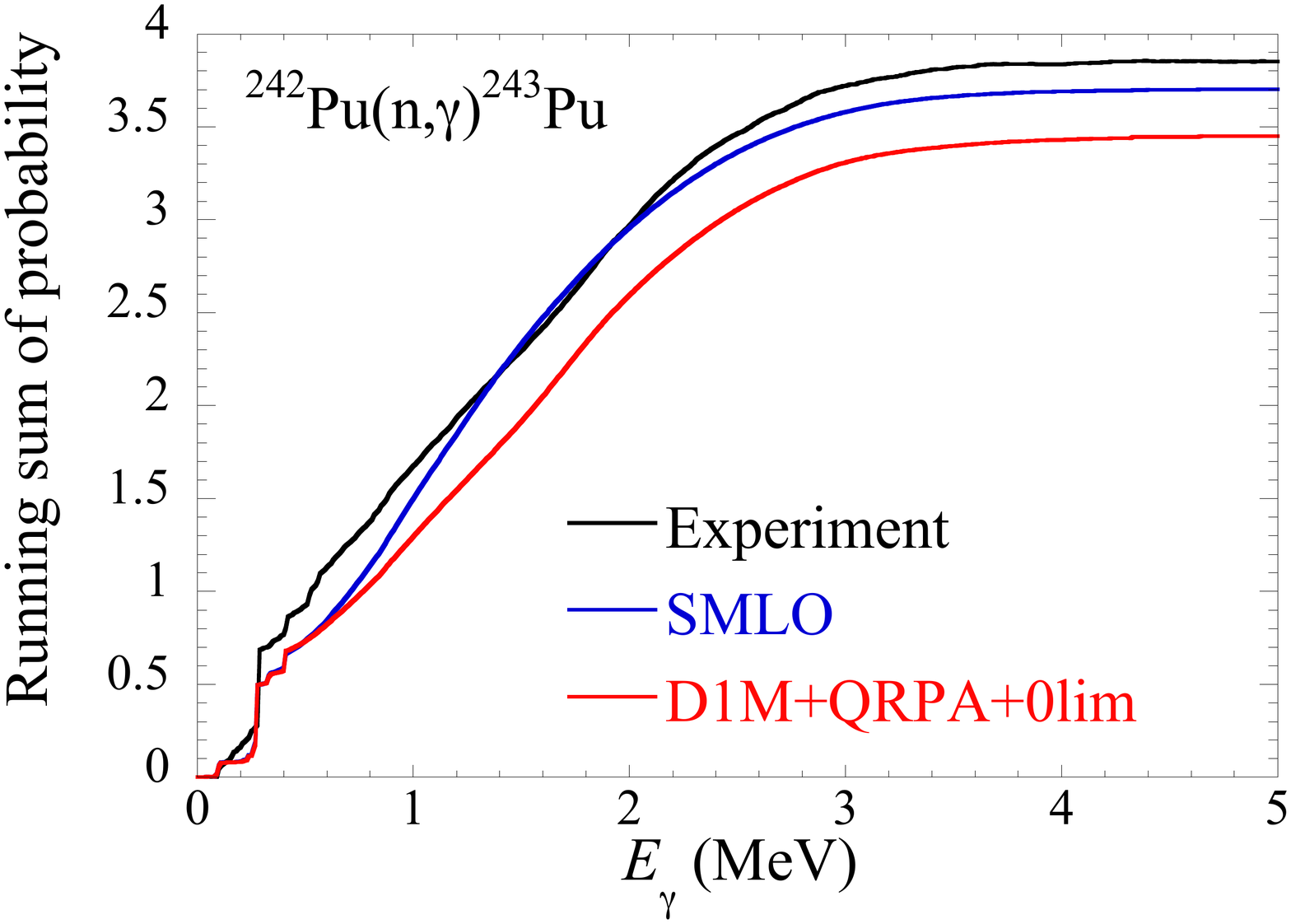}
\caption{(Color online) Running sum of probability corresponding to the $^{242}$Pu(n,$\gamma$)$^{243}$Pu spectrum given in Fig.~\ref{fig_comp_nth_243Pua}. See text for more details.}
\label{fig_comp_nth_243Pub}
\end{figure}

An example of modelling singles $\gamma$-ray spectra is given in Fig.~\ref{fig_comp_nth_243Pua} for the $^{242}$Pu(n,$\gamma$)$^{243}$Pu reaction. 
The experimental spectrum is compared with simulations performed using D1M+QRPA+0lim and  SMLO models for the PSF, combined with the HFB plus combinatorial  NLD \cite{Goriely08b}.
The D1M+QRPA+0lim describes the shape of the experimental spectrum fairly well, though the low-energy part of the spectrum, including the highest 
intensity $\gamma$-lines, is underestimated. This can be clearly seen in Fig. ~\ref{fig_comp_nth_243Pub} where the running sum of probabilities are 
shown as a function of the photon energy. 
To achieve a better agreement for the running sum of probabilities, an additional low-energy $E1$ strength would be required to the PSF to enhance intensities in the  whole energy region. A shift of about 0.2~MeV on the $M1$ scissors mode responsible for the the 1.8~MeV bump  would also improve both the running sum of probabilities and the agreement with the spectra (Fig.~\ref{fig_comp_nth_243Pua}).
For the calculation based on the SMLO PSF, while the  running sum of probability is closer to experiment 
(Fig.~\ref{fig_comp_nth_243Pub}), surprisingly the $E1$ and $M1$ PSF components provide the same intensity pattern in the spectrum 
(see upper panel of Fig.~\ref{fig_comp_nth_243Pua}), leading to an overestimate of the $\gamma$-decay probability in the 1~MeV region. 
Another difference with the D1M+QRPA+0lim model is the missing $M1$ scissors mode bump at 1.8 MeV, which, for SMLO, is located at higher energy, around 2.8~MeV, where it has little influence on the high energy spectrum. 

Similar, or even better, agreements have been obtained for the  $^{72}$Ge(n,$\gamma$)$^{73}$Ge,  $^{73}$Ge(n,$\gamma$)$^{74}$Ge,  $^{77}$Se(n,$\gamma$)$^{78}$Se as 
well as  $^{113}$Cd(n,$\gamma$)$^{114}$Cd.  In the cases where two capture spins are available, {\it i.e.} for the odd-$A$ target, one of the capture spin gives a better description than the other. The final result can be obtained by a weighted sum of both spin contributions where the weighting factor can be obtained from the cross section contribution of resonances with the corresponding spin. The generally good description of the continuum part of the studied reactions supports the use of D1M+QRPA+0lim PSF in combination with the HFB plus combinatorial NLD for the description of the $\gamma$-decay.

\subsection{Comparison with multi- and two-step cascade spectra}
\label{sec_comp_msc}

Similarly to singles spectra from thermal (cold) neutron spectra, information on PSF can be obtained also from coincidence $\gamma$-ray spectra following radiative capture of slow neutrons. Two observables from the $\gamma$ decay of individual resonances are used to compare model predictions with simulations. Specifically, the observables are the MD and MSC spectra from $\gamma$ cascades that have deposited all the energy of the cascade in the detector (see Sec.~\ref{sec_exp_msc}). In reality, a range of sum-energies, typically about 1 MeV wide is used \cite{Krticka19}. Such cuts on the energy sum help to suppress possible contributions from the background and impurities in the target. In the case of MSC spectra, these cuts also make some of the structures observed in the spectra more pronounced.

The measured MD and MSC spectra are products of a complex interplay between the PSFs of different multipolarities, NLD and the detector response to individual cascades. The cascades derived from the different PSF and NLD models were generated with the help of the Monte-Carlo {\sc DICEBOX} code \cite{Becvar98,Krticka19a}. The code allows to treat the expected Porter-Thomas fluctuations of partial radiation widths via the concept of nuclear realizations, {\it i.e.} different sets of all levels and partial radiation widths in a simulated nucleus. Individual nuclear realizations yield different predictions of observables even for a fixed combination of the PSF and NLD models and the spin and parity of the capturing state. For each tested combination of the PSF and NLD models, 15 different nuclear realizations were simulated. The response of the DANCE detector to each simulated cascade was then determined using the Monte-Carlo GEANT4-based code \cite{Agostinelli03,Jandel07}. 

\begin{figure}[t]
\includegraphics[clip,width=0.49\columnwidth]{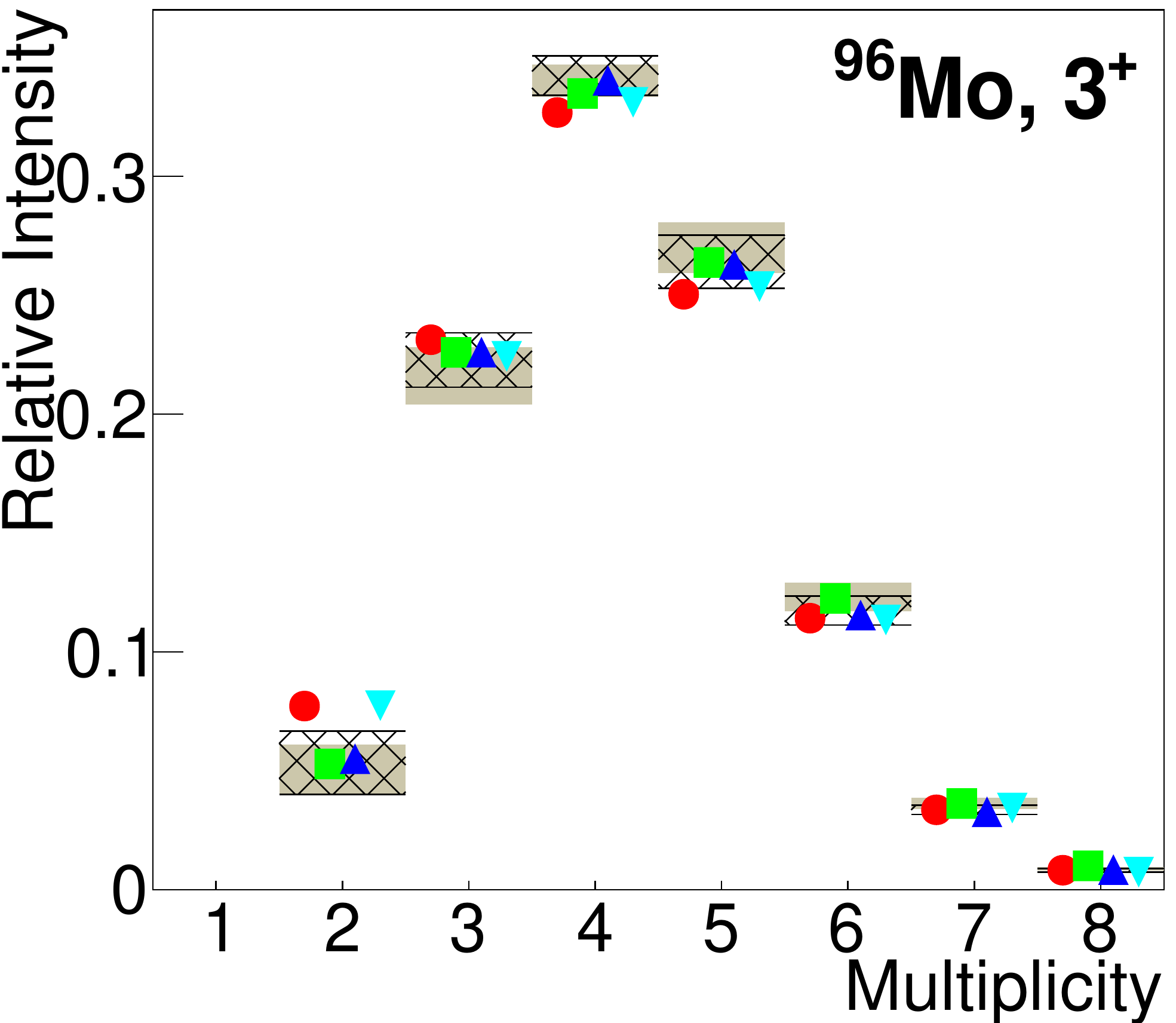}
\hfill
\includegraphics[clip,width=0.49\columnwidth]{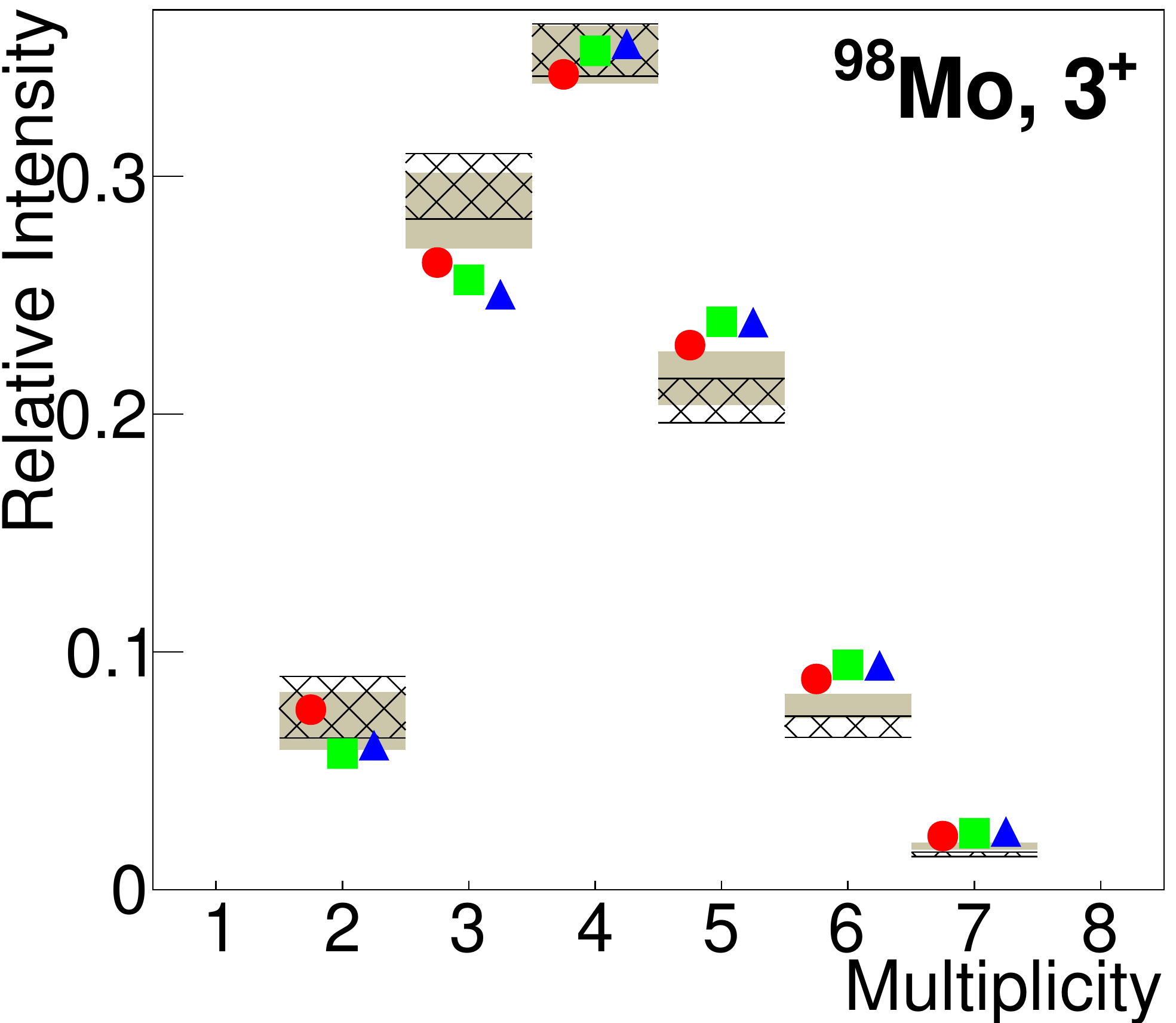}

\vspace*{2mm}
\includegraphics[clip,width=0.49\columnwidth]{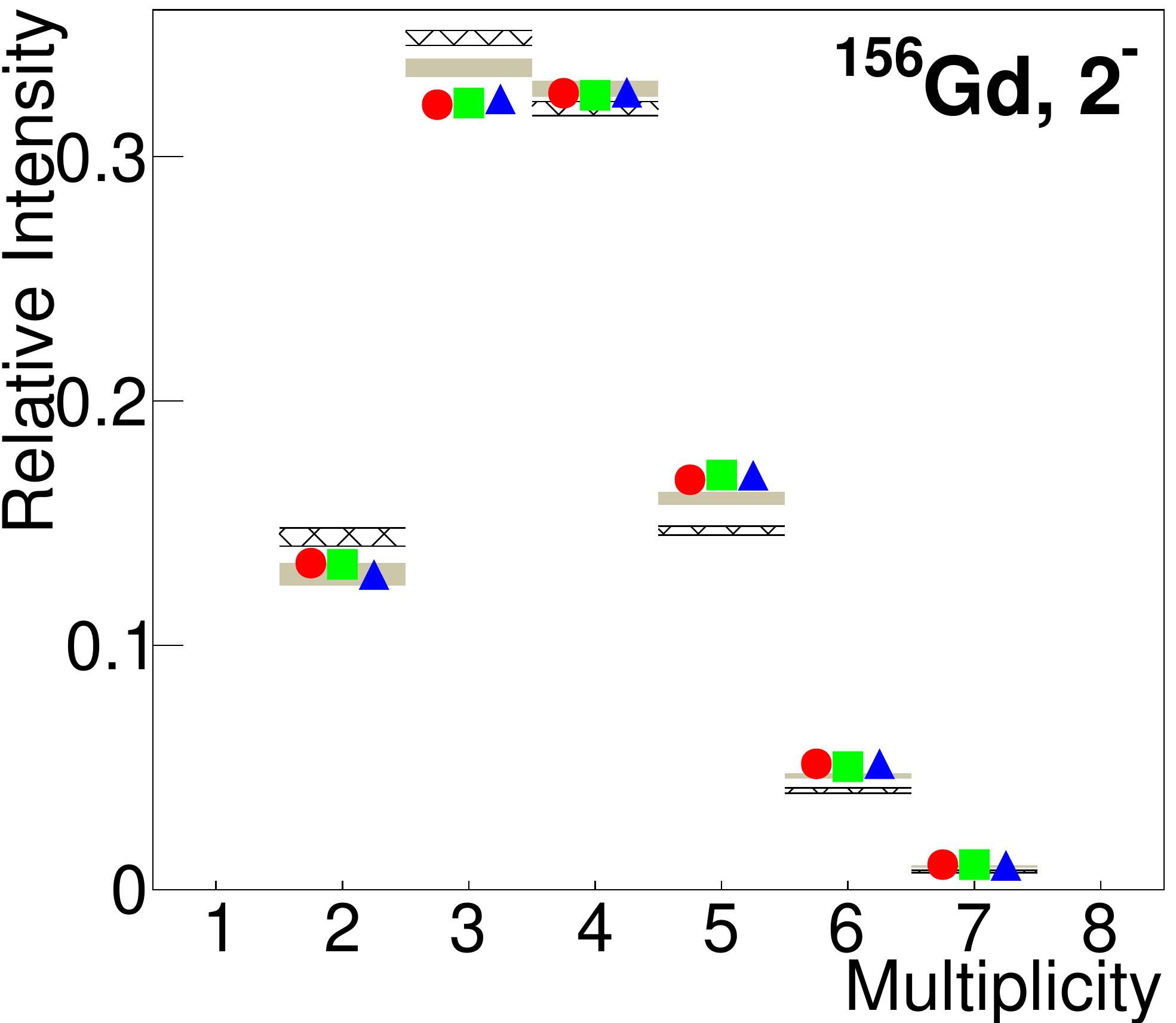}
\hfill
\includegraphics[clip,width=0.49\columnwidth]{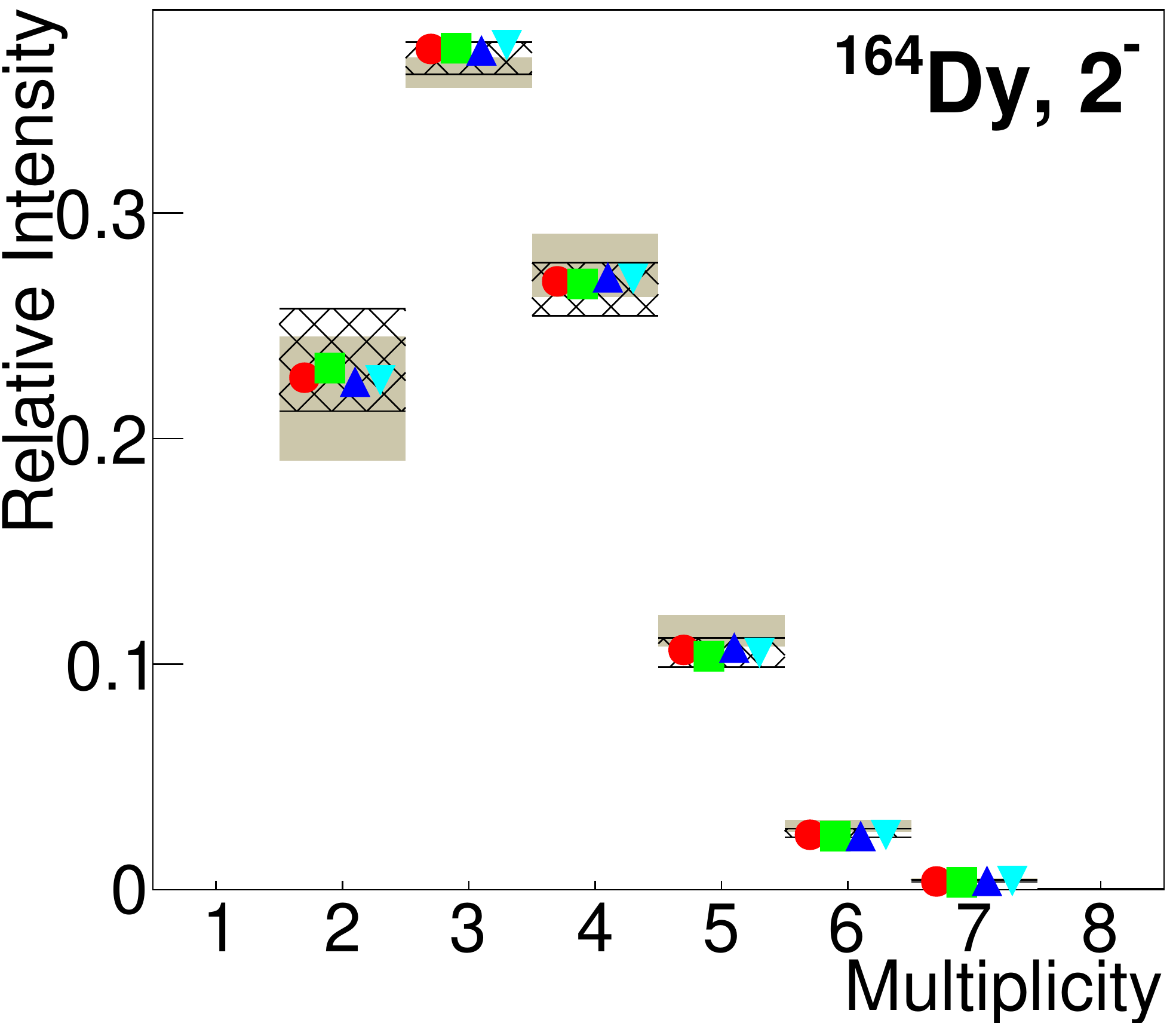}

\caption{(Color online) The MD from the decay of $3^+$ resonances in $^{96}$Mo and $^{98}$Mo, and $2^-$ resonances in $^{156}$Gd and $^{164}$Dy, as measured by the DANCE detector. The color symbols correspond to experimental data from different resonances, the black hatched area and the gray band to predictions from simulations (average $\pm$ one standard deviation) with SMLO and D1M+QRPA+0lim, respectively. The HFB plus combinatorial model of NLD is used for all simulations.}
\label{fig:md-mo96_s}
\end{figure}

\begin{figure}[t]
\includegraphics[clip,width=1.0\columnwidth]{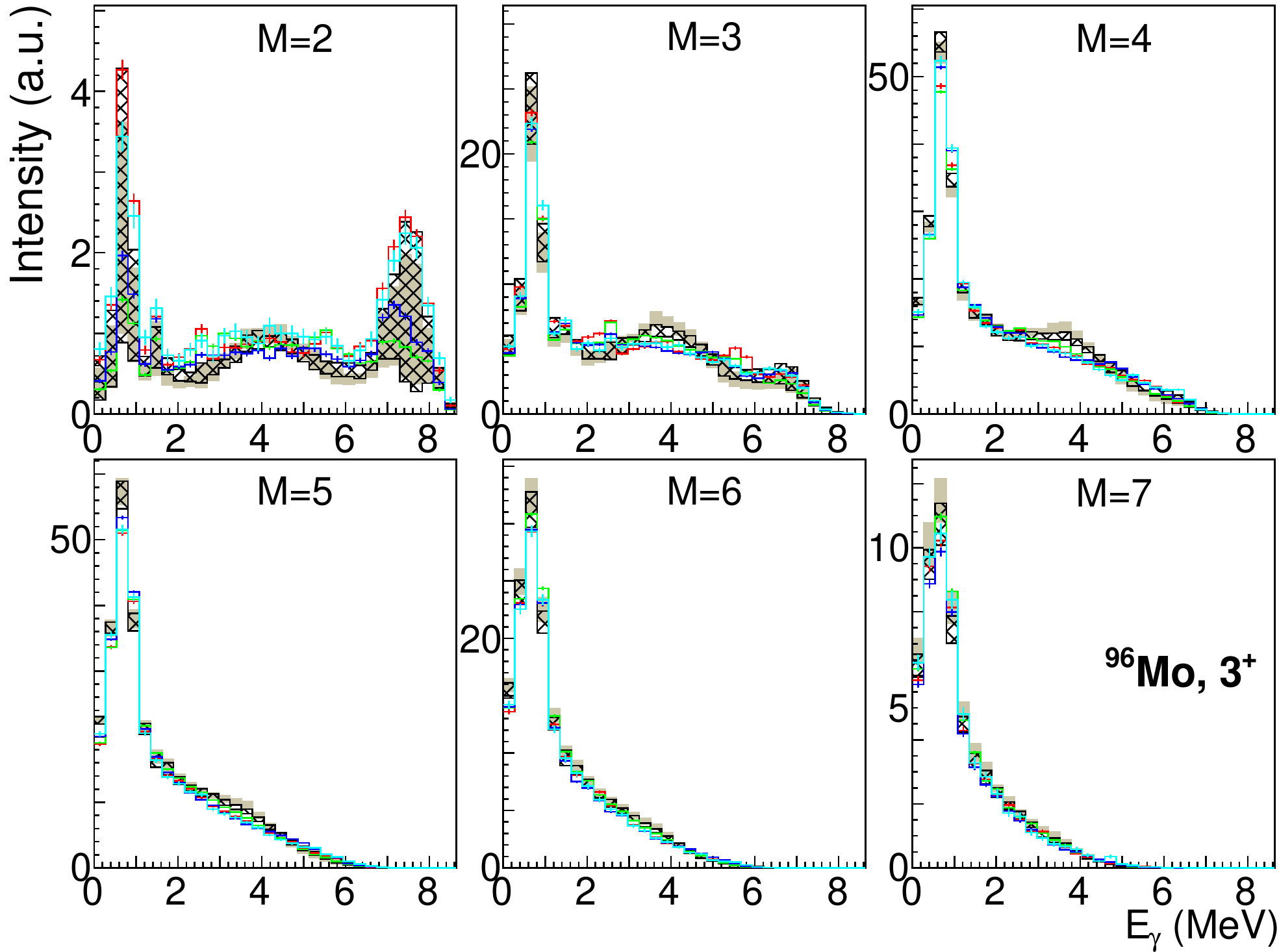}
\vspace*{2mm}
\includegraphics[clip,width=1.0\columnwidth]{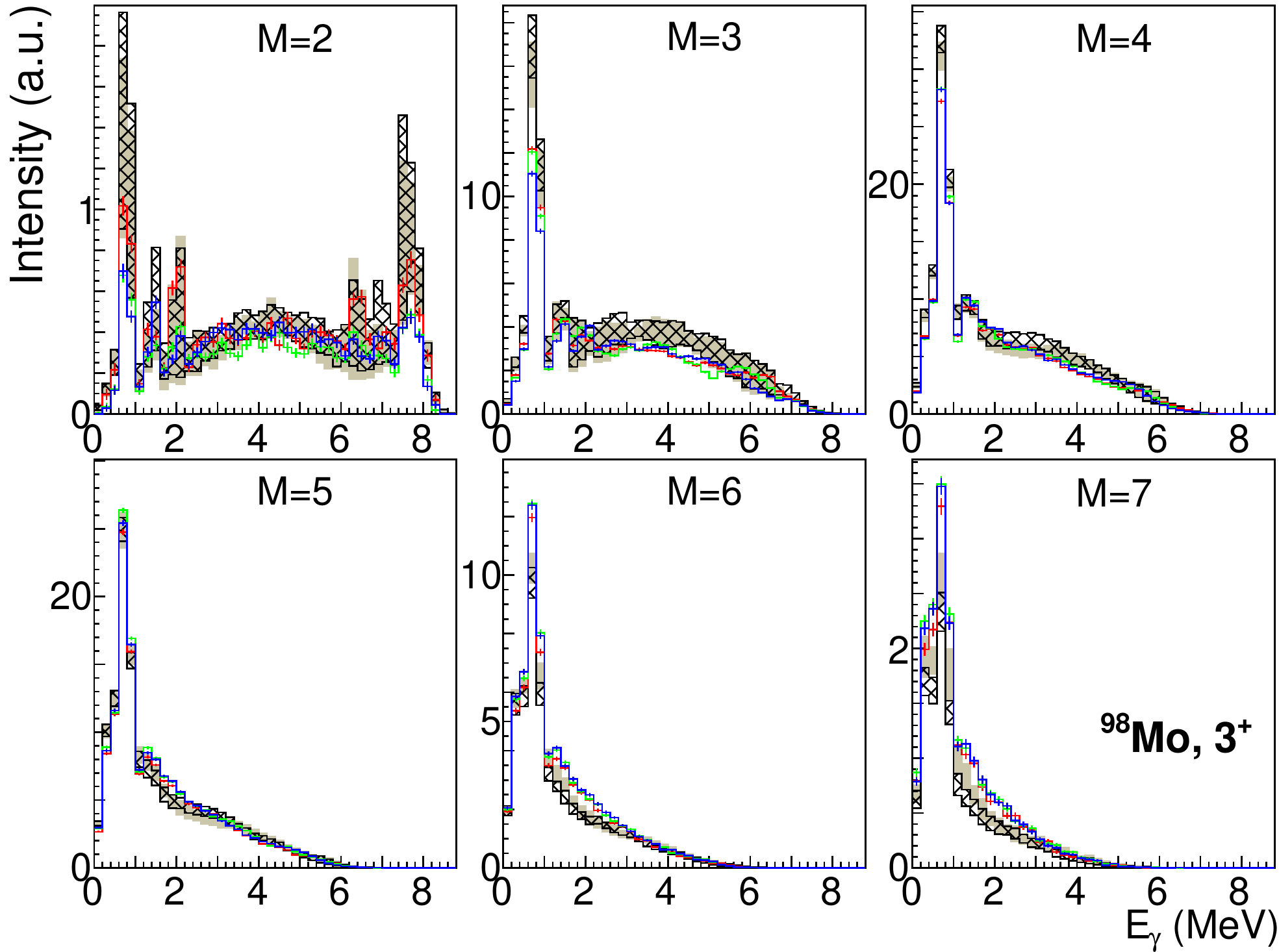}
\caption{(Color online) The MSC spectra from $3^+$ $^{96}$Mo and $^{98}$Mo resonances, as measured with the DANCE detector. The color lines correspond to experimental data, the black hatched area and the gray band to predictions from simulations (average $\pm$ one standard deviation) with the SMLO and D1M+QRPA+0lim, respectively. The HFB plus combinatorial NLD model is used in all simulations.}
\label{fig:msc-mo96_s}
\end{figure}

\begin{figure}[t]
\includegraphics[clip,width=1.0\columnwidth]{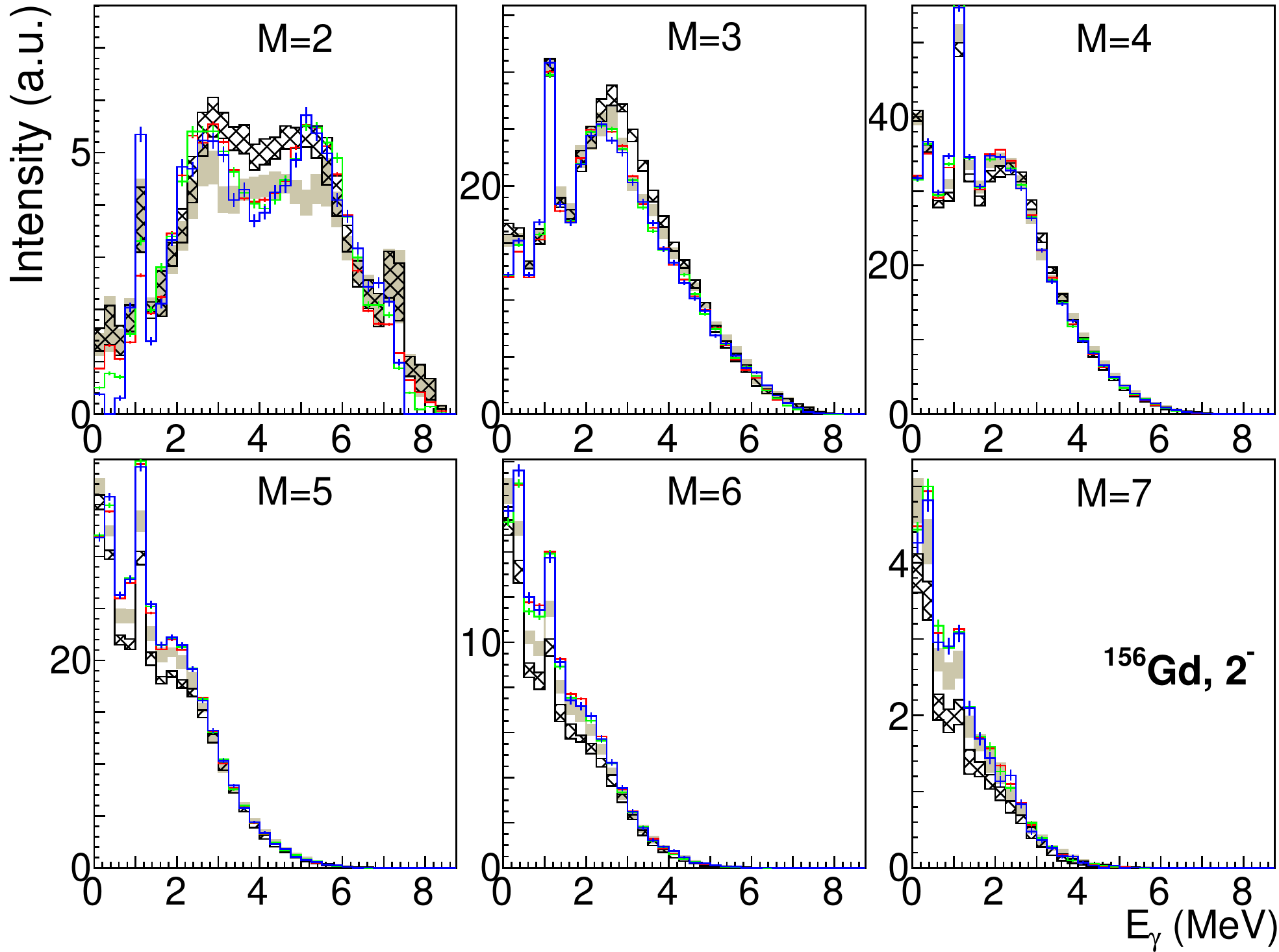}
\vspace*{2mm}
\includegraphics[clip,width=1.0\columnwidth]{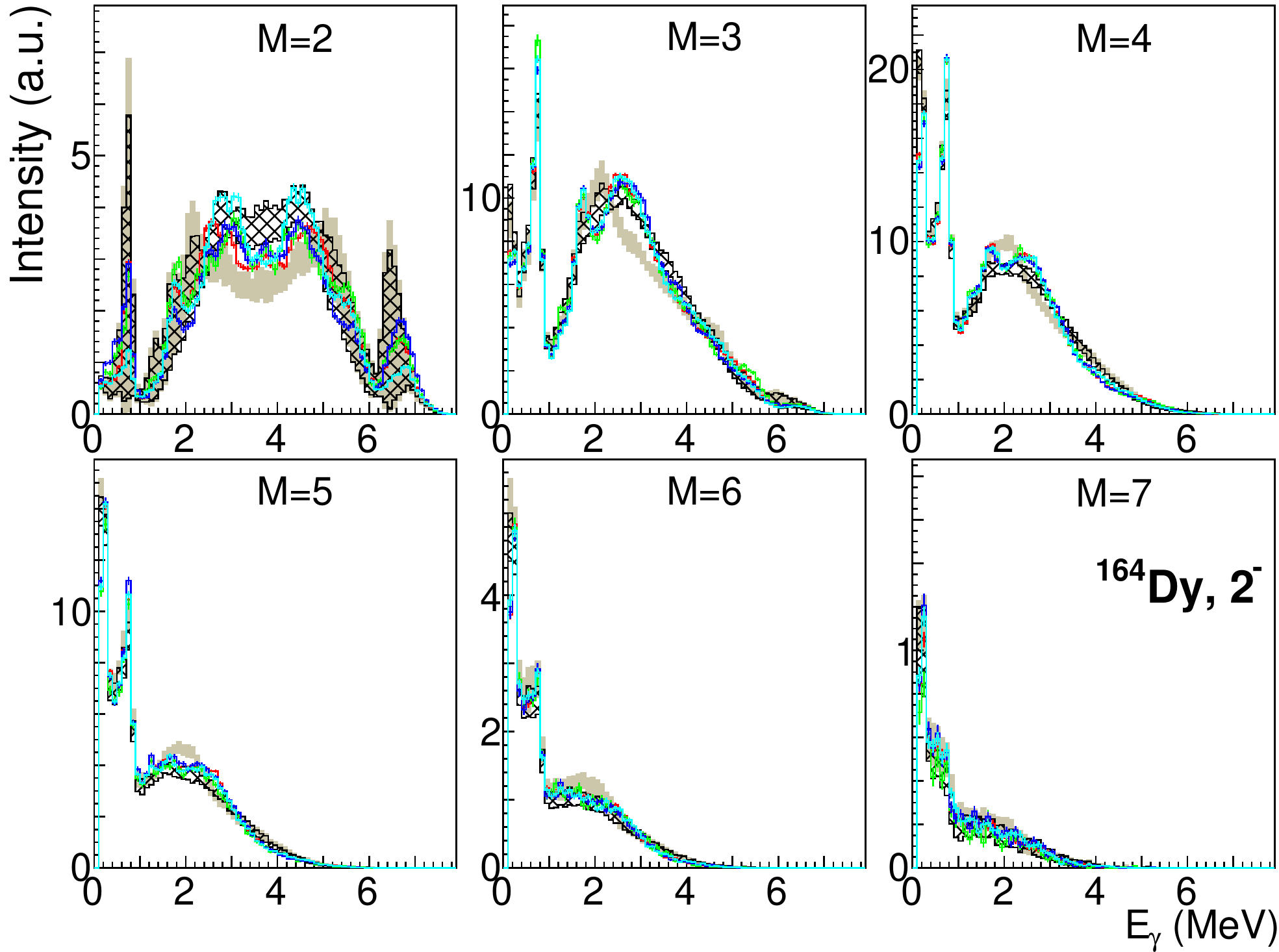}
\caption{(Color online) Same as in Fig. \ref{fig:msc-mo96_s} but for $2^-$ $^{156}$Gd and $^{164}$Dy resonances.}
\label{fig:msc-gd_s}
\end{figure}

The MD and MSC spectra were constructed separately from each simulated nuclear realization and were normalized to give the same area of the sum-energy spectra for multiplicities $M=2-7$ and the chosen sum-energy cut; for details see Ref. \cite{Krticka19} and references therein. The range of predictions corresponding to two standard deviations (average $\pm$ standard deviation) from individual nuclear realizations is illustrated in Figs. \ref{fig:md-mo96_s}-\ref{fig:msc-gd_s}. Simulated MSC spectra are normalized to experimental ones using one (common for all $M$) normalization factor, again to give the same area in the sum-energy spectra for $M=2-7$. The absolute 
scale on the vertical axes of MSC spectra is arbitrary but the relative contributions of different $M$ are kept.

A detailed comparison of simulations obtained with different ingredients in the D1M+QRPA+0lim PSFs was perfor\-med in Ref. \cite{Krticka19} for spectra from 15 different nuclei, sometimes from resonances with different spins (and parities). In the simulations of Ref.~\cite{Krticka19}, the size of the $M$1 zero-$E_\gamma$ amplitude $C$ (Eq.~\ref{eq_qrpa2}), the $M1$ broadening width $\Gamma_{M1}$ and the $E1$ low-$E_\gamma$ limit given by the $f_0$ and $U$ parameters (Eq.~\ref{eq_qrpa2}) were varied. 
In addition, two significantly different NLD models were used in the comparison, namely the microscopically based HFB plus combinatorial model \cite{Goriely08b} and the phenomenological CT plus Fermi gas model  \cite{Koning08} recommended by the RIPL-3 Library \cite{Capote09}. It was found that the predicted spectra show sensitivity to the adopted NLD model.

Although the agreement between experimental and simulated spectra using the global D1M+QRPA+0lim model is not always perfect, and, not surprisingly, better agreement can be reached with locally adjusted models or parameters (see original papers with an analysis of MSC spectra listed in Ref. \cite{Krticka19}), the predictions of the global model are acceptable, as discussed in details in Ref.~\cite{Krticka19}. The smearing width of the $M1$ PSF, that need to be applied to D1M+QRPA calculations should be smaller than about 1 MeV. Further, the influence of different proposed low-$E_\gamma$ $E1$ parametrizations remains small. This is not that surprising, as the $M1$ contribution for $E_\gamma$ below about 3~MeV is expected to be significantly higher than tested $E1$ contribution.

The sensitivity to the NLD models is reflected predominantly in the uncertainty affecting the low-$E_\gamma$ $M1$ enhancement. As a result of this sensitivity, a different enhancement is required to describe experimental data with different NLD models. In addition, some NLD models seem not to be able to describe the spectra for some nuclei, particularly in the case of rare-earth nuclei \cite{Krticka19}. 
In any case, the comparison of MSC spectra indicates that the contribution of the scissors mode is reasonably well described  by the D1M+QRPA+0lim model and that data seem to be compatible with a non-negligible $M1$ low-$E_\gamma$ strength in all the tested nuclei. 

A phenomenological parametrization was proposed for the low-energy limit C of the D1M+QRPA+0lim model from the comparison between predictions and experiment \cite{Krticka19}. As discussed in Sec.~\ref{sec_th_qrpa}, the form of the parametrization is  partly based on the expectation of the shell-model calculations.
A comparison for more nuclei with $A<100$ would however be needed to get more reliable information on the low-$E_\gamma$ limit for these nuclei.

Similar comparisons have been made also for predictions of the SMLO model. In this case, we fixed the parameters of the $M1$ spin-flip mode, the scissors mode as well as of the low-$E_\gamma$ limit to the values proposed in Sec. \ref{sec_th_smlo} (Eqs.~\ref{eq_smlo_m1}-\ref{eq_smlo_m2}).

Examples of the agreement between the MD spectra predicted with both the D1M+QRPA+0lim and SMLO models and experiment can be found in Fig. \ref{fig:md-mo96_s} for four nuclei. The MSC spectra for the same nuclei are compared in Figs. \ref{fig:msc-mo96_s} and \ref{fig:msc-gd_s}. The overall quality of the description of the experimental spectra with the SMLO predictions seem to be slightly worse than that with the D1M+QRPA+0lim model. 
The influence of the adopted NLD models on predictions was similar to that found with the D1M+QRPA+0lim model; the same two NLD models used as in combination with the D1M+QRPA+0lim model were tested. 

The proposed systematics of the low-$E_\gamma$ limit in the SMLO model in combination with the HFB plus combinatorial NLD model leads to a MD slightly shifted toward lower values if compared to experiment. 
The trend is even more pronounced if the SMLO PSF model is combined with the CT plus Fermi gas model of NLD. In addition, the position of the scissors mode in U isotopes (around 3~MeV in the SMLO model) leads to the absence of bumps near 2 MeV in $M=3-4$ spectra.

\begin{figure}[t]
\includegraphics[clip,width=0.98\columnwidth]{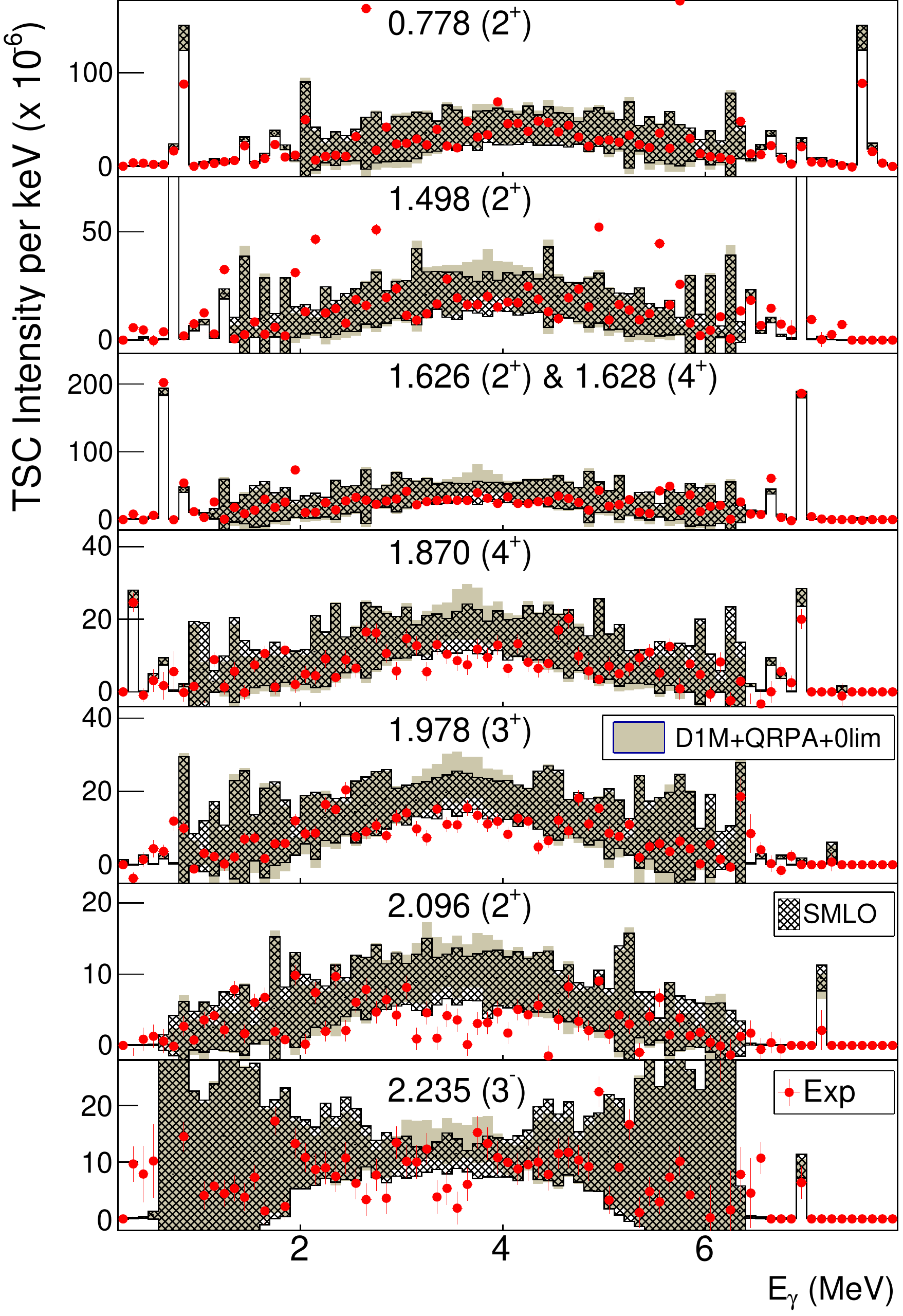}
\caption{(Color online) TSC spectra for several final levels in $^{96}$Mo. Red points correspond to experimental data, the black hatched area and the gray band to predictions from simulations (average $\pm$ one standard deviation) with the SMLO and D1M+QRPA+0lim, respectively. The HFB plus combinatorial NLD model is used. Energies (in MeV) of final levels are indicated.}
\label{fig:tsc_mo96}
\end{figure}
\begin{figure}[t]
\includegraphics[clip,width=0.98\columnwidth]{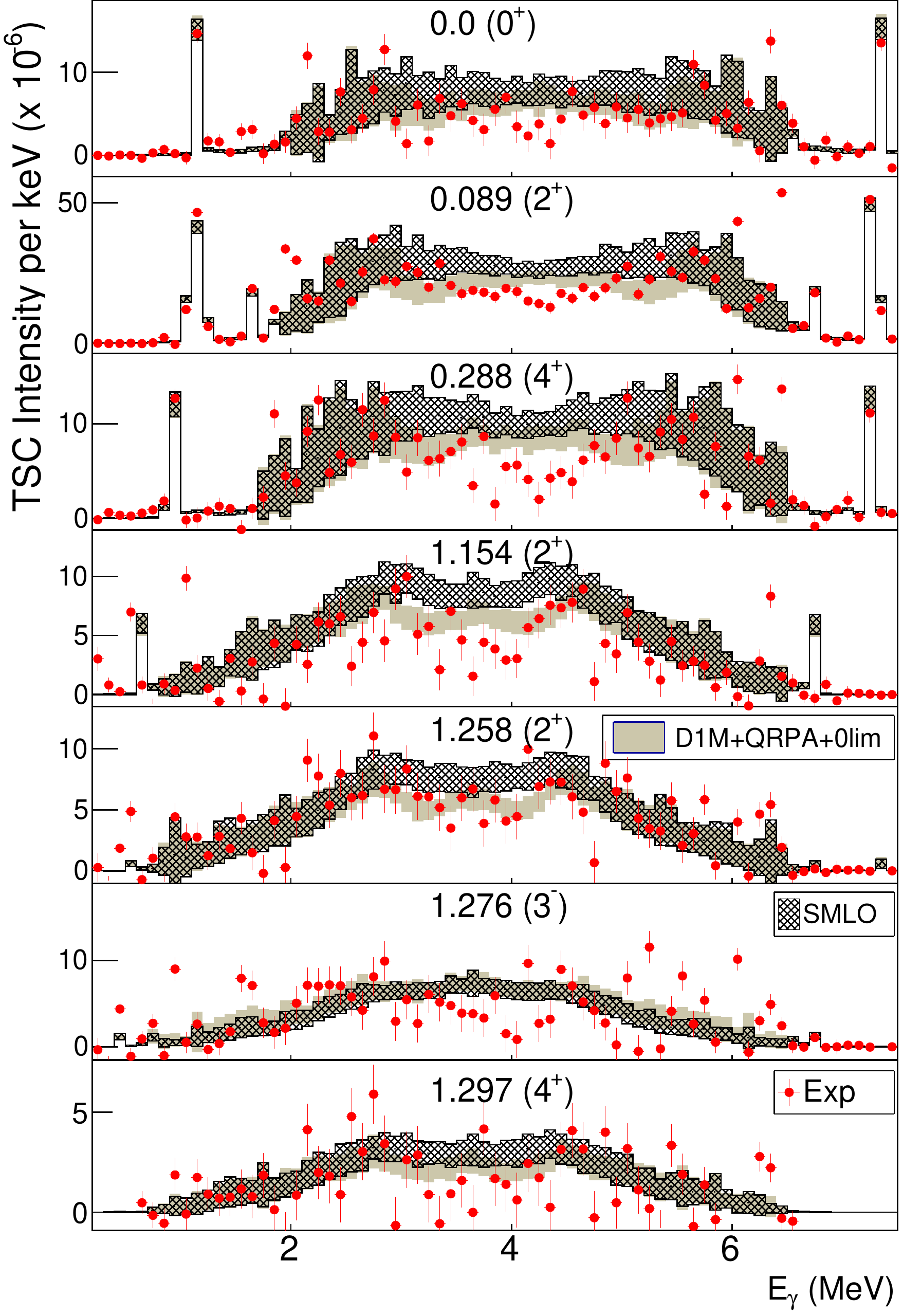}
\caption{(Color online) Same as in Fig. \ref{fig:tsc_mo96} but for $^{156}$Gd.}
\label{fig:tsc_gd156}
\end{figure}

Predictions from the statistical model were also compared for TSC spectra following thermal neutron capture for $^{96}$Mo and $^{156}$Gd nuclei, as shown in 
Figs.~\ref{fig:tsc_mo96} and \ref{fig:tsc_gd156}, respectively. Simulation of $\gamma$ cascades was again performed with the {\sc DICEBOX} code.

Similarly to MD and MSC spectra discussed above, the reproduction of experimental spectra is not perfect, as can be expected from global models. The D1M+QRPA+0lim model seems to give a better reproduction of $^{156}$Gd spectra than the SMLO  model which indicates that the scissors mode for this nucleus is better described by the D1M+QRPA+0lim model.

\subsection{Average radiative width}
\label{sec_comp_gamgam}

\begin{figure}
\centering
\includegraphics[width=\columnwidth, angle = 0]{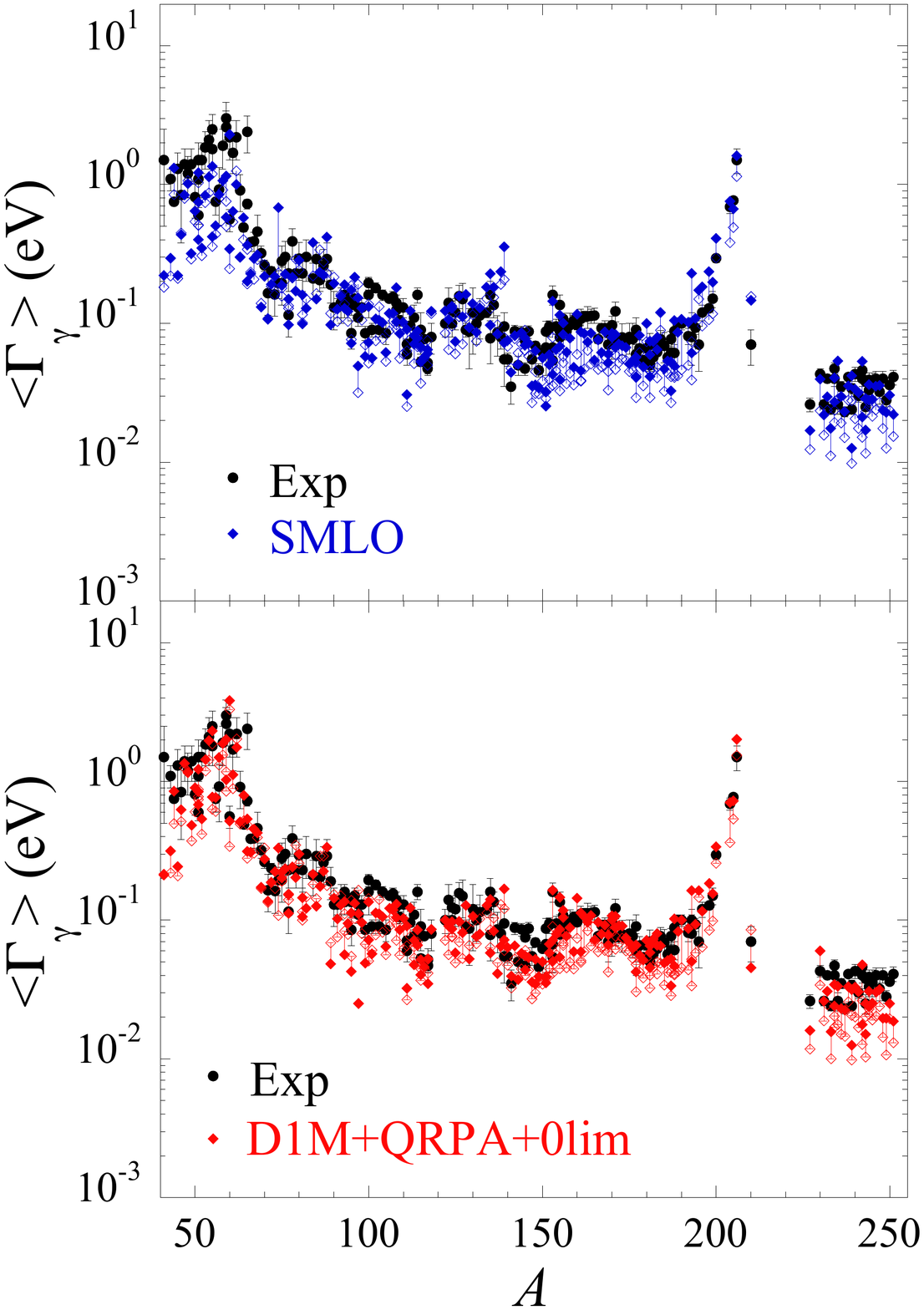}
\caption{(Color online) Upper panel: Comparison between the 228 experimental (black circles) \cite{Capote09} and SMLO (colored diamonds) average radiative width $\langle\Gamma_\gamma \rangle$ as a function of $A$. The NLD adopted here is from  HFB plus combinatorial model (full diamonds) \cite{Goriely08b} or the CT plus Fermi gas model (open diamonds) \cite{Koning08}. Lower panel: Same where the theoretical $\langle\Gamma_\gamma \rangle$ values are obtained with the D1M+QRPA+0lim model.}
\label{fig_comp_gamgam}
\end{figure}

The average radiative width, as given by Eq.~\ref{eq_gamgam}, is an additional source of indisputable information on global PSFs behavior below $S_n$.
It has been a long-standing problem that the phenomenological SLO models for the $E1$ strength overestimate the average radiative width 
$\langle\Gamma_\gamma \rangle$ significantly, while an improved and widely used version of these phenomenological models, the so-called GLO model \cite{Capote09,Kopecky90}, 
underestimates it. Such deviations are discussed in detail in Ref.~\cite{Goriely18a}. In Fig.~\ref{fig_comp_gamgam}, we compare the 
228 experimental average radiative widths for nuclei lying between $^{33}$S and $^{251}$Cf with the SMLO and D1M+QRPA+0lim predictions. The 
low-energy $M1$ components, {\it i.e.} the scissors mode for deformed nuclei and the upbend, are found to have non-negligible contributions to the 
$\langle\Gamma_\gamma \rangle$ integral (Eq.~\ref{eq_gamgam}), though they were not taken into account in the traditional Lorentzian approach 
\cite{Capote09,Kopecky90}; this explains why the GLO model tends to underestimate the experimental $\langle\Gamma_\gamma \rangle$. Both the  SMLO 
and D1M+QRPA+0lim PSF models are found to be globally in agreement with experimental data, as shown in Fig.~\ref{fig_comp_gamgam} and seen from 
the root mean square (rms) deviations given in Table~\ref{tab_rms}, where the $\varepsilon_{rms}$ and $f_{rms}$ deviation factors are defined as
\begin{eqnarray}
\varepsilon_{rms}&=&{\mathrm \exp} \left[\frac{1}{N_e} \sum_{i=1}^{N_e} \ln
r_i \right] \label{eq_rms1}\\
f_{rms}&=&{\mathrm \exp} \left[\frac{1}{N_e} \sum_{i=1}^{N_e} \ln^2
r_i \right]^{1/2} \label{eq_rms2}  ,
\label{eq_rms2}
\end{eqnarray}
where $N_e$ is the number of experimental data and $r_i$ is, for each data point $i$,  the ratio of theoretical to experimental 
$\langle\Gamma_\gamma \rangle$ which takes into account the experimental uncertainties $\delta_{{\rm exp}}$ affecting 
$\langle\Gamma_\gamma \rangle$. More precisely, the ratio $r$ is calculated as follows 
\begin{eqnarray}
r=&&\frac{\langle\Gamma_\gamma \rangle_{th}}{\langle\Gamma_\gamma \rangle_{{\rm exp}} - \delta_{{\rm exp}}} \quad {\rm if} \quad \langle\Gamma_\gamma \rangle_{th} < \langle\Gamma_\gamma \rangle_{{\rm exp}} - \delta_{{\rm exp}} \nonumber \\
=&&\frac{\langle\Gamma_\gamma \rangle_{th}}{\langle\Gamma_\gamma \rangle_{{\rm exp}} + \delta_{{\rm exp}}}\quad {\rm if} \quad \langle\Gamma_\gamma \rangle_{th} > \langle\Gamma_\gamma \rangle_{{\rm exp}} + \delta_{{\rm exp}} \nonumber \\
=&&1 \quad \hskip 2.1cm{\rm otherwise.}  
\label{eq_rms3}
\end{eqnarray}

As seen from Eq.~\ref{eq_gamgam}, the average radiative width is also sensitive to the spin- and parity-dependent NLD. For this reason, 
the average radiative widths shown in Fig.~\ref{fig_comp_gamgam} have been obtained with two fundamentally different NLD models, namely the HFB 
plus combinatorial model \cite{Goriely08b} and the CT plus Fermi gas \cite{Koning08}. The sensitivity to the NLD is depicted by the error bars 
in Fig.~\ref{fig_comp_gamgam}. The HFB plus combinatorial NLD model is seen to give rise to larger values of the average radiative width with 
respect to the CT plus Fermi gas model. This result is also reflected in the average rms deviations given in Table~\ref{tab_rms}. Globally, as reflected by the $f_{rms}$ factors, the SMLO model is found to reproduce experimental $\langle\Gamma_\gamma \rangle$ within 45--60\% while the D1M+QRPA+0lim is slightly more accurate with approximately 30\% accuracy.

\begin{table}
\centering
\caption{$\varepsilon_{rms}$ and $f_{rms}$ for the theoretical to experimental ratios of both  $\langle\Gamma_\gamma \rangle$ and the MACS $\langle\sigma \rangle$. Theoretical estimates are obtained with the present SMLO and D1M+QRPA+0lim recommended strengths. Both the CT plus Fermi gas \cite{Koning08} and the HFB plus combinatorial (Comb) \cite{Goriely08b} models of NLD are considered.}
\vskip 0.3cm
\begin{tabular}{l c c c c }
 \hline
 \hline
& \multicolumn{2}{c}{$\langle\Gamma_\gamma \rangle$ } & \multicolumn{2}{c}{$\langle\sigma \rangle$} \\
& $\varepsilon_{rms}$ & $f_{rms}$ & $\varepsilon_{rms}$ & $f_{rms}$ \\
 \hline
SMLO (Comb) & 0.90 & 1.45 & 1.11 & 1.47 \\
SMLO  (CT) &  0.74 & 1.62 &  0.98 & 1.40  \\
\hline
D1M+QRPA+0lim (Comb) &1.02 & 1.27 & 1.30 & 1.55 \\
D1M+QRPA+0lim  (CT) &  0.90 & 1.32  &  1.15 & 1.40  \\
\hline
\hline
\end{tabular}
\label{tab_rms}
\end{table}

\subsection{Maxwellian-averaged cross sections at 30~keV}
\label{sec_comp_macs}

\begin{figure}
\centering
\includegraphics[width=\columnwidth, angle = 0]{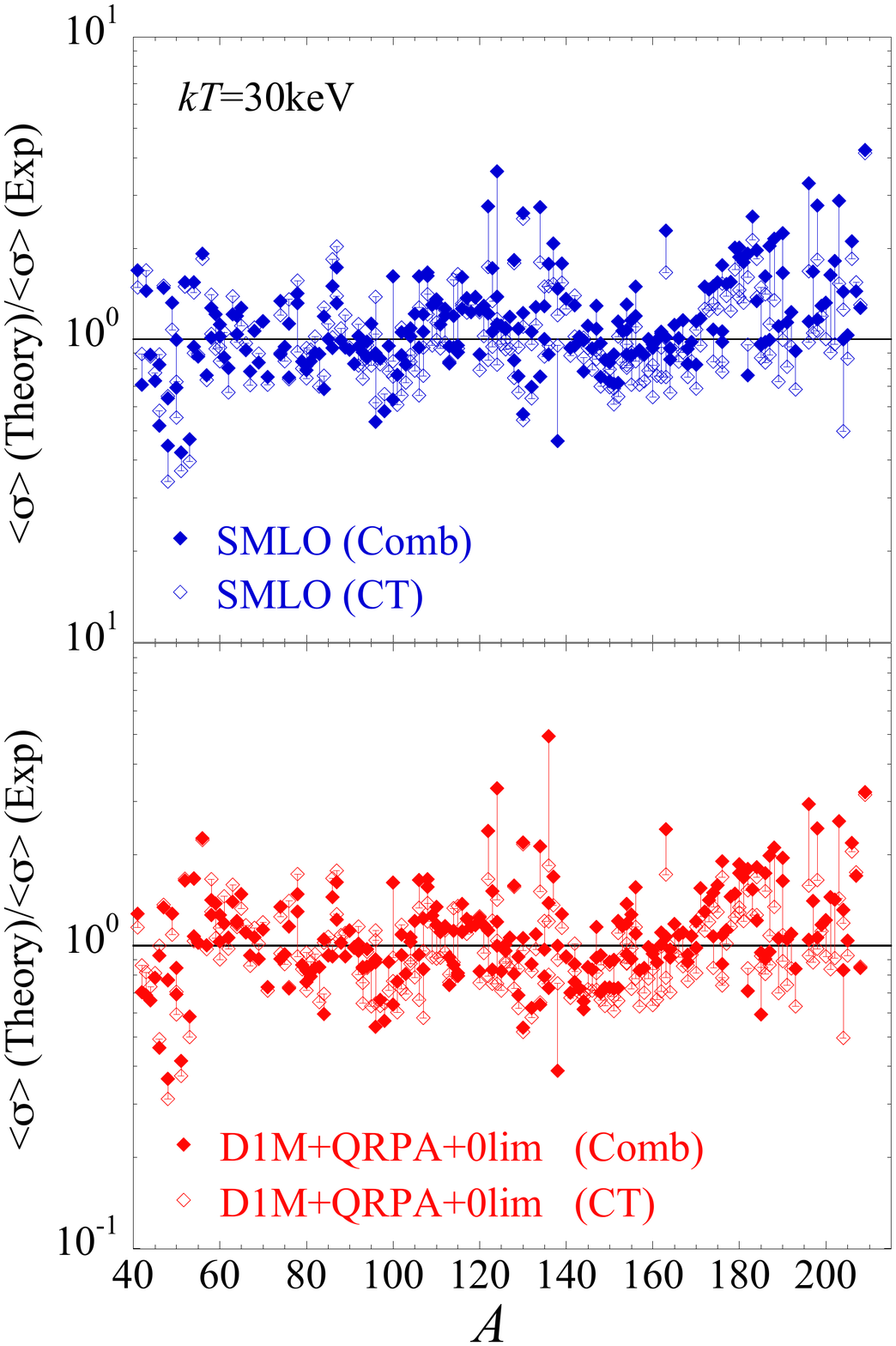}
\caption{(Color online) Upper panel: Ratio of the theoretical to experimental MACS at  $kT=30$~keV as a function of the atomic mass $A$ for all nuclei between Ca and Bi for which experimental MACS exist  \cite{Dillmann06,Dillmann14}. The theoretical MACS are obtained with the present SMLO model for the $E1$ and $M1$ strengths. The full symbols are calculations with the HFB plus combinatorial model of NLD \cite{Goriely08b} and the open symbols with the CT plus Fermi gas model  \cite{Koning08}. Lower panel: Same when the MACS are calculated with the D1M+QRPA+0lim model.}
\label{fig_signg_exp}
\end{figure}

Similarly to the average radiative width, the Maxwellian-averaged cross section also yield global information on PSFs below the neutron separation energy.
In Fig.~\ref{fig_signg_exp} we compare, for nuclei with $20\le Z \le 83$, the 240 experimental neutron-capture Max\-wellian-averaged  cross 
sections (MACS) \cite{Dillmann06,Dillmann14} at a neutron energy of $kT=30$~keV (assuming the target in its ground state only)  with the  
MACS estimated by the TALYS \cite{Koning12} reaction code using either  the SMLO or the D1M+QRPA+0lim PSF models. Both the 
HFB plus combinatorial and CT plus Fermi gas models of NLD are considered, as in previous sections. Note that in the TALYS calculation of the 
MACS $\langle \sigma\rangle$, the PSF is not renormalized so as to reproduce the experimental average radiative width; 
the capability of the models to reproduce experimental average radiative widths is described in the previous Sec.~\ref{sec_comp_gamgam}. 
Only nuclei with $Z\ge 20$ are considered in the comparison to ensure the validity of the Hauser-Feshbach approach,  the cross section for 
lighter nuclei being affected by contributions from direct reactions \cite{Xu14} and by the resolved resonance regime \cite{Rochman17} at the 
30~keV neutron energies considered here.

The deviation from experimental data can be characterized by the same rms factors, $\varepsilon_{rms}$ and $f_{rms}$, as defined for the 
average radiative width (Eqs.~\ref{eq_rms1}-\ref{eq_rms2}), with $r_i=\langle \sigma\rangle^i_{\rm th} / \langle\sigma\rangle^i_{\rm exp}$. 
In this case, the experimental error bars  are usually rather small (a few percent)  \cite{Dillmann06,Dillmann14}, so that 
the uncertainties $\delta_{\rm exp}$ have a small impact on the calculation of the rms factors. As shown in Table~\ref{tab_rms}, the MACS are 
well reproduced by both the SMLO and D1M+QRPA+0lim PSFs with an accuracy of 40--50\% ({\it i.e.} $f_{rms}\simeq 1.40-1.50$). MACS, like 
average radiative widths, are sensitive to the NLD model adopted in the calculations. The sensitivity to the NLD model is illustrated in 
Fig.~\ref{fig_signg_exp} and Table~\ref{tab_rms} where both the CT plus Fermi gas and HFB plus combinatorial models are considered in the 
calculation of the MACS. 

\section{IAEA PSF reference database}
\label{sec_data}

The IAEA PSF database aims to be an internationally recognised database of experimental, recommended and theoretical PSFs. As such, 
it contains all the available PSF data measured via the experimental techniques described in this report (Secs.~\ref{sec_exp} and 
\ref{sec_ass}) as well as the theoretical PSFs described in Sec.~\ref{sec_th}.

\subsection{The experimental PSF database} 
\label{sec_database_exp}

\begin{figure*}
\includegraphics[scale=0.6]{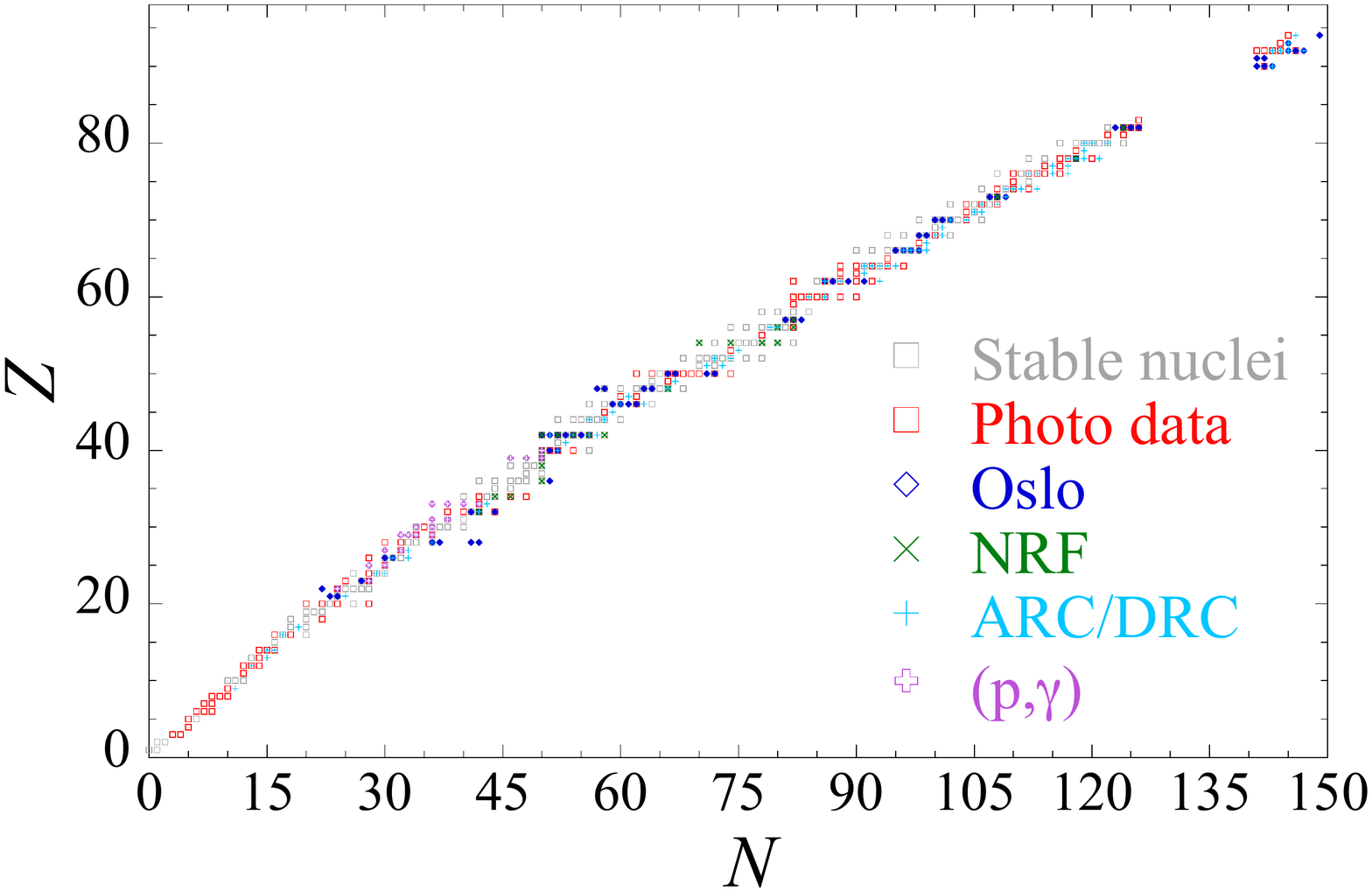}  
\vskip -1.cm
\caption{(Color online)  Representation in the ($N,Z$) plane of all the main experimental data used to extract the PSFs, namely photodata as well as Oslo, NRF, ARC/DRC or (p,$\gamma$) methods. Stable and long-lived nuclei are depicted by grey squares. }
\label{fig_allexp_NZ}
\end{figure*}

For the experimental PSF database, data are stored in separate data files per nuclide and for each measured multipolarity $XL$, if available, or total $L$. 
The data files contain information on the nuclide (Z,A), multipolarity $XL$,  units and data format, as well as the 
reference of the publication. Each data file is accompanied by a README file which contains information on 
the measurement technique, the assumptions and models used as input in the data analysis, the uncertainty budget, energy cut-offs 
and data omitted in the data files where needed, {\it e.g.} data points below $S_n + \Delta\varepsilon$ and above $\varepsilon_{QD}$ in photonuclear data. 
The database contains the following PSF data (described in Sec.~\ref{sec_ass_compil}):
\begin{itemize}
\item NRF measurements for 23 nuclei with $Z=32-78$  (total of  31 files);
\item charged-particle reaction data with the Oslo method for 72 nuclei with $Z=21-94$ (total of 113 files);
\item ARC/DRC measurements for 88 nuclei with $Z = 9 - 94$ (total of 193 files);
\item thermal neutron capture data for 209 nuclei with $Z=3 - 96$;
\item (p,$\gamma$) measurements for 22 nuclei with $Z=22-40$ (total of 38 files);
\item ratio method measurement for 1 nucleus, $^{95}$Mo (1 file).
\item (p,p$^\prime$) measurements for 3 nuclei, $^{96}$Mo, $^{120}$Sn and $^{208}$Pb (total of 5 files);
\item $E1$ photodata for 159 nuclei with $Z =3-94$ (total of 465 files).
\end{itemize}
An overview of the PSF data extracted from the various sources and compiled in the database is shown in Fig.~\ref{fig_allexp_NZ} which illustrates 
the extent of the experimental effort performed up to now. 

\begin{figure*}
\begin{center}
\includegraphics[scale=0.55]{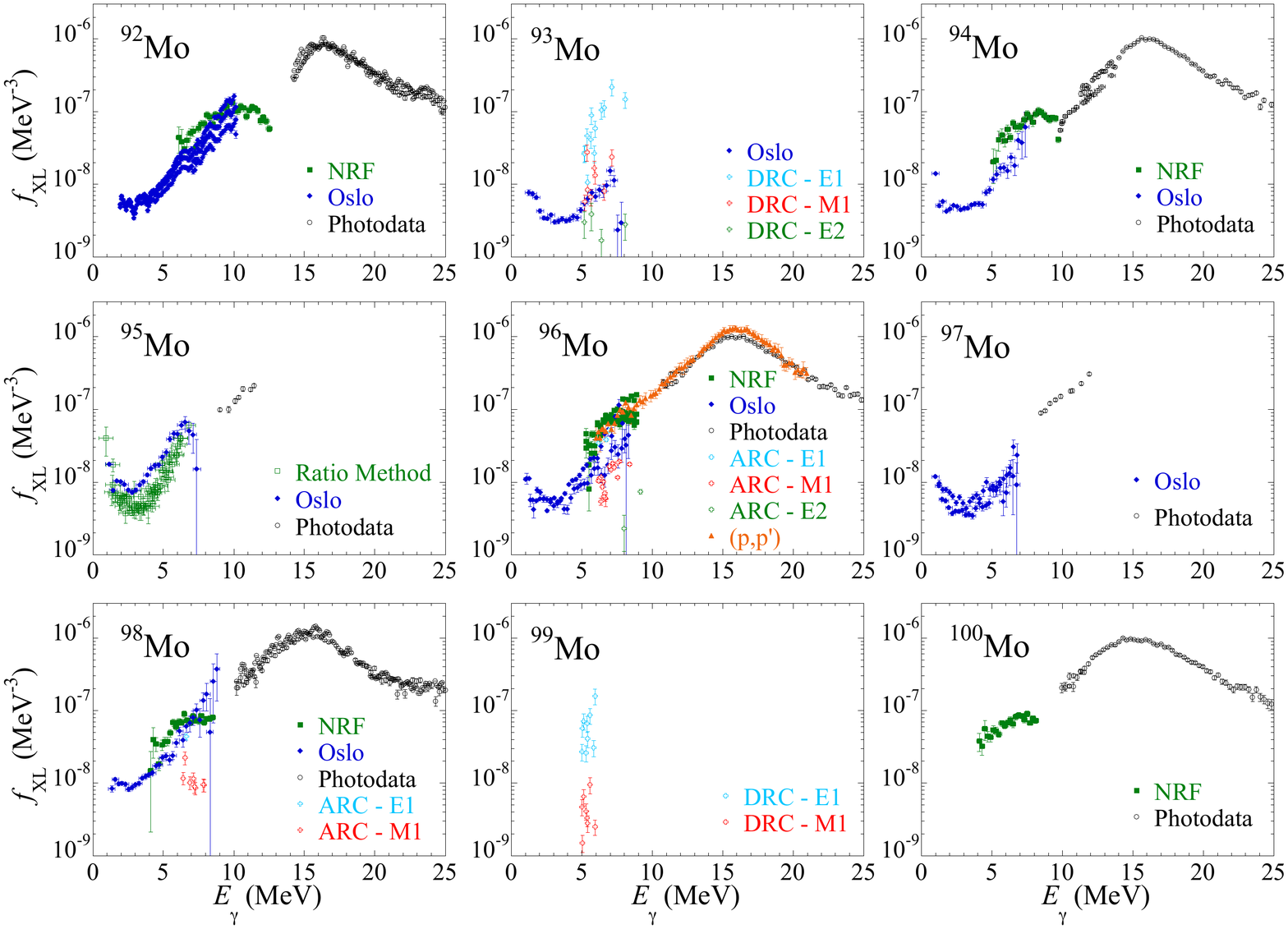}  
\caption{(Color online) Experimental PSFs of the Mo isotopes extracted from different types of measurements, including photodata as well as the Oslo, NRF, ARC/DRC, (p,p$^\prime$) and ratio methods, as described in Sec.~\ref{sec_ass_compil}. Data can be found in Refs.~\cite{Rusev08,Rusev09} for NRF measurements, \cite{Tveten16,Guttormsen05,Chankova06,Utsunomiya13} for Oslo data, \cite{Wiedeking12,Krticka16} for Ratio method data, \cite{Martin17} for (p,p$^\prime$) data, \cite{Varlamov03,Varlamov14,Ishkhanov70,Beil74,Utsunomiya13,Banu19} for photodata and \cite{Kopecky17,Kopecky17b,Kopecky18} for ARC and DRC data. Data correspond to the total dipole PSFs, except ARC and DRC data which yield $E1$, $M1$ and $E2$ PSFs separately.}
\label{fig_allexp_Mo}
\end{center}
\end{figure*}

To illustrate the variety of data, we show in Fig.~\ref{fig_allexp_Mo} the extracted PSFs for the 9 Mo isotopes for which data exist. Below the neutron separation energy, deviations 
between the different Oslo and NRF methods can be observed. These, together with the uncertainties associated to each of these methods, are discussed in Sec.~\ref{sec_ass_unc}. 
NRF measurements usually lead to higher PSFs below $S_n$, but may be compatible with the Oslo data when all systematic uncertainties are included. A non-negligible spread can also be seen 
with respect to ARC data, for example for nuclides $^{96}$Mo and $^{98}$Mo. A relatively satisfactory smooth transition between data below and above the neutron threshold can be observed, 
despite the $1-2$~MeV energy gap that can exist between PSFs extracted from photonuclear data and other methods. 

Recent (p,p$^\prime$) data for $^{96}$Mo \cite{Martin17} are seen to bridge this gap around the neutron threshold, and are also found to be in agreement with NRF data. Future analysis of the 
model-dependent methods could take advantage of this alternative experimental method to further constrain the measurements.

\begin{figure*}
\begin{center}
\includegraphics[scale=0.5]{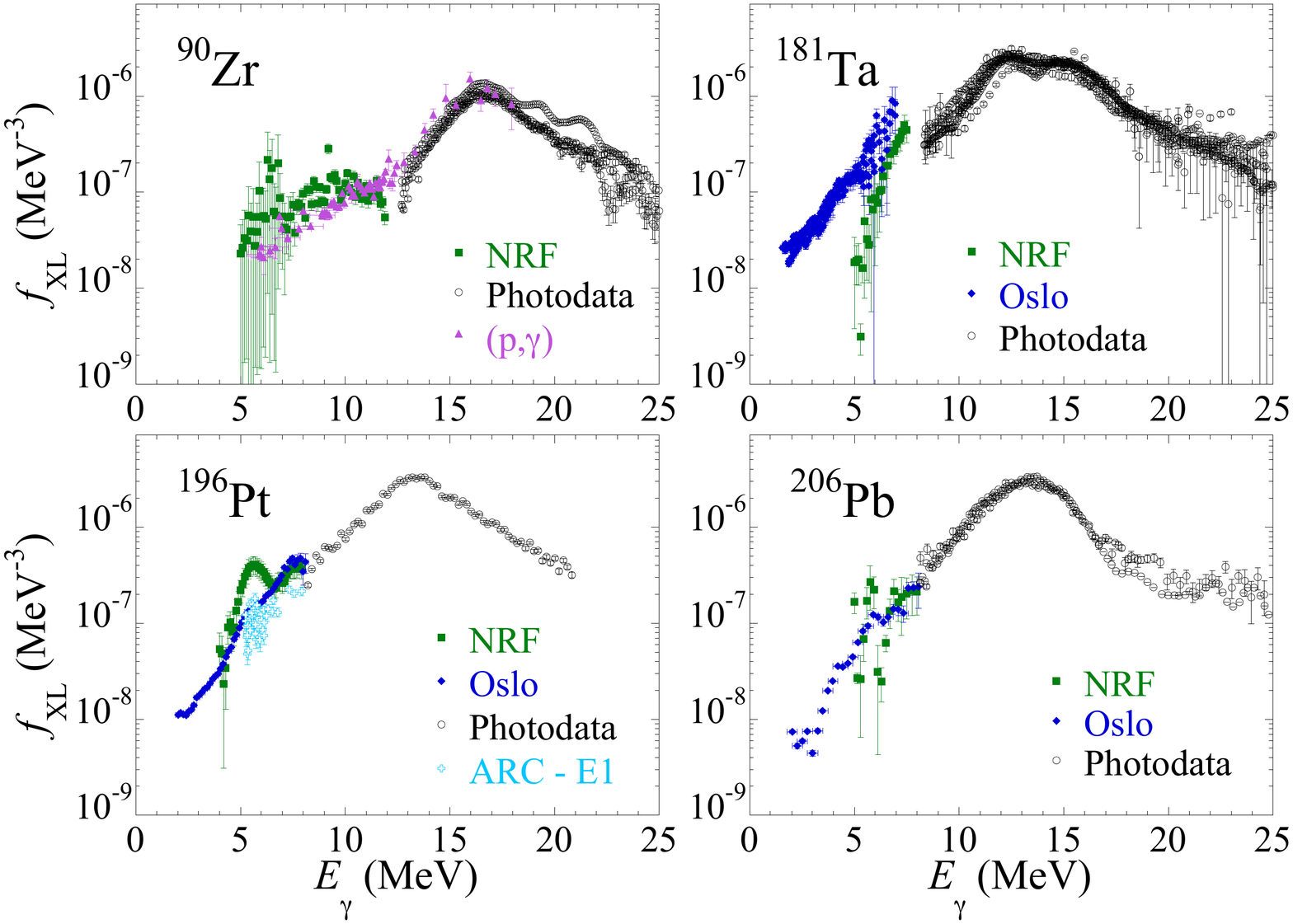}  
\caption{(Color online) Experimental PSFs for 4 nuclei, $^{90}$Zr, $^{181}$Ta, $^{196}$Pt and $^{206}$Pb, for which data from different methods exist. Data can be found in Refs.\cite{Schwengner08,Makinaga14,Massarczyk13,Shizuma18} for NRF data, \cite{Giacoppo15,Syed09,Brits19} for Oslo data, \cite{Fuller58,Bramblett63,Harvey64,Berman67,Antropov68,Bergere68,Lepretre71,Askin73,Sorokin73,Gurevich81,Utsunomiya03,Varlamov03,Varlamov14,Goko06,Kondo12,Netterdon15,Banu19} for photodata and  \cite{Szeflinska79,Szeflinski83} for (p,$\gamma$) data. Data correspond to the total dipole PSFs, except ARC data which yield $E1$ PSFs.}
\label{fig_allexp_other}
\end{center}
\end{figure*}

In Fig.~\ref{fig_allexp_other}, we illustrate the available PSF data for $^{90}$Zr, $^{181}$Ta, $^{196}$Pt and $^{206}$Pb for which different methods have been used to extract the PSFs. As with the Mo isotopes, 
discrepancies are observed among the PSFs extracted from NRF, Oslo method, ARC and (p,$\gamma$) methods below $S_n$. Differences also appear among the 
many photonuclear data at energies above the GDR peak. 

In cases, where multiple measurements yield discrepant PSFs, it is highly desirable to recommend 
the most reliable data set to the user community. In the particular case of photodata PSFs,  
several photoabsorption measurements can exist that may lead to rather different PSFs especially at energies above the GDR peak, as illustrated for example by the $^{181}$Ta case in 
Fig.~\ref{fig_allexp_other}. For each nucleus, a single PSF data set 
is therefore recommended  based on the recently re-evaluated photonuclear reaction cross sections~\cite{Kawano19}. The new evaluations were performed by the CRP for the ``Updated IAEA Photonuclear Data Library''~\cite{Kawano19}. 
In cases where a CRP evaluation was not performed, the recently published recommendations of Ref. \cite{Plujko18} are adopted. These recommendations 
are based on an assessment of the experimental data using the method of Varlamov {\it et al.} \cite{Varlamov14} and on systematics of neighbouring nuclei. 

As all other methods used to extract PSFs below $S_n$ depend strongly 
on the models used to calculate input parameters or reference data used for normalization, a full uncertainty analysis is required before an evaluation and recommendation 
can be made. Such an analysis has only partially been performed within the CRP for 2 cases, $^{89}$Y and $^{139}$La, for which both NRF and Oslo data exist and the required 
information was available. 
Despite the effort however, we were unable to find problems with any of the available data sets and could not reconcile the data within the partial uncertainty ranges. 
The recommendation of a single data set is thus not possible, with the exception of $^{74}$Ge, where both the NRF and Oslo data give compatible results and thus both 
data files can be recommended to the community. The user is advised to consult the global model calculations described in Sec.~\ref{sec_th} which have reasonable predictive power or at least could set the systematic trend. 
On the other hand, new measurements of the PSF for at least $^{89}$Y and $^{139}$La are strongly encouraged, using possibly other methods and/or reactions as well, 
to help resolve the discrepancies observed in these nuclei and many others, and allow us to recommend a unique PSF data set.

\subsection{Theoretical PSF database}
\label{sec_database_th}

Calculations of $E1$, $M1$, and total dipole PSFs have been made using the D1M+QRPA+0lim and SMLO models described in Sec.~\ref{sec_th} for all 
the nuclei for which experimental PSF data sets exist and are compiled in the experimental database as mentioned in the previous section. Globally, 
both models have proven their capacity to reproduce the bulk of experimental data. As detailed in Sec. \ref{sec_th}, while the SMLO PSF has been fitted directly on photoabsorption cross section data, the D1M+QRPA+0lim PSFs have been only globally adjusted. 
The photoabsorption D1M+QRPA calculations for  both the $E1$ and $M1$ PSFs are stored in tabulated form in data files per element for energies from 0 to 30 MeV by step of 0.1~MeV. Similarly, the $E1$ SMLO PSFs have been estimated at 11 different temperatures and tabulated on the same energy grid in separate columns in the data files. 
The SMLO photoabsorption $M1$ PSFs, {\it i.e.} the sum of scissors and spin-flip modes as given by Eq.~\ref{eq_smlo_m1}, are also provided in table format for energies ranging between 0 and 30~MeV. In contrast, the low-energy enhancement for the deexcitation $M1$ PSF given by Eq.~\ref{eq_smlo_m2} in the SMLO model and for the $E1$ and $M1$ PSFs given by Eqs.~\ref{eq_qrpa1}-\ref{eq_qrpa2} in the D1M+QRPA+0lim are not included in the tabulated files.

\subsection{Experimentally unknown nuclei}
\label{sec_sys}

\begin{figure*}
\centering
\includegraphics[scale=0.5, angle = 0]{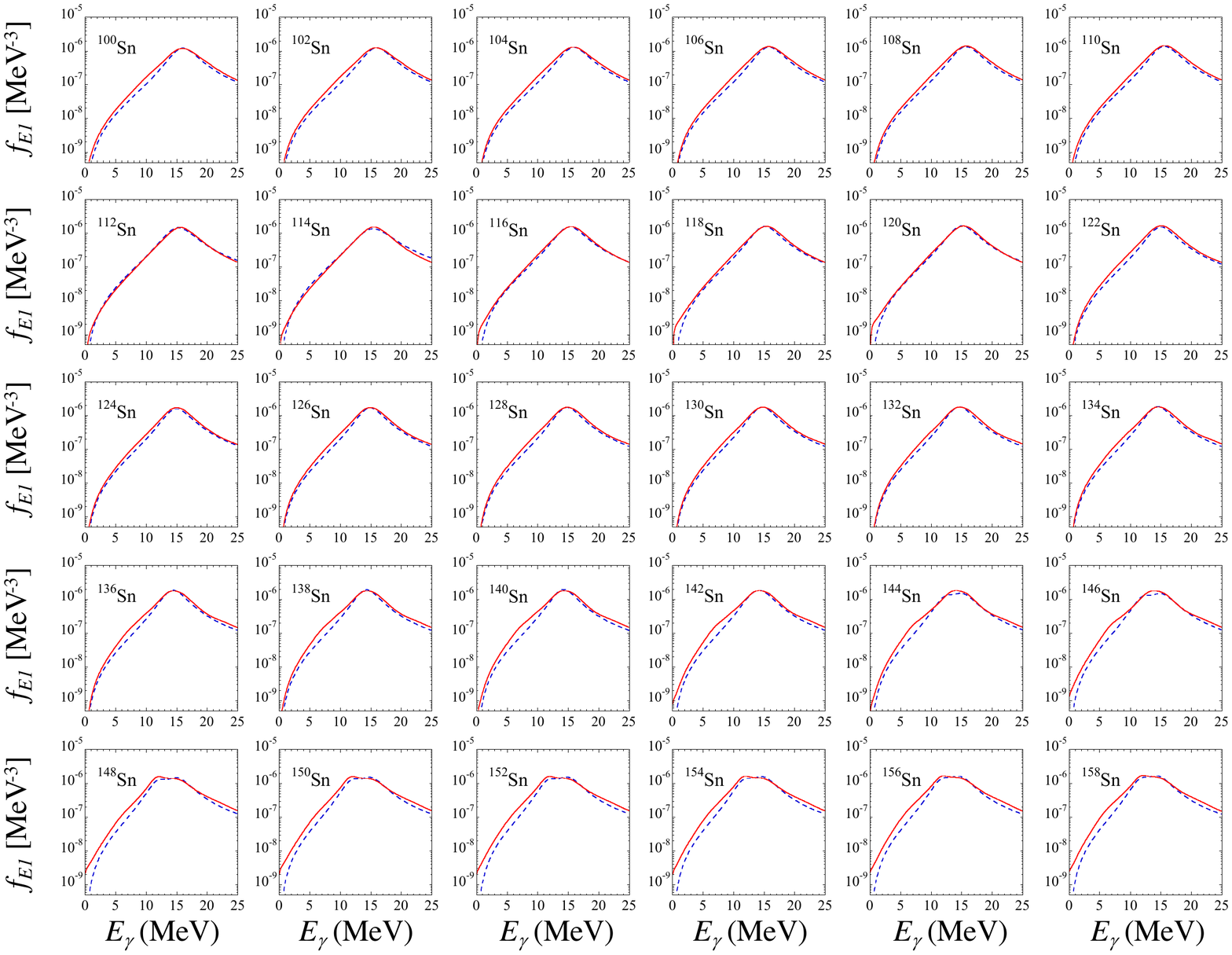}
\caption{(Color online) Comparison of the $E1$ photoabsorption PSFs obtained within the D1M+QRPA (red solid lines) and SMLO (blue dashed lines) models for Sn ($Z=50$) isotopes.} 
\label{fig_extra_E1_Z50}
\end{figure*}

\begin{figure*}
\centering
\includegraphics[scale=0.5, angle = 0]{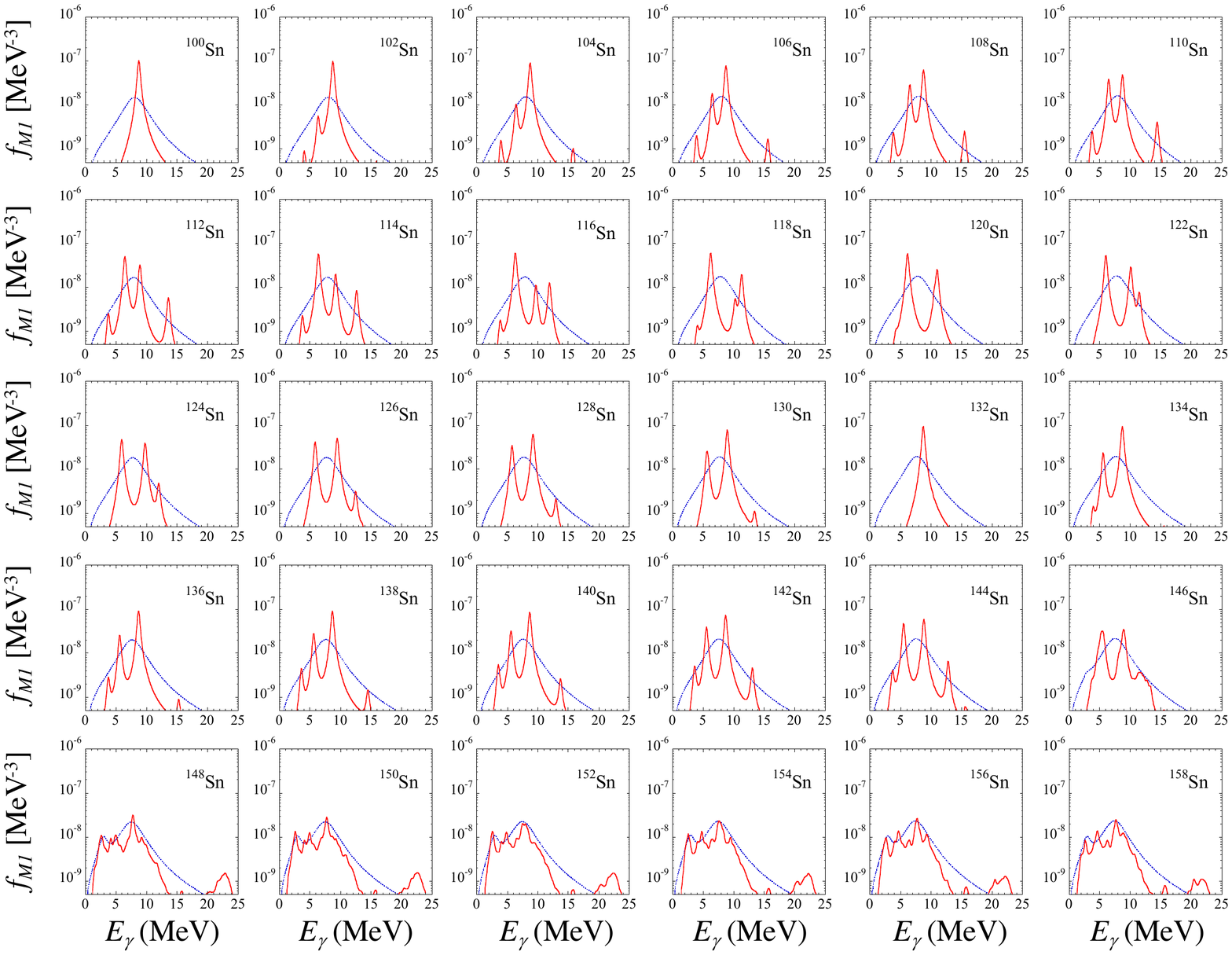}
\caption{(Color online) Comparison of the $M1$ photoabsorption PSFs obtained within the D1M+QRPA (red solid lines) and SMLO (blue dashed lines) models for Sn ($Z=50$) isotopes.} 
\label{fig_extra_M1_Z50}
\end{figure*}

\begin{figure*}
\centering
\includegraphics[scale=0.5, angle = 0]{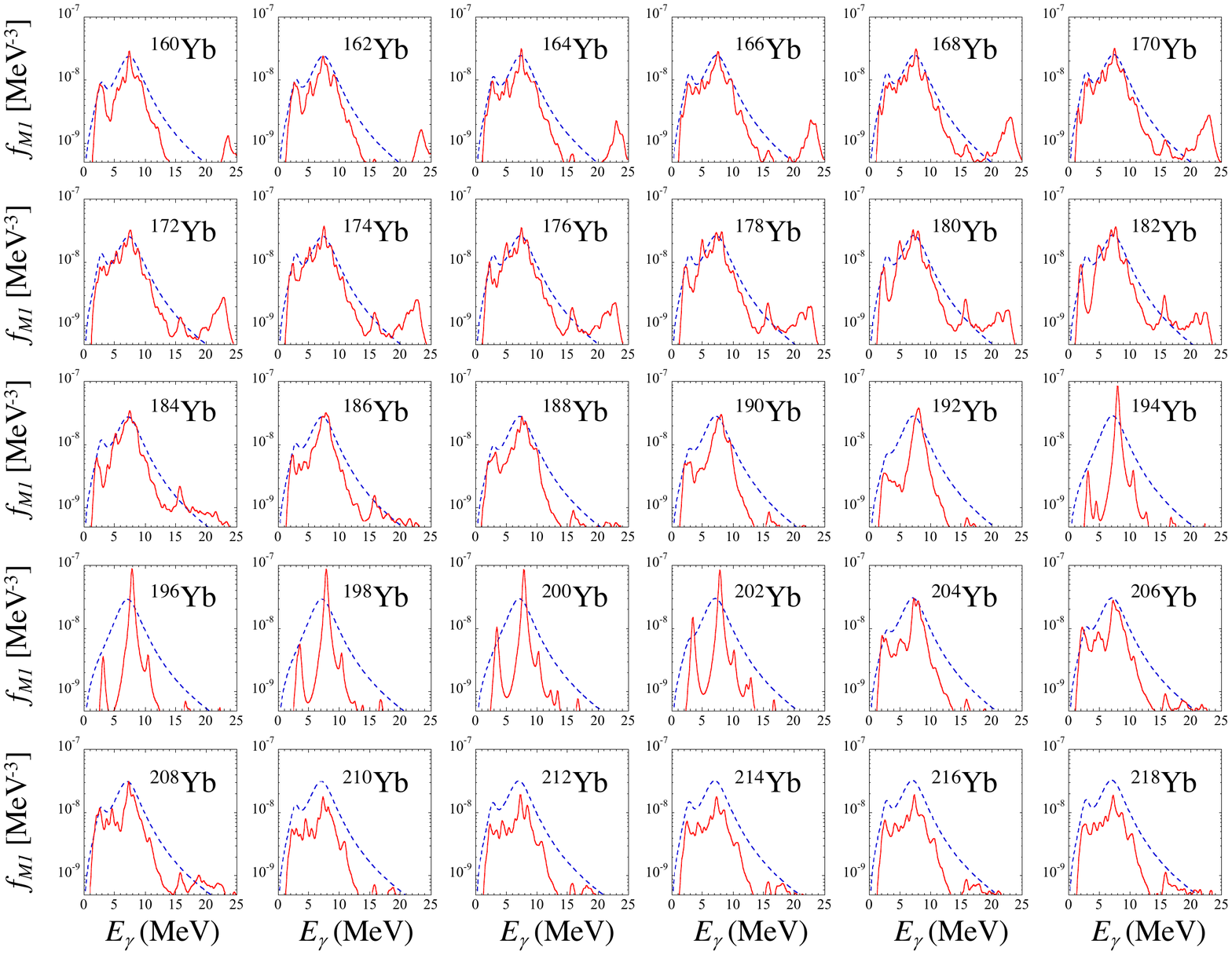}
\caption{(Color online) Comparison of the $M1$ photoabsorption PSFs obtained within the D1M+QRPA (red solid lines) and SMLO (blue dashed lines) models for  Yb ($Z=70$) isotopes.} 
\label{fig_extra_M1_Z70}
\end{figure*}

As shown in Sec.~\ref{sec_comp}, both the SMLO and D1M+QRPA+0lim models of PSF have shown their capability to reproduce experimental data of various sorts. Despite the variety of experimental data on a large number of different species, this information is available only for nuclei within the valley of stability. It is therefore of interest to see to what extent the two models recommended here predict similar PSFs when considering exotic neutron-rich or neutron-deficient nuclei. To illustrate such an extrapolation, we show in Figs.~\ref{fig_extra_E1_Z50} and \ref{fig_extra_M1_Z50} 
the $E1$ and $M1$, respectively, photoabsorption PSFs given by the SMLO (at $T=0$) and D1M+QRPA models for the even-even Sn isotopes. Note that the low-energy contributions corresponding to the non-zero temperature case for the $E1$ strength or the upbend component for the $M1$ strength, are not included in this comparison. The $E1$ predictions are seen to be rather similar from the most neutron-deficient to the most neutron-rich nuclei. In contrast, the $M1$ PSFs may somehow differ due to the presence of sharp peaks in the D1M+QRPA approach. However, the overall agreement is rather good as the maximum of the $M1$ spin-flip mode is at similar energies. Since the Sn isotopic chain is dominantly spherical, we also show in Fig.~\ref{fig_extra_M1_Z70} a similar comparison of $M1$ PSFs between the SMLO and D1M+QRPA models but this time for the rather deformed isotopic chain of Yb ($Z=70$). Globally, both the scissors  and spin-flip modes give rise to a total $M1$ PSF that is quite similar for both models, even for the most neutron-deficient $^{160}$Yb and most neutron-rich $^{218}$Yb nuclei. Both SMLO and D1M+QRPA $M1$ strengths are not fully independent since the $A$ and $\beta_2$ dependence in the $M1$ SMLO model has been inspired by the D1M+QRPA systematics \cite{Goriely18b} (see Sec.~\ref{sec_th_smlo}). Figs.~\ref{fig_extra_M1_Z50}-\ref{fig_extra_M1_Z70} illustrate the efficiency of the SMLO to describe in a simple phenomenological model the global trends obtained by the microscopic D1M+QRPA model.

\begin{figure}
\centering
\includegraphics[scale=0.29, angle = 0]{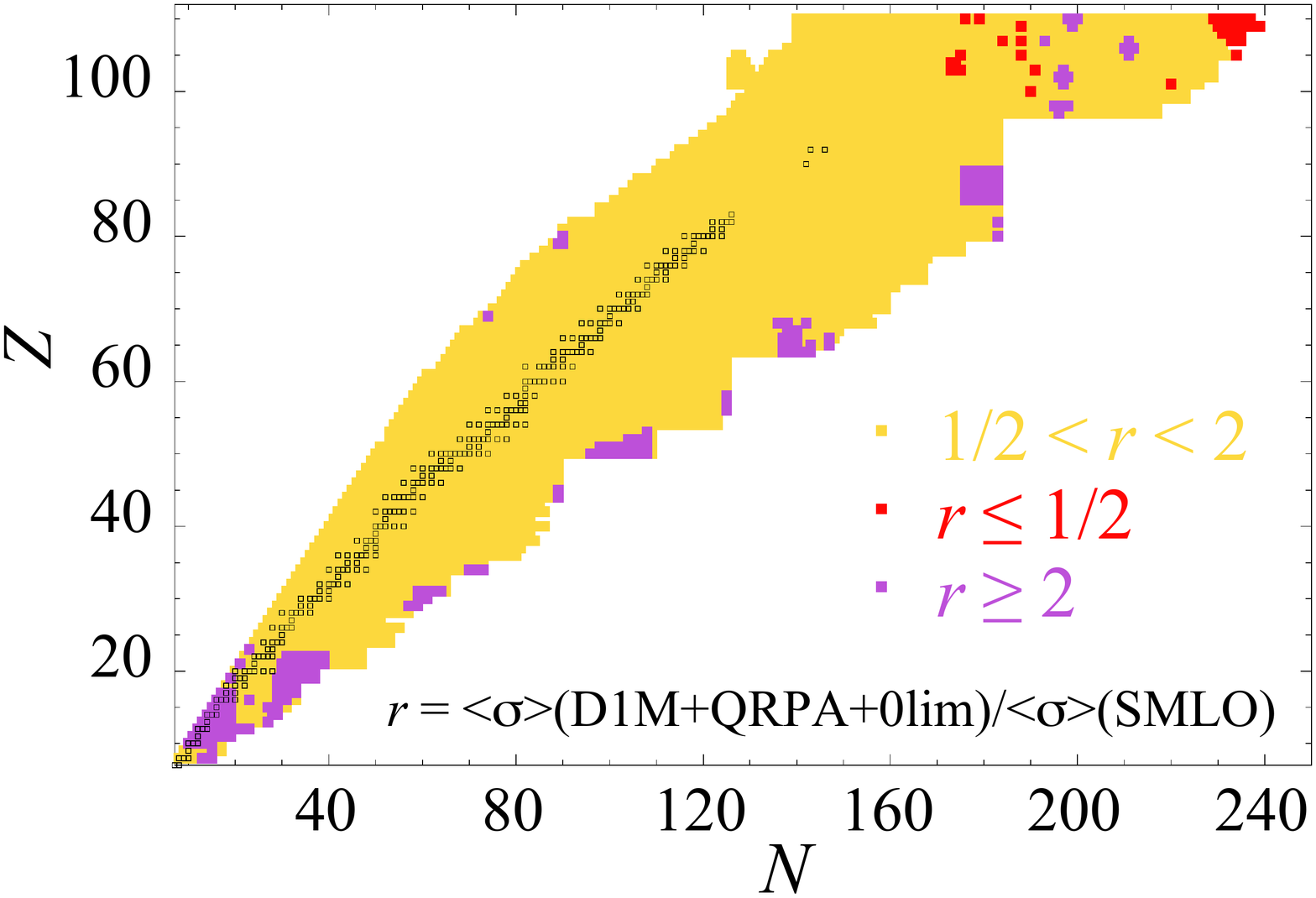}
\caption{(Color online) Representation in the ($N$,$Z$) plane of the ratio $r$ of the MACS obtained with the TALYS reaction code and the D1M+QRPA+0lim PSFs to those obtained with SMLO PSFs at  $T=10^9$~K. All nuclei between the D1M proton and neutron drip lines are included. The open squares indicate stable nuclei.} 
\label{fig_sigv_D1M_vs_SMLO_NZ}
\end{figure}

To further illustrate the potential impact of the differences between the SMLO and D1M+QRPA+0lim PSFs, we 
compare in Fig.~\ref{fig_sigv_D1M_vs_SMLO_NZ} the radiative neutron capture MACS obtained with both models. The MACS are calculated with the TALYS reaction code at a temperature of $T=10^9$~K relevant in nucleosynthesis applications. 
More specifically, the ratio between both MACS are plotted and seen to lie within a factor of 2 for the vast majority of nuclei lying between 
the proton and the neutron drip lines.

The database contains $E1$ and $M1$ PSFs calculated with the D1M+QRPA+0lim model for 7380 nuclei with $8\le Z\le 110$ from the proton to the neutron drip lines, and correspondingly, SMLO $E1$ and $M1$ PSFs are available for 8980 nuclei with $8\le Z\le 124$. The data files have the same format as described in Sec.~\ref{sec_database_th}.

\subsection{Database interface}
\label{sec_database_interf}

All the PSF data files contained in the IAEA PSF database can be accessed through an online interface at \textit{www-nds.iaea.org/PSFdatabase}. The interface offers the possibility to filter the database by nucleus $(Z,A)$ and method.
The selected data files, experimental or theoretical can be downloaded and plotted on-the-fly. The entire content of the experimental or theoretical database can also be downloaded in 
a compressed folder.

\section{Conclusions }
\label{sec_conc}

The present contribution summarizes a coordinated and systematic effort to compile and assess all the existing experimental data on PSFs extracted from many different experimental techniques at $\gamma$-ray energies below about 20~MeV. Further, two PSF models, one of them purely phenomenological, the other one semi-microscopic, have been systematically compared with all experimental data available and are recommended to globally describe the $E1$ and $M1$ PSFs for nuclei between the proton and neutron drip lines.

The CRP has produced a complete and comprehensive compilation of experimental PSFs.  The different methods that are used to extract PSF data from 
photon-,  neutron, and charged-particle-induced reactions have been described and assessed. Photonuclear PSFs have been recommended on the basis of 
a separate re-evaluation of photonuclear data performed within the CRP leading to a new IAEA photonuclear data library \cite{Kawano19}. For the other methods used to extract PSFs below the neutron separation 
energy $S_n$, where there is only one measurement available, then that data set  is recommended to the user. In cases where more than one different method was used to measure the PSFs, however, it was not possible to evaluate systematically the data due to lack of sufficient information on the model-dependent uncertainties that are inherent to all the considered methods. 
As all of these methods depend strongly on model calculations for normalization and/or disentanglement from level densities, 
a full uncertainty analysis taking into account the impact of the various models
and parameter uncertainties is a prerequisite for an evaluation to be performed. In the absence of such a full uncertainty analysis, or in the presence of only a partial analysis as the one mentioned below, we are unable to recommend experimental PSF data below the neutron separation energies in cases where the different methods provide discrepant results.

The partial uncertainty analysis  performed in both the 
NRF and Oslo methods  on two cases, $^{89}$Y and $^{139}$La, has led to the realization that there is a problem with the data of $^{139}$La so this nuclide needs to be re-measured while for $^{89}$Y both methods agree within the uncertainties for most part of the energy region, except for the energies around 6.5 MeV. It is plausible that if a full uncertainty analysis had been performed for $^{89}$Y, both data sets would have been in agreement within the uncertainties in the whole energy range. One major conclusion of this work is that proper consideration and documentation of all the sources of uncertainties including model-dependent ones, is needed when comparing the different experimental techniques and trying to recommend the best data.  It is therefore a strong recommendation of this CRP that a  standardisation of the treatment 
of uncertainties and model dependencies for all the methods affected by them should be a priority for the scientific community involved actively in measurements of PSFs below $S_n$. Establishing and promoting 'best practices' in extracting PSFs from raw data will allow the present and future generation of experimentalists to produce consistent data that can be fully exploited in basic and applied sciences.

With the existing information at hand, we can at best propose  new measurements for cases with discrepant data, using more than one method if possible and combining with a full uncertainty analysis,  to allow for verification of the results and a better understanding of the methods and the part of the electromagnetic response they probe.

Global semi-microscopic (D1M+QRPA+0lim) and empirical (SMLO) models have been developed for electric and magnetic dipole PSFs. Both models have been compared with the existing experimental data as well as other indirect methods based on neutron capture spectra (singles or coincident ones). The global models have also been validated against measured average radiative widths and Maxwellian-averaged cross sections. As a result, they are expected to perform reasonably well when extrapolated to unknown nuclei and could also be used as a guide in cases where the existing experimental data are discrepant. However, it is clear from the global comparisons that these models also have some limitations and that going beyond the QRPA is the direction to take in the future.

The PSF database, complete with compilation, recommendations and global calculations, is readily available to the broad scientific community
at the online IAEA server (URL:www-nds.iaea.org/PSFdatabase).

\begin{acknowledgement}
 This work was performed within the IAEA CRP on ``Updating the Photonuclear Data Library and Generating a Reference Database for Photon Strength Functions'' (F41032). Contributors from the following institutes are grateful to the IAEA for financial support within the CRP: CIAE, KAERI, IFIN-HH, Taras Shevchenko National University, Lomonosov Moscow State University, Centre for Energy Research (HAS), Charles University (Prague), and iThemba LABS.
We wish to thank A.J. Koning, P. Oblo\v{z}insk\'y, A.C. Larsen, K.L. Malatji, A. Tonchev, M. Guttormsen, V. Ingeberg, P. von Neumann-Cosel, G. M. Tveten, T. Renstr\o m, R. Capote, O. Gorbachenko, K. Solodovnyk and Yuan Tian for fruitful discussions and for providing relevant material and data.
This work was also partially supported by the Fonds de la Recherche Scientifique - FNRS and the Fonds Wetenschappelijk Onderzoek - Vlaanderen (FWO) under the EOS Project No O022818F,  
by the National Research Foundation of South Africa Grant No 118846,
by the Czech Science Foundation Grant No. 19-14048S, 
the Research Council of Norway under the Project Grant No 263030, 
and the EU FP7 CHANDA project No 605203  for (n,$\gamma$) measurements. 
H.U. acknowledges support extended to the PHOENIX collaboration for the IAEA CRP by the Premier Project of Konan University.
\end{acknowledgement}

\bibliographystyle{epj}
\bibliography{iaea_psf_resub_final}
\end{document}